\documentclass[aps,prd,superscriptaddress,amsmath,amsfont,amssymb,preprint,nofootinbib]{revtex4}
\pdfoutput=1
\allowdisplaybreaks
\usepackage{graphicx}
\usepackage{ bbold }
\usepackage{subfigure}
\usepackage{slashed}
\usepackage{mathtools}
\usepackage[english]{babel}
\newcommand{\op}{{\cal O}}

\newcommand{\C}{{\cal C}}
\newcommand{\Q}{{\cal Q}}
\newcommand{\todo}[1]{{\color{red} \ifmmode\else[todo]\fi #1}}
\usepackage[usenames,dvipsnames]{xcolor}
     \definecolor{hgreen}{rgb}{0,.3,0}
     \definecolor{hred}{rgb}{.3,0,0}
     \definecolor{hblue}{rgb}{0,0,.3}
     \definecolor{LightGray}{gray}{0.95}
\usepackage[backref=page,
            colorlinks=true,
            linkcolor=hblue,
            citecolor=hgreen,
            filecolor=hblue,
            urlcolor=hred]{hyperref}
\usepackage{fancyvrb}

\renewcommand*{\backref}[1]{}

\newcommand{\re}[0]{\mathrm{Re}\,}

\newcommand{\beq}{\begin{equation} }
\newcommand{\eeq}{\end{equation}} 
\newcommand{\bi}{\begin{itemize} }
\newcommand{\ei}{\end{itemize} }

\definecolor{Red}{rgb}{1.,0.,0.}
\definecolor{Grn}{rgb}{0.,0.75,0.}
\definecolor{Blu}{rgb}{0.,0.,1.}

\DeclareMathOperator{\Tr}{Tr}

\DeclareMathOperator{\Re1}{Re}
\DeclareMathOperator{\Im1}{Im}

\newcommand{\ilrpartial}{\negmedspace \stackrel{\leftrightarrow}{i \partial}}

\definecolor{pan624}{rgb}{0.482,0.635,0.588} 
\definecolor{pan576}{rgb}{0.412,0.569,0.231} 
\definecolor{pan129}{rgb}{0.961,0.812,0.278}
\definecolor{pan5405}{rgb}{0,0.129,0.278} 
\definecolor{shadecolor}{rgb}{0.482,0.635,0.588}
\definecolor{mygray}{HTML}{666666}
\definecolor{x11steelblue}{HTML}{4682B4}
\definecolor{x11firebrick}{HTML}{B22222}
\definecolor{x11forestgreen}{HTML}{228B22}

\newcommand{\CE}{CE$\nu$NS~}
\usepackage{tikz}
\tikzstyle{every picture}+=[remember picture]
\usetikzlibrary{shapes.geometric}
\usetikzlibrary{calc}
\usetikzlibrary{decorations.markings}
\usetikzlibrary{decorations.text}
\usetikzlibrary{patterns}
\usetikzlibrary{backgrounds}
\usetikzlibrary{positioning}
\tikzstyle arrowstyle=[scale=2]
\tikzstyle directed=[postaction={decorate,decoration={markings,
		mark=at position 0.6 with {\arrow[arrowstyle]{>}}}}]
\tikzstyle rarrow=[postaction={decorate,decoration={markings,
		mark=at position 0.999 with {\arrow[arrowstyle]{>}}}}]
		
\graphicspath{{./figs/}}

\newcommand{\ncdot}{\negthinspace \cdot \negthinspace}

\begin{document}

\title{Non-standard neutrino interactions and low energy experiments}

\def\Cincy{Department of Physics, University of Cincinnati, Cincinnati, Ohio 45221,USA}
\def\CERN{CERN, Theory Division, CH-1211 Geneva 23, Switzerland}
\def\ucsc{Santa Cruz Institute for Particle Physics, Santa Cruz, CA 95064, USA}

\author{\textbf{Wolfgang Altmannshofer}}
\email{waltmann@ucsc.edu}
\affiliation{\ucsc}

\author{\textbf{Michele Tammaro}}
\email{tammarme@mail.uc.edu}
\affiliation{\Cincy}

\author{\textbf{Jure Zupan}} 
\email{zupanje@ucmail.uc.edu}
\affiliation{\Cincy}

\date{\today}

\begin{abstract}
We formulate an Effective Field Theory (EFT) for Non Standard neutrino Interactions (NSI) in elastic scattering with light quarks, leptons, gluons and photons, including all possible operators of dimension 5, 6 and 7. We provide the expressions for the cross sections in coherent neutrino-nucleus scattering and in deep inelastic scattering. Assuming single operator dominance we constrain the respective Wilson coefficient using the measurements by the COHERENT and CHARM collaborations. We also point out the constraining power of future elastic neutrino-nucleus scattering experiments. Finally, we explore the implications of the bounds for SMEFT operators above the electroweak breaking scale. 
\end{abstract}

\pacs{--pacs--}

\maketitle
\tableofcontents


\section{Introduction}
\label{sec:Intro}
In the SM the neutrinos interact with matter through exchanges of $W$ and $Z$ bosons. In addition, in the presence of new physics, the neutrinos could interact with matter via new mediators. Such Non-Standard neutrino Interactions (NSI) were contemplated already 40 years ago by L.~Wolfenstein  in the seminal paper  on neutrino oscillations in matter~\cite{Wolfenstein:1977ue}. 
Since then the NSI were studied extensively, but with a strong focus on neutrino oscillations, see, e.g.,  \cite{Miranda:2015dra,Abe:2011sj,Fukuda:1998mi,GonzalezGarcia:2007ib,Bergmann:1999rz,Coloma:2017ncl,Flores:2018kwk,Esteban:2018ppq,Denton:2018xmq,Farzan:2017xzy}. 
The bounds from neutrino oscillations are limited in scope, since they are sensitive only to a subset of possible NSI. 
The common NSI effective Lagrangian 
relevant for neutrino oscillations contains only dimension 6 operators, see, e.g., \cite{Ohlsson:2012kf},
\beq \label{eq:LNSI}
{\cal L}_{\rm NSI}' = \frac{G_F}{\sqrt{2}} \sum_{f,\alpha,\beta} \left(\bar \nu_\alpha \gamma_\mu P_L \nu_\beta \right) \left(\varepsilon_{\alpha\beta}^{fV} \bar f \gamma^\mu f + \varepsilon_{\alpha\beta}^{fA} \bar  f \gamma^\mu \gamma_5 f \right).
\eeq 
The dimensionless coefficients $\varepsilon_{\alpha\beta}^{fV},  \varepsilon_{\alpha\beta}^{fA}$ parametrize the strength of the NSI relative to the SM weak force, controlled by the Fermi constant, $G_{\rm F}\simeq1.167 \times 10^{-5} {\rm~GeV}^{-2}$. The indices $\alpha$ and $\beta$ run over the three neutrino flavours, and $f$ over light charged fermions, $f=e, u,d,s$.

Eq.~\eqref{eq:LNSI} does contain all possible dimension 6 NSI operators. Still, these are not all the possible NSI. In these paper we list a complete basis of NSI operators up to and including dimension 7.
The  additional dimension 5 and dimension 7 operators either do not contribute to neutrino oscillations, because they lead to zero forward scattering matrix elements, or give contributions that are additionally suppressed by neutrino masses (tensor operators may be relevant for neutrino oscillations in polarized matter \cite{Bergmann:1999rz}). The dimension 5 and 7 NSI can be probed through neutrino inelastic scattering, and by precise measurements of solar neutrino scattering rates. 

A qualitatively new set of NSI probes is opening up through the coherent neutrino scattering measurements. The first measurement of coherent neutrino scattering on nuclei was achieved by the COHERENT collaboration roughly a year ago, Ref.~\cite{Akimov:2017ade}.
This result, and similar measurements in the future,
now make it possible to probe a wide variety of NSI at low momenta exchanges.

 The aim of present manuscript is to perform a systematic study of such NSI. We assume the NSI are described by an Effective Field Theory (EFT), i.e., that the new mediators are heavier than about ${\mathcal O}(100{\rm~MeV})$. In the analysis we include operators up to and including dimension 7, covering all possible chirality structures for neutrino currents. Our work extends previous NSI analyses of coherent neutrino scattering, where a subset of EFT operators were discussed \cite{Farzan:2018gtr,Billard:2018jnl,AristizabalSierra:2018eqm,Kosmas:2017tsq,Dent:2017mpr,Liao:2017uzy,Dent:2016wcr,Lindner:2016wff}. For projections of bounds on NSI from neutrino scattering in DUNE see  \cite{Falkowski:2018dmy,Bischer:2018zcz}, while the potential of dark matter direct detection experiments for probing NSI using solar neutrinos was discussed in \cite{Harnik:2012ni,Cadeddu:2018izq,Huang:2018nxj,Shoemaker:2018vii,AristizabalSierra:2017joc,Gonzalez-Garcia:2018dep,Dutta:2017nht,Bertuzzo:2017tuf,Dent:2016wcr,Cerdeno:2016sfi,Coloma:2014hka,Pospelov:2013rha,Pospelov:2012gm}. For the potential of superbeam experiments to probe NSI, see \cite{Kopp:2007ne}. For bounds on the neutrino dipole moment portal to heavy right-handed neutrino, see \cite{Magill:2018jla}.

The paper is organized as follows. In Section \ref{sec:nuEFT} we formulate the EFT for coherent elastic neutrino-nucleus scattering (CE$\nu$NS\footnote{While not all of the NSI scatterings will be coherently enhanced we keep the, by now standard, \CE terminology.}) in the presence of NSI. The EFT valid at $\mu\sim 2$ GeV in which neutrinos couple to light quarks, gluons and photons, is nonperturbatively matched onto an EFT with nonrelativistic nucleons in Section \ref{sec:Basis}, with the resulting \CE cross sections given in Section \ref{sec:Connection}. 
 Section \ref{sec:oscillation} reviews bounds on NSI from neutrino oscillations, and  Section \ref{sec:DISintro} the deep inelastic scattering (DIS) probes of NSI, while Section \ref{sec:Constraints} contains our numerical analysis.
In Section \ref{sec:ew:matching} we explore the connection with physics above the scale of electroweak symmetry breaking, and draw our conclusions in Section \ref{sec:conclusions}. Appendix \ref{app:formfactors} contains the definitions of nucleon form factors, Appendix \ref{app:plots} the predictions for the differential rates for various NSI operators, and Appendix \ref{app:ratios} the numerical predictions for differential rates as functions of NSI Wilson coefficients.

\section{Operator basis for NSI}
\label{sec:nuEFT}
We are interested in the experiments where momenta exchanges are $q\lesssim {\mathcal O}(100 {\rm MeV})$, and thus well below the electroweak scale. 
The interactions of neutrinos with matter, i.e., with quarks, gluons, photons, electrons and muons, are described by an effective Lagrangian, obtained by integrating out the heavy degrees of freedom. These are the heavy SM particles: $t,b,c$ quarks, $\tau$ lepton, $W$ and $Z$ bosons and the Higgs, as well as any heavy new physics particles. 

The interaction Lagrangian for $\nu_\alpha\to \nu_\beta$ transition  is given by a sum of non-renormalizable operators,
\begin{equation}
\label{eq:lightDM:Lnf5}
{\cal L}_{\nu_\alpha\to \nu_\beta}=\sum_{a,d=5,6,7}
\hat \C_{a}^{(d)} {\cal Q}_a^{(d)}+{\rm h.c.}
+\cdots, 
\qquad {\rm where}\quad 
\hat \C_{a}^{(d)}=\frac{\C_{a}^{(d)}}{\Lambda^{d-4}}\,.
\end{equation}
Here the $\C_{a}^{(d)}$ are dimensionless Wilson coefficients, while
$\Lambda$ can be identified, for ${\mathcal O}(1)$ couplings, with the mass of the new physics mediators. We consider a complete basis of EFT operators up to and including dimension seven.  The sum in \eqref{eq:lightDM:Lnf5} runs over operator dimensions, $d=5,6,7$, and operator labels, $a$, while in the notation we suppress the dependence on neutrino flavors $\alpha, \beta$. 
The renormalization scale is fixed to $\mu=2$ GeV, unless specified otherwise. 
  
We first write down the full basis of EFT operators assuming neutrinos are Dirac fermions, and then comment below on what changes are needed, if neutrinos are Majorana.  We use four-component notation, following the conventions of Ref. \cite{Dreiner:2008tw}. There is one
dimension-five operator for each $\nu_\alpha\to \nu_\beta$ transition,\footnote{We use the phase convention in which the QED covariant derivative is $D_\mu\psi=(\partial_\mu +i e Q_\psi A_\mu)\psi$, with $Q_\psi$ the electric charge of $\psi$. For Majorana neutrinos, for $\alpha=\beta$, one needs to include in the definitions of the operators an extra factor of $1/2$.}
\begin{equation}
\label{eq:dim5:nf5:Q1:light}
{\cal Q}_{1}^{(5)} = \frac{e}{8 \pi^2} (\bar \nu_\beta \sigma^{\mu\nu}P_L\nu_\alpha)
 F_{\mu\nu} \,,
\end{equation}
where $F_{\mu\nu}$ is the electromagnetic field strength tensor.  The dimension-six operators
are
\begin{align} 
{\cal Q}_{1,f}^{(6)} & = (\bar \nu_\beta \gamma_\mu P_L \nu_\alpha) (\bar f \gamma^\mu f),
&{\cal Q}_{2,f}^{(6)} & = (\bar \nu_\beta \gamma_\mu P_L \nu_\alpha)(\bar f \gamma^\mu \gamma_5 f)\,.\label{eq:dim6EW:Q1Q2:light}
\end{align}
The basis of dimension seven operators can be chosen such that there are four operators coupling neutrinos to photon or gluon field strengths, 
\begin{align}
 \label{eq:dim7:Q1Q2:light}
 {\cal Q}_1^{(7)} & = \frac{\alpha}{12\pi} (\bar \nu_\beta P_L \nu_\alpha) F^{\mu\nu}
 F_{\mu\nu},
& {\cal Q}_2^{(7)} & = \frac{\alpha}{8\pi} (\bar \nu_\beta P_L \nu_\alpha) F^{\mu\nu}\widetilde
 F_{\mu\nu},
\\
 \label{eq:dim7:Q3Q4:light} 
{\cal Q}_3^{(7)} & = \frac{\alpha_s}{12\pi} (\bar \nu_\beta P_L \nu_\alpha) G^{a\mu\nu}
 G_{\mu\nu}^a,
& {\cal Q}_4^{(7)} & = \frac{\alpha_s}{8\pi} (\bar \nu_\beta P_L \nu_\alpha) G^{a\mu\nu}\widetilde
 G_{\mu\nu}^a\,, 
\end{align}
three types of operators with chirality-flipping quark currents, 
\begin{align}
\label{eq:dim7:Q5Q6:light} 
{\cal Q}_{5,f}^{(7)} & = m_f (\bar \nu_\beta P_L \nu_\alpha)( \bar f f)\,, 
&{\cal Q}_{6,f}^{(7)} & = m_f (\bar \nu_\beta P_L \nu_\alpha) (\bar f i \gamma_5 f)\,, 
 \\
{\cal Q}_{7,f}^{(7)} & = m_f (\bar \nu_\beta \sigma^{\mu\nu} P_L \nu_\alpha) (\bar f \sigma_{\mu\nu} f)\,, 
\label{eq:dim5:Q7:light} 
\end{align}
and four types of operators with additional derivatives on the neutrino currents,
\begin{align}
\label{eq:dim7:Q8Q9:light} 
{\cal Q}_{8,f}^{(7)} & =  (\bar \nu_\beta  \negmedspace \stackrel{\leftrightarrow}{i \partial}_\mu \negmedspace P_L \nu_\alpha)( \bar f \gamma^\mu f)\,, 
&{\cal Q}_{9,f}^{(7)} & = (\bar \nu_\beta \negmedspace \stackrel{\leftrightarrow}{i \partial}_\mu \negmedspace P_L \nu_\alpha) (\bar f \gamma^\mu \gamma_5 f)\,, 
 \\
{\cal Q}_{10,f}^{(7)} & = \partial_\mu (\bar \nu_\beta \sigma^{\mu\nu} P_L \nu_\alpha) (\bar f \gamma_\nu f)\,, 
&{\cal Q}_{11,f}^{(7)} & = \partial_\mu (\bar \nu_\beta \sigma^{\mu\nu} P_L \nu_\alpha) (\bar f \gamma_\nu \gamma_5 f)\,. 
\label{eq:dim5:Q10Q11:light} 
\end{align}
Here $G_{\mu\nu}^a$ is the QCD field strength
tensor, $\widetilde G_{\mu\nu} = \frac{1}{2}\varepsilon_{\mu\nu\rho\sigma}
G^{\rho\sigma}$ its dual (and similarly for QED, $\widetilde F_{\mu\nu} = \frac{1}{2}\varepsilon_{\mu\nu\rho\sigma}
F^{\rho\sigma}$), and $a=1,\dots,8$ the adjoint color
indices.
 The fermion label, $f=u,d,s,e,\mu$, denotes the light quarks, electrons or muons, while $(\bar \nu \negthickspace \stackrel{\leftrightarrow}{i \partial}_\mu \negthickspace\nu)=(\bar \nu {i \partial}_\mu \nu)- (\bar \nu \negthickspace\stackrel{\leftarrow}{i \partial}_\mu \negthickspace\nu)$. We assume flavor conservation for charged fermions, while we do allow changes of neutrino flavor. 

For Dirac neutrinos the dimension 5 and dimension 7 operators, Eq. \eqref{eq:dim5:nf5:Q1:light} and Eqs. \eqref{eq:dim7:Q1Q2:light}-\eqref{eq:dim5:Q10Q11:light}, have a chirality flipping neutrino current. An incoming left-handed neutrino of flavor $\nu_\alpha$ is converted to a right-handed neutrino of flavor $\nu_\beta$. In contrast, the dimension 6 operators, Eq. \eqref{eq:dim6EW:Q1Q2:light}, preserve the chirality of the incoming neutrino. For Dirac neutrinos there are then two additional dimension 6 operators, ${\cal Q}_{1,q}^{(6)'}, {\cal Q}_{2,q}^{(6)'}$, obtained from \eqref{eq:dim6EW:Q1Q2:light} through $P_L\to P_R$ replacements. These operators cannot be well tested in neutrino experiments, since the production of right-handed neutrinos through SM weak interactions is neutrino mass suppressed. We therefore do not consider the operators ${\cal Q}_{1,q}^{(6)'}, {\cal Q}_{2,q}^{(6)'}$ further in our analysis. 

In the case of Majorana neutrinos the dimension 5 and dimension 7 operators, Eq. \eqref{eq:dim5:nf5:Q1:light} and Eqs. \eqref{eq:dim7:Q1Q2:light}-\eqref{eq:dim5:Q7:light},  violate lepton number by two units (note that we use the conventions of Ref. \cite{Dreiner:2008tw} also for Majorana neutrinos). Furthermore, for a Majorana neutrino the operators ${\cal Q}_{1}^{(5)}$ in \eqref{eq:dim5:nf5:Q1:light}, ${\cal Q}_{7,f}^{(7)} $ in 
\eqref{eq:dim5:Q7:light}, and  ${\cal Q}_{10,f}^{(7)}, {\cal Q}_{11,f}^{(7)}$ in  
\eqref{eq:dim5:Q10Q11:light} vanish identically for $\alpha=\beta$, and thus only mediate flavor changing transitions. 
Finally, for $\alpha=\beta$ we include in the definitions of the operators an extra factor of $1/2$ to compensate for the additional Wick contraction so that our results for cross sections and the bounds on Wilson coefficients can be used without change (cf. App A of Ref. \cite{Bishara:2017nnn} for explicit normalization of such operators, albeit for DM interactions).

Note that in general the above operators are not Hermitian, and thus can have complex Wilson coefficients, $\C_a^{(d)}$. The exception are dimension 6 operators with $\alpha=\beta$, in which case the operators are Hermitian, and thus the corresponding Wilson coefficients are real (for these operators the ``h.c.'' in Eq. \eqref{eq:lightDM:Lnf5} should be dropped).

\begin{figure}
\includegraphics[scale=0.68]{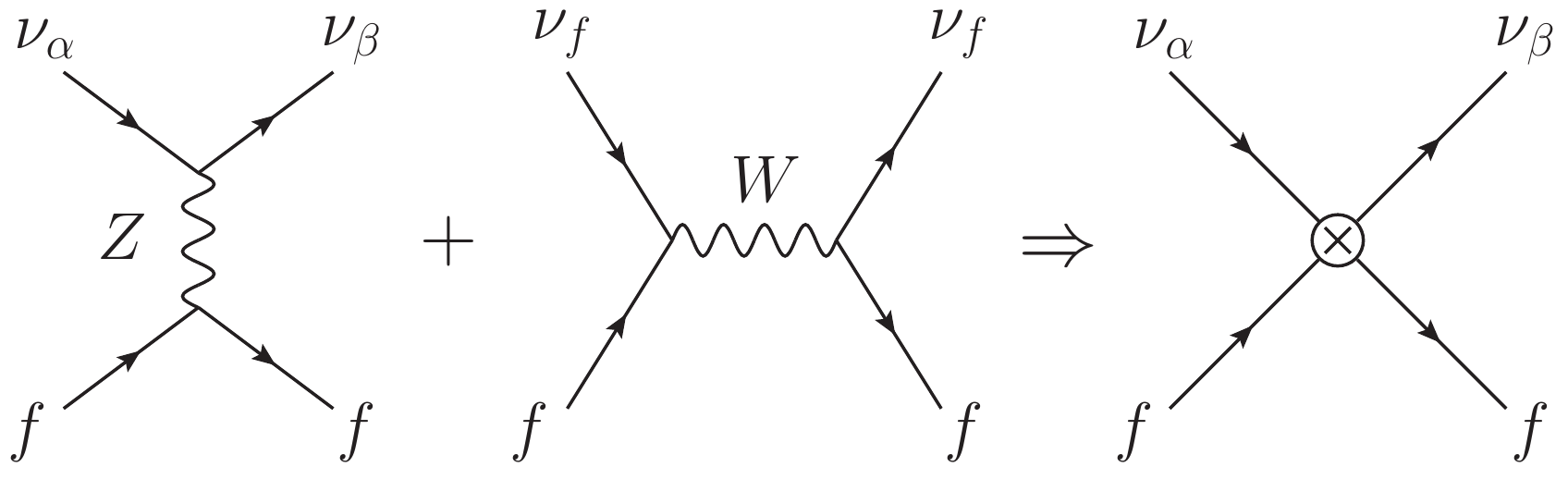}
\caption{Integrating out the $Z$ and $W$ bosons at tree level NC and CC (left) generates the effective four-fermions interaction (right). The blob vertex indicates an operator insertion of ${\cal Q}_{1,f}^{(6)}$ and ${\cal Q}_{2,f}^{(6)}$. 
}
         \label{fig:ZWeft}
\end{figure}

Note that the SM neutrino interactions with quarks are also described by the effective Lagrangian \eqref{eq:lightDM:Lnf5}, though not all the operators are generated.  
The SM neutral currents (NC), i.e., due to the tree level $Z$ exchanges, and the SM charged currents (CC), due to the tree level $W$ exchanges, generate the operators ${\cal Q}_{1,f}^{(6)}$ and ${\cal Q}_{2,f}^{(6)}$, see Fig. \ref{fig:ZWeft}. 
 Integrating out the $Z$ and $W$ bosons gives for the Wilson coefficient relevant for neutrino scattering on matter, i.e., on light quarks and electrons,
\begin{align} \label{eq:c6udsSM}
\hat \C_{1,u(d,s)}^{(6)}\big|_{\rm SM}&= \mp \frac{G_F}{\sqrt{2}}\left(1 - \frac{8(4)}{3}s_W^2\right)\delta_{\alpha\beta}, 
&\hat \C_{2,u(d,s)}^{(6)}\big|_{\rm SM}&= \pm \frac{G_F}{\sqrt{2}} \delta_{\alpha\beta}
 \\
 \hat \C_{1,e}^{(6)}\big|_{\rm SM}&=\frac{G_F}{\sqrt{2}}\left( (1 - 4 s_W^2)\delta_{\alpha\beta} - 2\delta_{\alpha e}\delta_{\beta e} \right) , 
 &\hat \C_{2,e}^{(6)}\big|_{\rm SM}&=-\frac{G_F}{\sqrt{2}} \left(\delta_{\alpha\beta} - 2\delta_{\alpha e}\delta_{\beta e} \right) , \label{eq:c6eSM}
\end{align}
where $s_W^2 \equiv \rm{sin}^2\theta_W \simeq 0.2223$ with $\theta_{\rm{W}}$ the weak mixing angle.
The second terms in Eq. \eqref{eq:c6eSM} are due to CC, cf. Fig.~\ref{fig:ZWeft}. 
In the presence of NSI
the above SM Wilson coefficients are modified to
\beq \label{eq:c6NSI}
\hat \C_{1(2),f}^{(6)}= \hat \C_{1(2),f}^{(6)}\big|_{\rm SM}+  \hat \C_{1(2),f}^{(6)}\big|_{\rm NSI}, \quad \text{~~with~~~~}
 \hat \C_{1(2),f}^{(6)}\big|_{\rm NSI}= \frac{G_F}{\sqrt{2}}\varepsilon_{\alpha\beta}^{fV(A)},
\eeq
where in the last equality we used the $\varepsilon$ notation of 
the NSI Lagrangian, Eq.~\eqref{eq:LNSI}. 

In the SM the dimension 5 and dimension 7 EFT operators, Eq.~\eqref{eq:dim5:nf5:Q1:light} and Eqs. \eqref{eq:dim7:Q1Q2:light}-\eqref{eq:dim5:Q10Q11:light}, are suppressed by the neutrino masses and thus negligible for all practical purposes. In this case an appreciable Wilson coefficient would immediately signal the existence of NSI.

\section{NSI and elastic scattering}
\label{sec:NSIcoh}
This section describes the nuclear response to the elastic neutrino scattering on nucleus A at low energies, $\nu A\to \nu A$,  due to either the SM and/or NSI interactions. 
The calculation is done in several steps. In Section \ref{sec:Basis} we first match onto an EFT describing neutrino interactions with non-relativistic protons and neutrons. 
The corresponding nuclear response to elastic neutrino scattering (\CE) is given in Section \ref{sec:Connection}.
For ease of comparison we also give the naive dimensional analysis (NDA) estimates for \CE cross sections induced by each of the EFT operators, while leaving the detailed numerical analysis for Section \ref{sec:Constraints}.

\subsection{Interactions of neutrinos with nonrelativistic nucleons}
\label{sec:Basis}
The neutrons and protons inside nuclei are non-relativistic and their interactions are well described by a chiral EFT with nonrelativistic nucleons. The momentum exchange, $q$, in \CE scattering is small so that nuclei remain intact, while neutrons and protons are non-relativistic throughout the scattering event. For instance, in the COHERENT experiment~\cite{Akimov:2017ade}, the typical momentum exchange is $q\sim 30-70$ MeV. This is well below the cut-off of chiral EFT, $\Lambda_{\rm ChEFT}\sim {\mathcal O}(1 {\rm ~GeV})$, so that the effective neutrino interactions in Eq. \eqref{eq:lightDM:Lnf5} can be included in the chiral EFT framework.
We work at leading order in the chiral expansion for each of the EFT operators in \eqref{eq:lightDM:Lnf5}, counting the light pseudoscalar masses to be parametrically of the order $m_\pi \sim {\mathcal O}(q)$. At leading chiral order the neutrino interacts only with a single nucleon, while interactions of a neutrino with two nucleons are suppressed by powers of $q/\Lambda_{\rm ChEFT}$. The exception to this rule are the dimension seven Rayleigh operators, Eq. \eqref{eq:dim7:Q1Q2:light}, which we discuss separately in Section \ref{sec:Rayleigh:scatt}.

\begin{figure}
\includegraphics[scale=0.70]{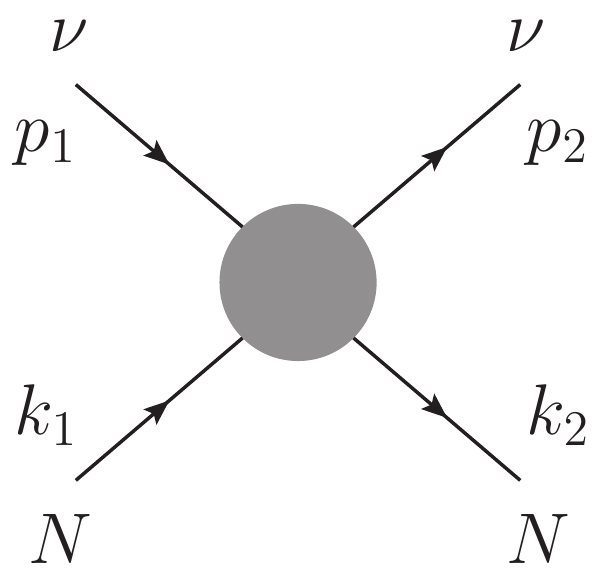}
\caption{The kinematics of neutrino scattering on nucleons,
  $\nu(p_1)N(k_1)\to \nu(p_2) N(k_2)$. 
  }
         \label{fig:scattering_kin}
\end{figure}

The effective Lagrangian describing neutrino interactions with non-relativistic nucleons is given by
\beq \label{eq:Lint}
{\cal L}_{\rm NR}=\sum_{i,N} c_{i,N}^{(d)}(q^2) \op_{i,N}^{(d)}(+{\rm h.c.}),
\eeq
where $N=n,p$, while $d$ counts the number of derivatives in the operator, which gives the suppression of the operator in terms of soft momenta, ${\mathcal O}(q^d)$. The momentum exchanged, $q^\mu=(q^0,\vec q)$, is given by, 
\begin{equation}
q^\mu = k_2^\mu-k_1^\mu=p_1^\mu - p_2^\mu.
\end{equation}
with $k_{1(2)}, p_{1(2)}$ the incoming(outgoing) nucleon and neutrino momenta, respectively, cf. Fig.~\ref{fig:scattering_kin}. The nuclear recoil energy, $E_R=\vec q^2/2 m_A$, can, for fixed neutrino energy, be anywhere between $E_{R,{\rm min}}=0$ for forward scattering, to a maximal value of $E_{R,{\rm max}}\simeq 2 E_\nu^2/ m_A$ obtained in the case of neutrino back-scattering.

\begin{figure}
\includegraphics[scale=0.75]{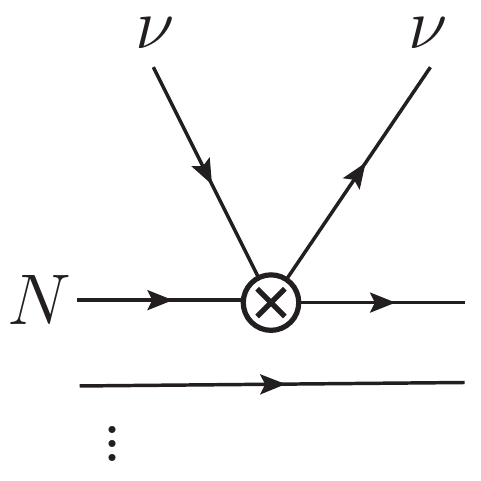}\hspace*{1.3cm}
\includegraphics[scale=0.75]{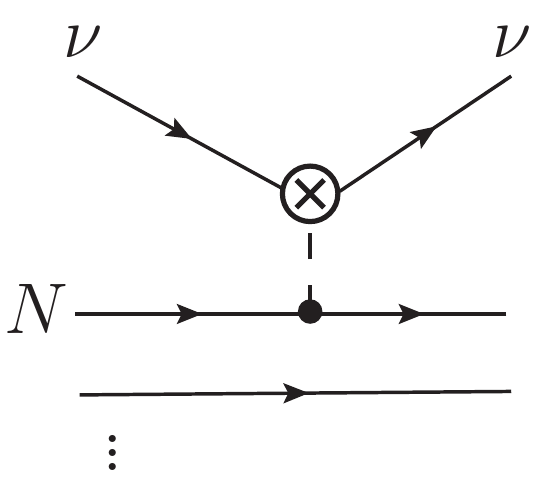}\hspace*{1.3cm}
\includegraphics[scale=0.75]{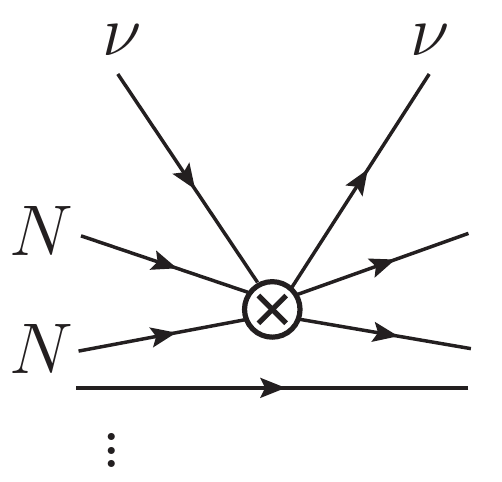}
\caption{The chirally leading diagrams for the neutrino-nucleus scattering (the first and second diagrams), and a representative diagram for two nucleon scattering (the third diagram). The effective neutrino--nucleon and neutrino--meson interactions are denoted by a
  circle, the dashed line denotes a pion, and the dots represent the remaining $A-2$ nucleon lines.}
         \label{fig:LOChPT}
\end{figure}

The matching of quark and gluon currents onto nonrelativistic nucleon currents is performed using heavy baryon chiral perturbation (HBChPT) theory \cite{Jenkins:1990jv,Bishara:2016hek}, while neutrino currents maintain their relativistic form. In this way one can explicitly show that the chiraly leading interactions of neutrinos are with a single nucleon current.
We write the non-relativistic operators in the Lagrangian (\ref{eq:Lint}) using the heavy nucleon formalism of HBChPT, where the nucleon mass is effectively integrated out. To the order we are working, the heavy nucleon field, $N_v$, is given by
\begin{equation}
N=e^{-i m_N v \cdot x} \Big(1 +\frac{i \slashed
  \partial_\perp}{i v\cdot \partial+2 m_N-i \epsilon}\Big)
N_v\,,\label{eq:chi:rel}
\end{equation}
where $v^\mu$ is the nucleon four-velocity, which we may take to coincide with the lab frame, $v^\mu=(1,0,0,0)$, while $\partial_\perp^\mu=\partial^\mu-v^\mu v\ncdot \partial $ is the soft momentum. The momentum due to the heavy nucleon mass, $m_Nv^\mu$, has been factored out from the definition of $N_v$ by the exponential prefactor.

The nonrelativistic operators in \eqref{eq:Lint} are, for proton and with $d=0$,
\begin{align}
\op_{1,p}^{(0)} & = (\bar \nu_\beta \gamma_\mu P_L \nu_\alpha) (v^\mu \bar p_v p_v),
& \op_{2,p}^{(0)} & = (\bar \nu_\beta \gamma_\mu P_L \nu_\alpha)(\bar p_v S^\mu_N p_v)\,,
  \label{eq:0der:O1O2}
  \\
  \op_{3,p}^{(0)} & = (\bar \nu_\beta P_L \nu_\alpha)(\bar p_v p_v)\,,
   &\op_{4,p}^{(0)} & = (\bar \nu_\beta \sigma_{\mu\nu} P_L \nu_\alpha)(\bar p_v \sigma_\perp^{\mu\nu} p_v)\,, 
   \label{eq::0der:O3O4} 
\end{align}
and a similar set of operators for neutrons with $p \rightarrow n$. Here we have defined $\gamma_\perp^\mu = \gamma^\mu - v^\mu \slashed v$, $\sigma_\perp^{\mu\nu} = \frac{i}{2} [\gamma_\perp^\mu,\gamma_\perp^\nu]$ and the spin operator $S_N^\mu = \gamma_5 \gamma_\perp^\mu/2$.
There are two relevant operators with a single derivative, $d=1$,  
\begin{align}
  \op_{1,p}^{(1)} & = (\bar \nu_\beta P_L \nu_\alpha)\Big(\bar p_v \frac{ i q\cdot S_N}{m_N}  p_v\Big)\,,
 & \op_{2,p}^{(1)} & = (\bar \nu_\beta P_L \nu_\alpha)\Big(\bar p_v \frac{ p_{12}\cdot S_N}{m_N}  p_v\Big)\,,
\label{eq::1der:O1} 
\end{align}
and one  relevant $d=2$ operator,
\begin{align}
  \op_{1,p}^{(2)} & = \frac{i q_\mu p_{12,\nu}}{m_N^2} (\bar \nu P_L \nu)(\bar p_v \sigma_\perp^{\mu\nu} p_v)\,,\label{eq::2der:O1} 
\end{align}
where $p_{12}^\nu=p_1^\nu+p_2^\nu$. We work in the isospin limit in which the proton and neutron masses are equal, so that $m_N=m_p=m_n\simeq  939$ MeV.  Above, the nucleon operators with $\sigma_\perp^{\mu\nu}$ are related to the nucleon spin through 
\begin{equation}\label{eq:sigma-to-epsilon-S}
\bar N_v \sigma_\perp^{\mu\nu}N_v=
-2 \epsilon^{\mu\nu\alpha\beta}v_{\alpha}\big(\bar N_v S_{N,\beta}N_v\big)\,,
\end{equation}
where $\epsilon^{\mu\nu\alpha\beta}$ is the totally antisymmetric
Levi-Civita tensor, with $\epsilon^{0123}=1$.
In \eqref{eq:Lint} the Hermitian conjugation is present in the sum for almost all the operators -- the exception are $\op_{1,p}^{(0)}, \op_{2,p}^{(0)}$ for $\alpha=\beta$, in which case the two operators are already Hermitian. The coefficients of these operators are thus real, 
while for the other operators they can be complex in general.

The nonrelativistic coefficients in \eqref{eq:Lint} are (summations are  over $q = u,d,s$), 
\begin{align} \label{eq:c1(0)}
c_{1,p}^{(0)} & = \sum_q F_1^{q/p} \hat \C_{1,q}^{(6)}, 
\\
c_{2,p}^{(0)} & =  2 \sum_q F_A^{q/p} \hat \C_{2,q}^{(6)}, 
\\
\label{eq:c56(0)}
c_{3,p}^{(0)} & =  F_G^{p} \hat \C_{3}^{(7)} + \sum_q \Big(-Q_q \frac{e^2}{4\pi^2}\frac{2\bar E_\nu}{q^2}F_1^{q/p} \hat \C_{1}^{(5)} +F_S^{q/p} \hat \C_{5,q}^{(7)}+ 2 \bar E_\nu F_1^{q/p}  \big(\hat \C_{8,q}^{(7)}+\hat \C_{10,q}^{(7)}\big)\Big),
\\
c_{4,p}^{(0)} & =  \sum_q F_{T,0}^{q/p} \hat \C_{7,q}^{(7)} 
\\
\label{eq:c1p1}
 c_{1,p}^{(1)} & =  F_{\tilde G}^{p} \hat \C_{4}^{(7)} + \sum_q F_P^{q/p} \hat \C_{6,q}^{(7)},
 \\
 c_{2,p}^{(1)} & = \sum_q  2 m_N   F_{A}^{q/p} \big( \hat \C_{9,q}^{(7)}+  \hat \C_{11,q}^{(7)}\big),
 \\ 
 \label{eq:c12(2)}
c_{1,p}^{(2)} & =   \sum_q Q_q \frac{e^2}{4\pi^2}\frac{m_N}{2 q^2}F_2^{q/p} \hat \C_{1}^{(5)},
\end{align}
where $\bar E_\nu=(p_1+p_2)\cdot v/2$ is the average energy of the neutrino before and after scattering. In COHERENT experiment the incoming neutrinos have energy $\sim$16-53 MeV, so that $\bar E_\nu \sim {\mathcal O}(q)\lesssim {\mathcal O}(m_\pi)$. The coefficients for neutrons are obtained through $p\to n$ replacement. 
The form factors, $F_i$, 
describe the hadronization of quark and gluon currents. They are functions of $q^2=-\vec q^2$ only, and their definitions are given in appendix~\ref{app:formfactors}.

In order to derive the above expressions for the nonrelativistic coefficients in Eqs. \eqref{eq:c1(0)}-\eqref{eq:c12(2)} we used the non-relativistic reduction of the nucleon currents summarized in appendix~\ref{app:formfactors}.
We keep only the leading terms in the $q/m_N$ expansion for each of the non-standard neutrino interaction operators, Eqs.~\eqref{eq:dim5:nf5:Q1:light}-\eqref{eq:dim5:Q10Q11:light}. The leading contributions start at different orders in $q$ expansion, depending on the structure of the NSI operators. For instance, the operators $Q_{4}^{(7)}, Q_{6,q}^{(7)}, Q_{9,q}^{(7)},$ and $Q_{11,q}^{(7)}$, all match onto non-relativistic operators with one derivative, and thus their contributions to the scattering amplitude start only at ${\mathcal O}(q)$. All the other operators have contributions already at ${\mathcal O}(q^0)$. 

There are two specific exceptions, where these leading contributions naturally vanish. For the QED dipole operator, $Q_1^{(5)}$, the leading scattering on neutrons comes from operators with two derivatives. This is despite the $Q_1^{(5)}$ also contributing to the ${\mathcal O}(q^0)$ nonrelativistic operator, see Eq. \eqref{eq:c56(0)}. The reason is that for the neutron $\sum_q Q_q F_1^{q/n}(0)=0$, so that in this case the contribution to \eqref{eq:c56(0)} vanishes. Similarly, the absence of valence strange quarks in nucleons gives $F_1^{s/N}(0)=0$, so that we need to keep the form factor $F_1^{s/N}(q^2)$ expanded to quadratic order. 
 Note as well, that the contributions from the axial current form factor $F_{P'}(q^2)$ are proportional to the neutrino masses and thus neglected, even though $F_{P'}(q^2)$ is $1/m_\pi^2$ enhanced due to the pion pole, corresponding to the middle diagram in Fig. \ref{fig:LOChPT}.

In summary, in Eqs. \eqref{eq:c1(0)}-\eqref{eq:c12(2)} most of the form factors are to be evaluated at $q^2 \rightarrow 0$, 
\beq
F_i^{q/N}(q^2)=F_i^{q/N}(0)+\cdots,
\eeq
since this gives the chirally leading contribution, and we neglect the $q^2/m_N^2$ suppressed contributions, denoted above with the ellipsis.  
The three exceptions are the form factors $F_{P}^{q/N}$, $F_{\tilde G}^{q/N}$, and $F_1^{q/N}$ where we keep the $q^2$ dependence. The chirally leading contributions to $F_{P}^{q/N}$ and $F_{\tilde G}^{q/N}$ have pion and $\eta$ pole contributions (corresponding to the second diagram in Fig.~\ref{fig:LOChPT}), 
\begin{align}
\label{eq:F_P}
F_{P}^{q/N}(q^2)&=\frac{m_N^2}{m_\pi^2-q^2} a_{P,\pi}^{q/N}+\frac{m_N^2}{m_\eta^2-q^2} a_{P,\eta}^{q/N}
+\cdots, 
\\
\label{eq:F_tildeG}
F_{\tilde G}^{N}(q^2)&=
\frac{q^2}{m_\pi^2-q^2} a_{\tilde G,\pi}^{N}+\frac{q^2}{m_\eta^2-q^2} a_{\tilde G,\eta}^{N}+b_{\tilde G}^{N}
+\cdots,
\end{align}
while for the vector form factor of the neutron we need to go to ${\mathcal O}(q^2)$, 
\beq \label{eq:F1exp}
F_1^{q/N}(q^2)=F_1^{q/N}(0)+F_1^{q/N}{}'(0)q^2+\cdots,
\eeq 
since for the strange quark $F_1^{s/N}(0)=0$. For simplicity we keep quadratic orders in $F_1^{q/N}(q^2)$ also for $q=u,d$.\footnote{Incidentally, in this way we also capture the first nonzero term in $c_{3,n}^{(0)}$ from $\hat \C_1^{(5)}$ for scattering on neutrons, cf. Eq.~\eqref{eq:c56(0)}. In that case the leading term in $c_{3,n}^{(0)}$ from $\hat \C_1^{(5)}$ cancels because neutron has zero electric charge. Note that the leading contribution from $\hat \C_1^{(5)}$ for neutrino scattering on neutrons is described by $c_{1,n}^{(2)}$, Eq. \eqref{eq:c12(2)}, while the contribution from $c_{3,n}^{(0)}$ is relatively ${\mathcal O}(q)$ suppressed.
}
The input values of the nonperturbative parameters are given in \cite{Bishara:2017pfq}.

\subsection{Nuclear response to nonstandard neutrino interactions}
\label{sec:Connection}

The NSI coupling neutrinos to nonrelativistic nucleon currents, Eq. \eqref{eq:Lint}, are of two types -- the neutrinos either couple to the nucleon number operator, $\bar N_v N_v$,  or to the nuclear spin, $\bar N_v S_N^\mu N_v$. In the notation of Ref. \cite{Fitzpatrick:2012ix} the effective Lagrangian is 
\beq
\label{eq:LNR}
\mathcal{L}_{\rm NR} =\big(\bar \nu l_0 P_L\nu\big)1_N +2 \big( \bar \nu \vec l_5 P_L \nu\big) \cdot \vec S_N +\cdots, 
\eeq
where the ellipsis denote terms with $P_R\nu$, irrelevant for our case where the incoming flux is due to left-handed neutrinos.\footnote{A contribution to \CE from right-handed neutrinos in the incoming flux requires two insertions of NSI interactions, one in the production and one in the scattering on nucleus, and is thus of second order in small perturbations. 
} The two Dirac structures are, 
\begin{align}
l_0 &=  c_{1,N}^{(0)} \slashed{v}  + c_{3,N}^{(0)} ,
\\
\begin{split}
l_5^\mu &= -\frac{1}{2}c_{2,N}^{(0)} \gamma^\mu  -c_{4,N}^{(0)}  \epsilon^{\mu\nu\alpha\beta}v_\nu  \sigma_{\alpha\beta} -\frac{iq^\mu}{2 m_N}c_{1,N}^{(1)} -\frac{ip_{12}^\mu}{2 m_N}c_{2,N}^{(1)} -  \epsilon^{\mu\nu\alpha\beta} v_\nu \frac{i q_\alpha p_{12,\beta}}{m_N^2} c_{1,N}^{(2)} .
\end{split}
\end{align}
Note that only the spatial three-vector components of $l_5^\mu$ enter the leading order nonrelativistic Lagrangian, \eqref{eq:LNR}. The EFT counting is such that all components of neutrino momenta count as the neutrino energy, $E_\nu$, while the nucleon currents are expanded in $q/m_N$, as discussed in the previous subsection. The results for the $c_{i,N}^{(d)}$ coefficients, that are in general functions of $q^2$, are given in \eqref{eq:c1(0)}-\eqref{eq:c12(2)}.

The cross sections  for the neutrino--nucleus scattering is 
\beq \label{eq:cross}
\frac{d\sigma_A}{dE_R}=2 m_A\frac{d\sigma_A}{d\vec q^2}= \frac{m_A}{\pi E_\nu^2}\overline{\mathcal{M}^2 },
\eeq
where $E_R$ is the recoil energy of the nucleus and the averaged amplitude square is given by~\cite{Anand:2013yka},
\beq
\label{eq:Msquared}
\begin{split}
\overline{\mathcal{M}^2 }=\frac{1}{2J_A + 1}\sum_{\rm spins} |\mathcal{M}|^{2} &= \frac{4\pi}{2J_A + 1} \sum_{\tau , \tau'=0,1} \Big(  R_{M}^{\tau\tau'} W_M^{\tau\tau'} 
 + R_{\Sigma''}^{\tau\tau'}W_{\Sigma''}^{\tau\tau'}+ R_{\Sigma'}^{\tau\tau'}W_{\Sigma'}^{\tau\tau'}  \Big),
\end{split}
\eeq
where $J_A$ is the spin of the target nucleus, the $W_{M,\Sigma',\Sigma''}(q)$ are the nuclear response functions and $\hat{q} \equiv \vec{q} / q$. The sum is over the isospin, $\tau,\tau'=0,1$, with the kinematic factors 
\begin{align} 
\begin{split}
R_{M}^{\tau\tau'}&=\Tr \big( P_L \slashed p_1 l_{0,\tau'}^\dagger \slashed p_2 l_{0,\tau}   \big),
\end{split} 
\\
\begin{split}
R_{\Sigma''}^{\tau\tau'}&= \Tr \big( P_L \slashed p_1  l_{5,\tau'}^{j\dagger}\slashed p_2 l_{5,\tau}^i \big) \hat{q}^i \,  \hat{q}^j ,
\end{split}
 \\
R_{\Sigma'}^{\tau\tau'}&= \Tr \big( P_L \slashed p_1  l_{5,\tau'}^{j\dagger} \slashed p_2 l_{5,\tau}^i  \big)\big(\delta^{ij}- \hat{q}^i \,  \hat{q}^j \big),
\end{align}
where the summations over spatial indices, $i,j=1,2,3$, are implied. In the evaluation of the kinematic factors we only keep the leading terms in $E_R/E_\nu$ and $E_\nu/m_N ,q/m_N$, which gives (note that $E_R=\vec q^2/(2m_A)$)
\begin{align} 
\begin{split} \label{eq:RM}
R_{M}^{\tau\tau'}=& \big(4 E_\nu^2-\vec q^2\big) c_{1,\tau}^{(0)}c_{1,\tau'}^{(0)*}
+\vec q^2 c_{3,\tau}^{(0)}c_{3,\tau'}^{(0)*},
\end{split} 
\\
\begin{split} \label{eq:RSpp}
R_{\Sigma''}^{\tau\tau'}=& \frac{\vec q^4}{4 m_N^2}c_{1,\tau}^{(1)}c_{1,\tau'}^{(1)*} + \frac{E_\nu^2 \vec q^4}{4 m_A^2 m_N^2}c_{2,\tau}^{(1)}c_{2,\tau'}^{(1)*}+
\frac{\vec q^2}{16m_A^2}\big(4 E_\nu^2-\vec q^2\big)c_{2,\tau}^{(0)}c_{2,\tau'}^{(0)*} +
16 E_\nu^2 c_{4,\tau}^{(0)}c_{4,\tau'}^{(0)*}
\\
&+2i \frac{E_\nu}{m_N} \vec q^2 \big( c_{1,\tau}^{(1)}c_{4,\tau'}^{(0)*}-c_{4,\tau}^{(0)} c_{1,\tau'}^{(1)*}\big)+
2i \frac{E_\nu^2 \vec q^2}{m_Am_N} \big( c_{2,\tau}^{(1)}c_{4,\tau'}^{(0)*}-c_{4,\tau}^{(0)} c_{2,\tau'}^{(1)*}\big)
\\
&+ \frac{E_\nu \vec q^4}{4 m_A m_N^2} \big( c_{1,\tau}^{(1)}c_{2,\tau'}^{(1)*}+c_{1,\tau}^{(1)} c_{2,\tau'}^{(1)*}\big),
\end{split}
 \\
 \begin{split} \label{eq:RSp}
R_{\Sigma'}^{\tau\tau'}=& \frac{1}{4} \big(4 E_\nu^2 +\vec q^2\big) c_{2,\tau}^{(0)}c_{2,\tau'}^{(0)*} +\big(4 E_\nu^2 -\vec q^2\big)\Big(4 c_{4,\tau}^{(0)}c_{4,\tau'}^{(0)*}+ \frac{\vec q^4}{m_N^4} c_{1,\tau}^{(2)}c_{1,\tau'}^{(2)*} \Big) 
\\
& -2 \big(4 E_\nu^2 -\vec q^2\big)\frac{\vec q^2}{m_N^2}\big(c_{1,\tau}^{(2)}c_{4,\tau'}^{(0)*}+c_{4,\tau}^{(0)}c_{1,\tau'}^{(2)*}\big)
\\
&+\frac{(4 E_\nu^2-\vec q^2)\vec q^2}{4 m_N^2}\Big(c_{2,\tau}^{(1)}c_{2,\tau'}^{(1)*}- 2i \frac{m_N}{m_A}
\big(c_{2,\tau}^{(1)}c_{4,\tau'}^{(0)*}-c_{4,\tau}^{(0)}c_{2,\tau'}^{(1)*}\big) \Big).
\end{split}
\end{align}
Here $\tau, \tau'=0,1$ denote the isospin so that, 
\begin{equation}
\label{eq:ci:isospinrel}
c_{i,0}^{(d)}= \frac{1}{2} \left( c_{i,p}^{(d)} + c_{i,n}^{(d)} \right), \qquad c_{i,1}^{(d)}= \frac{1}{2} \left( c_{i,p}^{(d)} - c_{i,n}^{(d)} \right).
\end{equation} 
The non-relativistic coefficients describing neutrino interactions with protons and neutrons, $c_{i,p}^{(d)}$ and  $c_{i,n}^{(d)}$, are listed in Eqs. \eqref{eq:c1(0)}-\eqref{eq:c12(2)}.

Before proceeding, we give NDA estimates for the neutrino--nucleus scattering cross section \eqref{eq:cross}, switching on a single NSI Wilson coefficient $\hat C_{1,q}^{(d)}$, Eqs.~\eqref{eq:dim5:nf5:Q1:light}-\eqref{eq:dim5:Q10Q11:light}, at a time. Subtracting the contribution induced by the SM neutrino interactions gives the correction to the scattering cross section due to the presence of NSI,
\beq
\label{eq:Delta:sigmaNSI}
\Delta \sigma_{\rm NSI}\equiv \sigma-\sigma_{\rm SM}\sim {\mathcal O}(R_\alpha W_\alpha).
\eeq
Note that $\Delta \sigma_{\rm NSI}$ can be negative, if NSI interfere with the SM. 
In the last equality in \eqref{eq:Delta:sigmaNSI} we show the parametric dependence on kinematical factors and nuclear response functions, $W_\alpha$, taking $E_R\sim {\mathcal O}(E_\nu^2/m_A)$. The subscript is any of $\alpha=M,\Sigma',\Sigma''$, depending on the Wilson coefficient $\hat C_{1,q}^{(d)}$. 

In the long wavelength limit, $q\to0$, the nuclear response functions, $W_\alpha$, have the following parametric sizes, 
\beq\label{eq:parametricW}
W_M^{\tau \tau'}\sim {\mathcal O}(A^2), \qquad W_{\Sigma'}^{\tau \tau'}\sim {\mathcal O}(1), \qquad W_{\Sigma''}^{\tau \tau'}\sim {\mathcal O}(1). 
\eeq
The response functions  $W_{\Sigma'}^{\tau \tau'}$ and $W_{\Sigma''}^{\tau \tau'}$ encode the response of nucleus to the transverse and longitudinal axial operators, and thus measure the spin content of the nucleus. The values of $W_{\Sigma', \Sigma''}^{\tau \tau'}$  depend critically on the details of the nuclear wave function and can be much smaller than \eqref{eq:parametricW} for nuclei with all protons and neutrons paired. The $W_M^{\tau \tau'}$ response functions  count, in the long wavelength limit, the number of nucleons inside nucleus. This leads to coherent enhancement, also present for neutrino scattering through SM interaction -- the tree level $Z$ exchange, Eq.~\eqref{eq:c6udsSM}. The $Z$ boson couples most strongly to neutrons, so that in the SM case the enhancement is ${\mathcal O}(N^2)$, where $N$ is the number of neutrons in the nucleus. Depending on the flavor structure the NSI can be due to couplings to proton or neutrons or both. 

The NSI operators, Eqs.~\eqref{eq:dim5:nf5:Q1:light}-\eqref{eq:dim5:Q10Q11:light}, fall into three categories: the operators that interfere with the SM contribution, the operators that do not interfere with the SM but still lead to coherently enhanced scattering, and the operators that are not coherently enhanced. The NDA estimates of the scattering cross sections for each of the three sets of operators are as follows. 

\underline{\em The operators that interfere with the SM contribution.} These are the operators with quark vector currents, $\Q_{1,q}^{(6)}$ in \eqref{eq:dim6EW:Q1Q2:light}. The SM contribution to the corresponding Wilson coefficient is given in Eq. \eqref{eq:c6udsSM}. The NDA estimate of the NSI correction to the scattering cross section is 
\beq \label{eq:cohNDA1}
\Delta \sigma_{\rm NSI}\sim \frac{E_\nu^2}{\Lambda^2} \big(\hat C_{1,q}^{(6)}\big)_{\rm SM} \big( C_{1,q}^{(6)}\big)_{\rm NSI} A^2, 
\eeq
where we used that the interference with the SM dominates over the purely NP contribution. 

\underline{\em Coherently enhanced but no interference with the SM.} The operators that lead to coherently enhanced scattering, but do not interfere with the SM contribution, are the ones that contribute to $c_{3,N}^{(0)}$ nonrelativistic coefficients. These are the dimension 5 magnetic dipole operator, $\Q_1^{(5)}$, Eq. \eqref{eq:dim5:nf5:Q1:light}, and the set of dimension 7 operators, the $\Q_3^{(7)}$ operator in \eqref{eq:dim7:Q3Q4:light} that couples the neutrino current to gluons,  the  operator $\Q_{5,q}^{(7)}$ in \eqref{eq:dim7:Q5Q6:light} that couples neutrino and quark scalar currents, and the operators $\Q_{8,q}^{(7)}$, $\Q_{10,q}^{(7)}$ in Eqs. \eqref{eq:dim7:Q8Q9:light}, \eqref{eq:dim5:Q10Q11:light} that involve derivatives on the neutrino currents. The corresponding modification of the scattering cross section is parametrically, 
\beq \label{eq:cohNDA2}
\Delta \sigma_{\rm NSI}\sim \frac{E_\nu^2}{\Lambda^4} \bigg[\frac{m_N}{\Lambda}\Big(\C_{3}^{(7)} +{\mathcal O}(0.05) \C_{5,q}^{(7)}\Big)+\frac{E_\nu}{\Lambda}\Big(\C_{8,q}^{(7)} +\C_{10,q}^{(7)}\Big)+\frac{\alpha_{\rm EM}}{4\pi}\frac{\Lambda}{E_\nu}\C_1^{(5)}\bigg]^2 A^2. 
\eeq
Here we counted $q\sim {\mathcal O}(E_\nu)$, and only show the parametric dependence, neglecting ${\mathcal O}(1)$ factors. All the Wilson coefficients are due to NP. Above we thus dropped the NSI subscripts on the Wilson coefficients.

\underline{\em No coherent enhancement.} The remaining operators do not receive coherent enhancement. The correction to the neutrino scattering cross section is then parametrically given by
\beq \label{eq:cohNDA3}
\Delta \sigma_{\rm NSI}\sim \frac{E_\nu^2}{\Lambda^2} \big(\hat C_{2,q}^{(6)}\big)_{\rm SM} \big( C_{2,q}^{(6)}\big)_{\rm NSI} +
\frac{E_\nu^4}{\Lambda^6} \bigg[\frac{E_\nu^2}{m_\pi^2}\C_{4}^{(7)} +\frac{m_N m_q}{m_\pi^2} \C_{6,q}^{(7)}+\frac{m_q}{E_\nu} \C_{7,q}^{(7)}\bigg]^2.
\eeq
To shorten the expression we did not include additional numerical suppressions present for the case of strange quarks. 

\subsection{Scattering from Rayleigh operators}
\label{sec:Rayleigh:scatt}
Finally, we include the estimates for the \CE induced by the Rayleigh operators, ${\cal Q}_1^{(7)}$ and ${\cal Q}_2^{(7)}$ in \eqref{eq:dim7:Q1Q2:light}. The CP even Rayleigh operator ${\cal Q}_1^{(7)}$ leads to a coherently enhanced cross section, given by  Eq. \eqref{eq:cross} with the following matrix element squared \cite{Ovanesyan:2014fha} (for earlier work see 
\cite{Weiner:2012cb,Barger:2010gv})
\beq
\label{eq:Msquared:Rayleigh}
\begin{split}
\overline{\mathcal{M}^2 }=\frac{4\pi}{2J_A + 1} \vec q\,{}^2  \big|\hat C_{1}^{(7)}\big|^2
&\biggr( \frac{\sqrt2}{6}\frac{\alpha^2}{\pi^{3/2}}Z(Z-1) Q_0  \bar F_{pp}(q)
\\
&\qquad+ \frac{1}{3}\Big(\frac{\alpha}{4\pi}\Big)^2 m_N \sqrt{\sum_{\tau , \tau'=0,1}   a_{F,\tau}^{(0)} a_{F,\tau'}^{(0)*} W_M^{\tau\tau'} }  \biggr)^2.
\end{split}
\eeq
The first term is due to a contribution from two-body currents, calculated in Ref.~\cite{Ovanesyan:2014fha}, where the two photon lines attach to two different protons in the nucleus, while the second contribution is due to both photon lines attaching to the same proton. The two contributions to the cross section scale as $\sigma \propto {\mathcal O}(Z^4 Q_0^2/m_N^2)$ and $\sigma \propto {\mathcal O}(Z^2)$, respectively (not showing the common factors and interference terms). For $Q_0/m_N\sim 0.1$ the two contributions are parametrically of the same size for light nuclei $Z\sim {\mathcal O}(10)$, while the first term dominates for heavy nuclei, $Z\sim {\mathcal O}(50)$. 

 For the 2-proton form factor we use the phenomenological ansatz from Ref. \cite{Ovanesyan:2014fha},
\beq
\label{eq:twobody:form}
\bar F_{pp}(q) =e^{-\bar q^2/2}\left[1+\frac{1}{4}c_1 -\frac{\pi^{3/2}}{2\sqrt{2}}\bar q+\left(\frac53-\frac{5}{12}c_1+c_2\right) \bar q^2\right],
\eeq
with $\bar q=|\vec q|/Q_0$ and $Q_0=(0.5{\rm ~GeV})\times (0.3+0.9A^{1/3})^{-1}$ the inverse of the charge radius of the target nucleus. In the numerics we set the unknown coefficients to $c_{1,2}=1$, while varying  them in the range $c_{1,2}\in [-1,1]$ does not change results appreciably. 

The single nucleon matrix elements of the di-photon operators are not well known. We parametrize them as 
\begin{align}
\langle N_v |F_{\mu\nu} F^{\mu\nu}|N_v\rangle&= \frac{\alpha}{4\pi}   a_{F, N}^{(0)} m_N \langle N_v |\bar N_v N_v |N_v\rangle,
\\
\label{eq:single:FFtilde}
\langle N_v |F_{\mu\nu} \tilde F^{\mu\nu}|N_v\rangle&= \frac{\alpha}{4\pi}  a_{\tilde F,N}^{(1)} \langle N_v |\bar N_v \tfrac{i q \cdot S_N}{m_N} N_v |N_v\rangle,
\end{align}
with $a_{3, N}^{(0)}(q^2)$ and $a_{1,N}^{(1)}(q^2)$ the form factors. 
The NDA estimates for their values are, for $q^2\sim 0$, 
\beq
\label{eq:NDA:Rayleigh}
a_{F, p}^{(0)}\sim {\mathcal O}(1), \quad a_{F, n}^{(0)}\sim 0, \qquad \text{and}~~  a_{\tilde F, p}^{(1)}\sim a_{\tilde F, n}^{(1)} \sim {\mathcal O}(1).
\eeq
With these definitions the contributions to neutrino scattering due to two photon exchanges with a single nucleon are obtained by setting the coefficient  of the $\op_{3,N}^{(0)}$ operator to $c_{3,N}^{(0)}=a_{F,N}^{(0)}  m_N  \alpha^2/(48 \pi^2)$ for contributions from the ${\cal Q}_1^{(7)}$ operator, while for the CP-odd Rayleigh operator, ${\cal Q}_2^{(7)}$, one can set the coefficient of the $\op_{1,N}^{(1)}$ non-relativistic operator to $c_{1,N}^{(1)}=a_{\tilde F,N}^{(1)} \alpha^2/(32 \pi^2) $.

The CP odd Rayleigh operator leads to spin-dependent interactions, both from the single nucleon matrix element, \eqref{eq:single:FFtilde}, as well as from the 2 nucleon contributions. The two nucleon contributions arise from one photon interacting with the proton charge, and the second photon with the magnetic moment of the other nucleon, be it proton or neutron. The single nucleon and two nucleon contributions to the cross section are parametrically $\sigma \propto {\mathcal O}((q/m_N)^2)$ and $\sigma\propto {\mathcal O}((Z q Q_0/m_N^2)^2)$ (not showing the common factors and interference terms). The two-body current contribution is thus expected to dominate for heavy nuclei, $Z\sim 50$, while for light nuclei, $Z\sim 10$ the single current contributions are important. The formalism for the two-body current contribution was worked out in \cite{Ovanesyan:2014fha}, however, without deriving estimates for the resulting form factor. In the phenomenological analysis we thus conservatively ignore the 2-nucleon term and take as nonzero only the $a_{\tilde F,N}^{(1)}$ coefficient, using the NDA estimate in \eqref{eq:NDA:Rayleigh}. While this estimate for the cross section does not capture the largest contribution for heavy nuclei, where the NDA suggest the cross section to be $\sim (Z Q_0/m_N)^2\sim 20$ times bigger for $Z\sim 50$, the resulting error on bounds on $\Lambda$ will be only $\sim (Z Q_0/m_N)^{2/3}\sim 3$, which suffices in view of other uncertainties in our estimates for this particular operator.

\section{NSI and neutrino oscillations in matter}
\label{sec:oscillation}

Neutrino oscillation data bound a subset of NSI -- the ones that result in a nonzero forward scattering amplitude. The forward scattering 
of neutrinos on electrons and nuclei gives rise to matter effects in neutrino oscillations, described by an effective potential~\cite{Blennow:2013rca}. 
In this section we review the simplest case -- electrically neutral unpolarized medium at rest. For extension to a polarized medium see \cite{Bergmann:1999rz}, while for extensions to sterile neutrinos see, e.g., \cite{Bergmann:1998rg,Dentler:2018sju,Capozzi:2018ubv}.

In the SM the effective potential receives contributions from both CC and NC. The NC contribution is neutrino flavor universal. It induces an overall phase shift in neutrino oscillation that is not observable (though it needs to be considered for oscillations into sterile neutrinos). The CC contributes to forward scattering of electron neutrinos on electrons. Electron neutrino scattering on an isotropic, homogeneous gas of unpolarized electrons is therefore described by the following effective Hamiltonian, see, e.g., \cite{Blennow:2013rca}, 
\beq
\overline{{\cal H}}_{\rm eff}^{\rm CC}\Big|_{\rm SM} = \sqrt{2} G_F n_e \big( \bar \nu_{eL} \gamma^0 \nu_{eL} \big),
\eeq
where $n_e$ is the number density of electrons. The resulting potential energy,
\beq\label{eq:VeffhSM}
{\cal V}_{\rm eff}^{(h)} = \langle \nu_\alpha (p,h) |\int_V d^3 x \overline{{\cal H}}_{\rm eff}^{\rm CC}| \nu_\alpha (p,h)  \rangle = \sqrt{2} G_F n_e \frac{ (E_\nu - h |\vec p|)}{2 E_\nu} \delta_{\alpha e},
\eeq
 leads to a change in the oscillation frequency for electron neutrinos.
Here $p$ is the neutrino momentum, $\alpha$ its flavor and $h$ its helicity. The integral is performed over a finite volume $V$ which is also included in the normalization of the states, $| \nu_\alpha (p,h)  \rangle = (2E_\nu V)^{-1/2} a^{(h) \dagger} (p) | 0 \rangle$, and thus drops out in the final result.  
Since weak interactions couple to left-handed fields, 
the $h=-1$ ultrarelativistic neutrinos obtain the effective potential energy
\beq
{\cal V}_{\rm eff}^{(-)} \simeq \sqrt{2} G_F n_e \delta_{\alpha e},
\eeq
while the positive helicity,  $h=+1$, neutrinos are exposed to a negligible effective potential, ${\cal V}_{\rm eff}^{(+)} \sim {\mathcal O}( m_\nu^2/E)$.

The above SM results are readily extended to NSI that couple neutrinos to the charged fermion vector current, i.e., the operators ${\cal Q}_{1,f}^{(6)}$ in Eq. \eqref{eq:dim6EW:Q1Q2:light}. These lead to nonzero forward scattering amplitudes and thus induce an  effective potential (the dependence on neutrino flavors is implicit)
\beq\label{eq:Veff:NSI}
{\cal V}_{\rm eff}^{(-)}\Big|_{\rm NSI} \simeq  -\hat\C_{1,e}^{(6)}\big|_{\rm NSI} n_e - \left( \hat\C_{1,u}^{(6)}\big|_{\rm NSI} + 2 \hat\C_{1,d}^{(6)}\big|_{\rm NSI}\right) n_n- \left(2\hat\C_{1,u}^{(6)}\big|_{\rm NSI} + \hat\C_{1,d}^{(6)}\big|_{\rm NSI}\right) n_p.
\eeq 
Here $n_p=n_e$ is the number density of protons, equal to the number density of electrons in an electrically neutral medium, and $n_n$ the number density of neutrons. 
The global fits to neutrino oscillation data then lead to severe bounds on the parameters $\epsilon_{\alpha\beta}$  (see Section  
\ref{sec:Constraints}). 

The other NSI operators are poorly constrained from neutrino oscillations. This is most easily seen by analyzing the effects of NSI using the nonrelativistic basis, Eqs.~\eqref{eq:0der:O1O2}-\eqref{eq::2der:O1}. The matrix elements of spin-dependent operators, ${\mathcal O}_{2,p}^{(0)}, {\mathcal O}_{4,p}^{(0)} , {\mathcal O}_{1,p}^{(1)}, {\mathcal O}_{2,p}^{(1)}, {\mathcal O}_{1,p}^{(2)}$, vanish for unpolarized medium, and thus are not bounded by global fits 
of neutrino oscillations. The two sets of operators that have non-vanishing forward scattering elements are the operators ${\cal O}_{1,N}^{(0)}$ and ${\cal O}_{3,N}^{(0)}$. The ${\cal O}_{1,N}^{(0)}$ leads to the effective potential in \eqref{eq:Veff:NSI},  while the operator ${\cal O}_{3,N}^{(0)}$ results in an effective Hamiltonian
\beq \label{eq:Heffnewop}
\overline{{\cal H}}_{\rm eff} \supset - c_{3,N}^{(0)} n_N (\bar \nu_{\alpha R} \nu_{\beta L}) ,
\eeq
where $n_N$ is the nucleon number density. This gives the effective potential  that is suppressed by the neutrino mass matrix, ${\cal V}_{\rm eff}^{(h)}\simeq - c_{3,N}^{(0)} n_N m_N (m_{\nu})_{\alpha \beta}$,  
and thus gives only extremely weak constraints on NSI.

\section{NSI and deep inelastic scattering}
\label{sec:DISintro}
For completeness we include the bounds on NSI that arise from deep inelastic  neutrino--nucleon scattering (DIS). While the DIS data were obtained at much higher momenta exchanges, $q\sim {\mathcal O}(10{\rm~GeV})$, the constraints are severe enough that the EFT description may still be valid at least in parts of the parameter space. 
 Throughout this section we thus assume that the EFT Lagrangian in Eq.~\eqref{eq:lightDM:Lnf5} is valid also for DIS. We comment on the validity of this assumption in Section \ref{sec:Constraints} where we confront predictions with data. 
 
\begin{figure}
\includegraphics[scale=0.7]{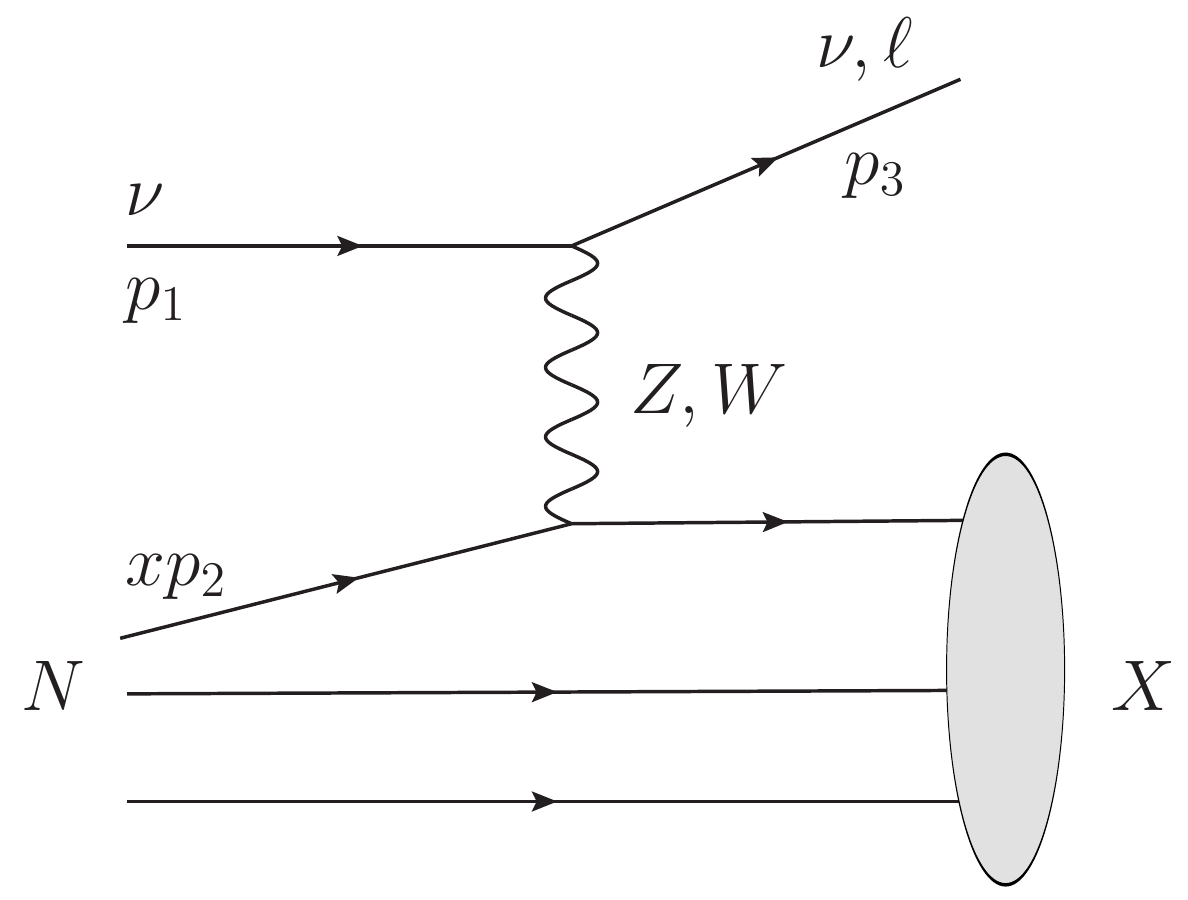}
\caption{Kinematics in neutrino Deep Inelastic Scattering (DIS). }
         \label{fig:nuDIS}
\end{figure}

In neutrino--nucleus DIS the typical momentum  exchange, $q$, is much larger than the inverse radius of the nucleus. The total cross section is therefore an incoherent sum of contributions from neutrino scattering on protons and neutrons inside the nucleus, 
\beq \label{eq:sigma_nu}
\sigma_\nu=Z \sigma_{\nu p} +(A-Z) \sigma_{\nu n},
\eeq
where
\beq \label{eq:disfact}
\sigma_{\nu N} = \int_0^1 dx \sum_{j}  \hat \sigma_{\nu j} (x) \,  f_{j/N}(x) +\cdots ,\qquad N=n,p.
\eeq
The ellipses indicate power suppressed corrections to the factorization \cite{Collins:1989gx}. The sum runs over the partons, $j=u,d,s,c,g,\gamma$, with $f_{j/N}(x)$ the corresponding parton distribution functions (PDF) for the nucleon $N$.

The kinematics of the process are shown in Fig.~\ref{fig:nuDIS}, with $p_1 (p_3)$ the incoming neutrino (outgoing lepton) momentum, $p_2$ the incoming nucleon momentum a fraction $x$ of which is carried by the parton, and we sum over the hadronic final state, $X$. 
We work in the limit of large neutrino energy, $E_\nu$, and  large momentum exchanged, $q^2$, i.e.,  $E_\nu \gg m_N$ and $q^2\gg m_N^2$. The usual DIS variables, the partonic center of mass energy squared, $\hat s$, and the fraction, $y$, of incident neutrino energy transferred to the hadronic system, are, 
\beq \label{eq:disvar}
\hat s = (p_1 + xp_2)^2 \simeq 2 xm_N E_\nu, \qquad y = 1 - \frac{E_{\nu(\ell)}'}{E_\nu},
\eeq 
where $E_{\nu(\ell)}'$ is the out-going neutrino (lepton) energy. 

The double-differential neutrino--nucleon DIS  cross section is, 
\beq
\label{eq:sigmanuN}
\frac{d^2 \sigma_{\nu N}}{dx dy} = \sum_{j}  \frac{|{\cal M}_{\nu j}|^2}{16 \pi x s} \,  f_{j/N}(x),
\eeq
where $\hat s = x s$. We neglect the intrinsic charm content of the proton, take the off-diagonal CKM matrix elements to zero, while $V_{ud},V_{cs}\to 1$. In $\nu q$ and $\bar \nu \bar q$ ($\bar \nu q$  and $\nu \bar q$) scattering, the total spin of the initial state is $J=0(1)$, so that the cross section is $y$ independent (has $\left(1 - y \right)^2$ angular dependence). 

We will be interested in neutrino DIS measurements by the  CHARM collaboration \cite{Dorenbosch:1986tb}, which used the target material that is to a very good approximation isoscalar, i.e., composed of equal number of protons and neutrons. 
The average neutrino--nucleon scattering cross section for an isoscalar target is given by,
\beq
\frac{d^2 \sigma_{\nu N}}{dx dy} = \frac{1}{2} \left( \frac{d^2 \sigma_{\nu p}}{dx dy} + \frac{d^2 \sigma_{\nu n}}{dx dy} \right).
\eeq
For an isoscalar target the SM predictions 
for the CC neutrino-nucleon scattering, $\nu N\to \ell^- X$ and $\bar \nu N\to \ell^+ X$, are,
\begin{align}
\label{eq:sigmanuNCC}
&\frac{d^2 \sigma_{\nu N}^{\rm (CC)}}{dx dy}\biggr|_{\rm SM} = \frac{G_F^2 m_N E_\nu}{\pi} x \Big\{u(x) + d(x) +2s(x)+ \left[ \bar u(x) + \bar d(x) \right] \left(1 - y \right)^2 \Big\}, 
\\
\label{eq:sigmanubarNCC}
&\frac{d^2 \sigma_{\bar \nu N}^{\rm (CC)}}{dx dy}\biggr|_{\rm SM} = \frac{G_F^2 m_N E_\nu}{\pi} x \Big\{\bar u(x) + \bar d(x) +2 \bar s(x)+ \left[u(x) + d(x) \right] \left(1 - y \right)^2 \Big\},
\end{align}
where we assumed isospin symmetry, 
\beq
u(x) \equiv f_{u/p}(x) = f_{d/n}(x), \qquad d(x) \equiv f_{d/p}(x) = f_{u/n}(x), \qquad s(x) \equiv f_{s/p}(x) = f_{s/n}(x).
\eeq
and similarly for antiquarks, $\bar q(x)\equiv f_{\bar q/p}(x)$. 
In \eqref{eq:sigmanuNCC}, \eqref{eq:sigmanubarNCC} we integrated out the $W$, and traded the $m_W$ dependence for the Fermi constant, $G_F$.

The NC neutrino--nucleon scattering is, in the SM, given by,
\beq
\begin{split}\label{eq:NC:SM}
\frac{d^2 \sigma_{\nu N}^{({\rm NC})}}{dx dy}= \frac{m_N E_\nu}{\pi} x  \Big\{ &\left( \hat C_{L,u}^2 + \hat C_{L,d}^2 \right)  \sum_{q=u,d} \big(q(x)+ \bar q(x) \left(1 - y \right)^2 \big) \\
+&\left( \hat C_{R,u}^2 + \hat C_{R,d}^2 \right)  \sum_{q=u,d} \big(q(x) \left(1 - y \right)^2 + \bar q(x) \big) 
\\
+&  2 \left[ \hat C_{L,s}^2  \big(s(x)+ \bar s(x) \left(1 - y \right)^2 \big)+ \hat C_{R,s}^2 \big(s(x) \left(1 - y \right)^2 + \bar s(x) \big)\right]\Big\},
\end{split}
\eeq
where 
\beq
\hat C_{R,q} = \frac{1}{\sqrt{2}} \left( \hat \C_{1,q}^{(6)} + \hat \C_{2,q}^{(6)}\right), \qquad \hat C_{L,q} =\frac{1}{\sqrt{2}} \left(\hat \C_{1,q}^{(6)} - \hat \C_{2,q}^{(6)}\right),
\eeq
with the Wilson coefficients given in \eqref{eq:c6udsSM}.
The antineutrino cross section is obtained by exchanging $L \leftrightarrow R$.

The NSI, Eq. \eqref{eq:lightDM:Lnf5}, only affect the NC scattering cross section in \eqref{eq:NC:SM}. 
The matrix elements squared, Eq.
\eqref{eq:sigmanuN}, come from a sum of the EFT operators, 
\beq
|{\cal M}_{\nu j}|^2 = \Big|\sum_{i,d} {\cal M}_{i;\nu j}^{(d)}\Big|^2 =  \sum_{i,d} \big|{\cal M}_{i; \nu j}^{(d)}\big|^2 + \,   \sum_{i\neq k,d,d'}2 \re \left({\cal M}_{i;\nu j} ^{(d)} \, {\cal M}_{k;\nu j}^{(d')*} \right)\,,
\eeq
where  ${\cal M}_{i; \nu j}^{(d)}$ is the matrix element of the operator ${\cal Q}_i^{(d)}$ for neutrino scattering on parton $j$. 
The dimension six operators only interfere among themselves, since the spin average of the axial-vector Dirac structure and the scalar or tensor currents vanishes. This gives, 
\beq\label{eq:c6matrel}
\left| {\cal M}_{\nu q}^{(6)} \right|^2 = 16 \hat s^2 \left[\left( \hat \C_{1,q}^{(6)} \right)^2 (1 + (1 - y)^2) + \left(\hat \C_{2,q}^{(6)} \right)^2 (1 + (1 - y)^2) + 2 y (2 - y)  \re \hat \C_{1,q}^{(6)}  \hat \C_{2,q}^{(6)*} \right],
\eeq
where the Wilson coefficients contain both the SM contributions and the NSI correction, cf.
eq.~\eqref{eq:c6NSI}.
The matrix elements squared for dimension 5 and dimension 7 operators are 
\begin{align}
|{\cal M}_{1; \nu \gamma}^{(5)}|^2  &= 8 \left| \frac{e}{8 \pi^2} \hat {\cal C}_{1}^{(5)} \right|^2\hat s y, \label{eq:c15matr}\\
|{\cal M}_{1; \nu\gamma (3; \nu g)}^{(7)}|^2  &= 16 \left|\frac{\alpha_{(s)}}{12 \pi} \hat {\cal C}_{1(3)}^{(7)} \right|^2 \hat s^3 y ^3,\\
|{\cal M}_{2; \nu\gamma (4; \nu g)}^{(7)}|^2  &= 16 \left|\frac{\alpha_{(s)}}{8 \pi} \hat {\cal C}_{2(4)}^{(7)} \right|^2 \hat s^3 y ^3,\\
|{\cal M}_{5(6),q;\nu q}^{(7)}|^2  & = 8 m_q^2 \left| \hat {\cal C}_{5(6),q}^{(7)} \right|^2 \hat s^2 y^2, \label{eq:c56matr}
\\ 
\label{eq:c7matr}
|{\cal M}_{7,q;\nu q}^{(7)}|^2 &= 64 m_q^2 \left| \hat {\cal C}_{7,q}^{(7)} \right|^2 \hat s^2\left( 2 \left(1 - y \right)^2  + 2 - y^2\right),\\
|{\cal M}_{8(9),q;\nu q}^{(7)}|^2 &= 64 \left| \hat {\cal C}_{8(9),q}^{(7)} \right|^2\hat s^3 y (y -1) ,\\
|{\cal M}_{10(11),q;\nu q}^{(7)}|^2 &= 32 \left| \hat {\cal C}_{10(11),q}^{(7)} \right|^2\hat s^3 y (y -1) .
\end{align}
The remaining non-zero interference terms are, 
\begin{align}
&\, \re \left({\cal M}_{5,q;\nu q}^{(7)} {\cal M}_{7,q;\nu q}^{(7)*}\right) = 16 \hat s^2 y (y - 2)m_q^2 \re \left(\hat {\cal C}_{5,q}^{(7)} \hat {\cal C}_{7,q}^{(7)} \right), \\
&\, \re \left({\cal M}_{8(9),q;\nu q}^{(7)} {\cal M}_{10(11),q;\nu q}^{(7)*}\right) = 64 \hat s^3 y (y - 1) \re \left(\hat {\cal C}_{8(9),q}^{(7)} \hat {\cal C}_{10(11),q}^{(7)} \right).
\end{align}
In order to calculate DIS cross sections, we use the {\tt ManeParse} package \cite{Clark:2016jgm} to get the quarks and gluon PDFs (we use the {\tt CT10} NLO pdf set),
while we take the photon PDF from \cite{Manohar:2017eqh}.


\section{Constraints on new neutrino interactions}
\label{sec:Constraints}
Utilizing the results from Sections \ref{sec:NSIcoh}-\ref{sec:DISintro} we now derive the bounds on the NSI Wilson coefficients, Eq.~\eqref{eq:lightDM:Lnf5}, from neutrino oscillations, \CE \cite{Akimov:2017ade}, DIS \cite{Allaby:1988bb} and from searches for neutrino dipole moment \cite{Borexino:2017fbd}. We also explore the reach of future \CE  measurements in reactor experiments \cite{Akimov:2012aya}. We restrict the analysis to interactions of $\nu_e$ and $\nu_\mu$ since these are the NSI probed in \CE. The results are summarized in Figs.~\ref{fig:chart-all} and~\ref{fig:chart-all2}.

\subsection{Constraints on NSI from oscillations}
The oscillation data constrain the NSI contributions to the operators ${\cal Q}_{1,f}^{(6)}$, Eq. \eqref{eq:dim6EW:Q1Q2:light}.
The global fits to oscillation data allow at 95\% C.L. \cite{Esteban:2018ppq} (in the notation of Eq. \eqref{eq:c6NSI})
\begin{align}
-0.182 < \,&\varepsilon_{ee}^{eV} < 0.264\,, &\quad -0.120 < \,&\varepsilon_{\mu\mu}^{eV} < 0.120\,, 
\\
\label{eq:eps:uV:oscill}
-0.008 < \,&\varepsilon_{ee}^{uV} < 0.618\,, &\quad -0.111 < \,&\varepsilon_{\mu\mu}^{uV} < 0.402\,, 
\\
\label{eq:eps:dV:oscill}
-0.012 < \,&\varepsilon_{ee}^{dV} < 0.361\,, &\quad -0.103 < \,&\varepsilon_{\mu\mu}^{dV} < 0.361\,.
\end{align}
There are also bounds on operators that change neutrino flavor, or involve $\nu_\tau$ (for details see Ref. \cite{Esteban:2018ppq}). The oscillations do not constrain NSI couplings to strange quarks, because the corresponding forward scattering matrix elements vanish.

The above ranges on $\varepsilon_{\alpha\beta}^{fV}$ imply the following lower bounds on the NP scale, for $f=e(u,d)$, setting $\C_{1,f}^{(6)}=1$ in Eq. \eqref{eq:lightDM:Lnf5},
\begin{align}
\nu_e\to \nu_e:&\qquad  \Lambda > 571\,(373,\,488) \,\rm GeV\,, 
\\
\nu_\mu\to \nu_\mu:& \qquad   \Lambda > 847\,(463,\,488) \,\rm GeV\,.
\end{align}

\subsection{Constraints on NSI from \CE}\label{subsec:cohlimit}
Roughly a year ago the COHERENT collaboration measured for the first time the cross section for coherent neutrino-nucleus scattering \cite{Akimov:2017ade} (see also data release in \cite{Akimov:2018vzs}). The target was 14.6 kg of CsI[Na], while stopped pion decays, $\pi^+ \rightarrow \nu_\mu \,(\mu^+ \rightarrow e^+ \,\nu_e \,\bar \nu_\mu)$, acted as a source of neutrinos.
The resulting time integrated neutrino fluxes per energy interval, $\phi_{\nu_i}$, are well known \cite{Liao:2017uzy}, 
\begin{align}
\label{eq:phi:nue}
\phi_{\nu_e}(E_\nu) &= {\cal N} \frac{192E_{\nu}^2}{m_\mu^3}\left(\frac{1}{2} - \frac{E_\nu}{m_\mu} \right)\,,
\\
\label{eq:phi:numubar}
\phi_{\bar\nu_\mu}(E_\nu) &= {\cal N} \frac{64E_{\nu}^2}{m_\mu^3}\left(\frac{3}{4} - \frac{E_\nu}{m_\mu} \right)\,,
\\
\label{eq:phi:numu}
\phi_{\nu_\mu}(E_\nu) &= {\cal N}\delta\left(E_\nu - \frac{m_\pi^2 - m_\mu^2}{2 m_\pi} \right)\,.
\end{align}
Here $m_\pi$ ($m_\mu$) is the charged pion (muon) mass, $E_\nu$ the energy of the neutrino, and ${\cal N}={r N_{\rm POT}}/({4\pi L^2})$ the time integrated neutrino flux, for each flavor, reaching the COHERENT detector. It depends on $N_{\rm POT} = 1.73\times10^{23}$, the delivered number of protons on target (POT), on $r=0.08$, the number of neutrinos per flavor produced for each POT, and on $L=19.3$m, the distance between the neutrino source and the detector.

The expected number of \CE events for each neutrino flavor, $\alpha = \nu_e, \nu_\mu, \bar \nu_\mu$, is
\beq \label{eq:totN}
\frac{d N_\alpha}{d E_R} = \sum_{N=n,p} n_N \int_{E_{\nu,{\rm min}}}^{E_{\nu,{\rm max}}}dE_\nu \phi_\alpha(E_\nu) \frac{d\sigma_A (E_\nu)}{dE_R},
\eeq 
where $E_R$ is the nuclear recoil energy, $E_\nu$ the energy of the incoming neutrino, and $n_N$ the number of nucleons of type $N=n,p,$ in the detector. The lower boundary in the integration over $E_\nu$ is given by $E_{\nu,{\rm min}} \approx \sqrt{M_A E_R/2}$, the minimal energy neutrinos need to have in order to induce nuclear recoil energy $E_R$.
The upper integration boundary, $E_{\nu, {\rm max}}$, is given by the highest energy in the incoming neutrino flux. 
The $\nu_e$ and $\bar \nu_\mu$ are produced in muon decay and thus have the maximal energy $E_{\nu,{\rm max}}=m_\mu/2$, while $\nu_\mu$ is produced in pion decay, and has $E_{\nu,{\rm max}}\approx 30$ MeV. 
The maximal nuclear recoil energy deposited by $\nu_e$ and $\bar\nu_\mu$ in the detector is thus $E_{R,{\rm max}}\simeq 47$ keV, while for $\nu_\mu$ it is  $E_{R,{\rm max}}\simeq 15$ keV. The differential elastic neutrino--nucleus scattering cross section, $d\sigma_A/dE_R$, is given in Eq.~\eqref{eq:cross}. 

The prediction for the total number of \CE events expected in the COHERENT experiment is obtained by integrating Eq.~\eqref{eq:totN} over $E_R\in[0,47]$~keV, convoluted with the signal acceptance fraction for COHERENT, given in Fig. S9 of  \cite{Akimov:2017ade} (which has an onset at about $4.25$ keV). The experimentally allowed difference from the SM prediction then translates into bounds on the Wilson coefficients for the NSI operators, $\hat \C_{a}^{(d)}$, Eq.~\eqref{eq:lightDM:Lnf5}, using Eqs.~\eqref{eq:Msquared},  \eqref{eq:RM}-\eqref{eq:ci:isospinrel}, and \eqref{eq:c1(0)}-\eqref{eq:c12(2)}.

\begin{table}
\begin{center} 
\begin{tabular}{cccccccc}
 \hline\hline
~~~~opers.~~~~& ~~~$ F_{i}$~~~ & ~~~$\sigma_{F_i}$~~~ & ~~~$\sigma_\alpha$~~~&
~~~~opers.~~~~ & ~~~$ F_{i}$~~~ & ~~~$\sigma_{F_i}$~~~ & ~~~$\sigma_\alpha$~~~\\
\hline
 $\hat\C^{(5)}_{1},\hat\C^{(6)}_{1,u/d},\hat\C^{(7)}_{8,u/d;10,u/d}$ & $F_1^{u,d/N}(0) $ & $0\%$ & 0.28 & 
 $\hat\C^{(7)}_{3}$ &$ F_{G}^{N}(0)$  & $2\%$ & 0.28
 \\
$\hat\C^{(6)}_{1,s},\hat\C^{(7)}_{8,s;10,s}$ &$ F_{1}^{s/N}{}'(0) $  & $50\%$ & 0.57 & 
$\hat\C^{(7)}_{4}$ & $ F_{\tilde G}^{N}(q)$  & 20\% & 0.39
\\
 $\hat\C^{(6)}_{2,u/d}, \hat\C^{(7)}_{9,u/d;11,u/d}$ &$ F_{A}^{u,d/N}(0)$  &  $3-7\%$ & 0.34 & 
$\hat\C^{(7)}_{7,u}$  & $F_{T,0}^{u/N}(0)$ & $2\%$  & 0.34
 \\
$\hat\C^{(6)}_{2,s},\hat\C^{(7)}_{9,s;11,s}$&$ F_{A}^{s/N}(0)$  & $16\%$ & 0.37 & 
$\hat\C^{(7)}_{7,d}$ & $F_{T,0}^{d/N}(0)$ & $4\%$  & 0.34
\\
$\hat\C^{(7)}_{5,u/d}$ &$ F_{S}^{u,d/N}(0)$  & $28 - 31 \%$  & 0.40 & 
$\hat\C^{(7)}_{7,s}$ & $F_{T,0}^{s/N}(0)$ & $270\%$ & 2.7
 \\
$\hat\C^{(7)}_{5,s}$ &$ F_{S}^{s/N}(0)$  & $18 \%$ & 0.33 & 
$\hat\C^{(7)}_{6,q}$ & $a_{P,(\pi/\eta)}^{q/N}$ & $4 - 11 \%$ & 0.35 
\\
\hline\hline
\end{tabular}
\caption{Form factors uncertainties, $\sigma_{F_i}$, and the resulting relative theoretical uncertainties on \CE cross sections,  $\sigma_\alpha$, for different Wilson coefficients $\hat\C_a^{(d)}$ (see the main text). Unless specified, the uncertainty is the same for all quark flavors and for both nucleons. For $F_{T,0}^{q/N}(0)$ the quark masses are fixed to $m_u=2.3$ MeV, $m_{d}=4.8$ MeV, $m_s=95$ MeV.
} 
\label{table:sigmaalpha}
\end{center}
\end{table}

In the numerical analysis we take only a single NSI Wilson coefficient at a time to be nonzero (apart from the SM contributions, Eq.~\eqref{eq:c6udsSM}). For simplicity we assume that the NSI affect either only  $\nu_e$ or only $\nu_\mu$. 
To derive the 90 \% C.L. allowed ranges for $\hat \C_{a}^{(d)}$ we follow the COHERENT collaboration \cite{Akimov:2017ade, Fogli:2002pt} and define,
\beq
\chi^2\left( \hat{\cal C}_{a}^{(d)}, \alpha\right) = \frac{\left( N_{\rm meas} - N_{\rm th}\big( \hat{\cal C}_{a}^{(d)} \big)(1 + \alpha) \right)^2}{\sigma_{\rm stat}^2} + \left( \frac{\alpha}{\sigma_\alpha} \right)^2,
\eeq
where $ N_{\rm meas} = 142$ is the number of detected \CE events, $\sigma_{\rm stat}=31$ its statistical uncertainty, and $N_{\rm th}\big( \hat {\cal C}_{a}^{(d)} \big)$ the number of \CE events when Wilson coefficent $\hat{\cal C}_{a}^{(d)}$ is taken to be nonzero. The theoretical uncertainties are taken into account by marginalizing $\chi^2$ over the parameter $\alpha$.  The relative theory error, $\sigma_\alpha$, on the prediction for $N_{\rm NSI}\big( \hat {\cal C}_{a}^{(d)} \big)$, depends on which Wilson coefficent $\hat{\cal C}_{a}^{(d)}$ we consider, and is a quadratic sum of errors from: the uncertainty on signal acceptance ($\pm5\%$), neutrino flux ($\pm10\%$), quenching factor ($\pm25\%$), from nuclear response functions $W_i$ (estimated conservatively, both for scattering on I and Cs, as $\pm 10\%$ for $W_M$, which multiplies the $F_1^{q/N}$, $F_G^N, F_S^{q/N}$ form factors, and $\pm 20\%$ for $W_{\Sigma, \Sigma'}$ response functions, multiplying the other form factors), and from the nucleon form factors ($\sigma_{F_i}$ listed in Table \ref{table:sigmaalpha}). The central values for the form factors and the uncertainties are taken from \cite{Bishara:2017pfq}.  The resulting $\sigma_\alpha$ are shown in Table \ref{table:sigmaalpha} for each of the Wilson coefficients.  For the dipole, $\hat\C_1^{(5)}$,  in general two NR operators contribute, ${\cal O}_{3,N}^{(0)}$ and ${\cal O}_{1,N}^{(2)}$. However, for heavy elements the latter is negligible, giving the estimate for $\sigma_\alpha$ in  Table \ref{table:sigmaalpha}. The central values for $W_i$ are taken from \cite{Anand:2013yka}. Our estimates for the theoretical errors on $W_i$ are educated guesses. While this suffices at present, since these are subleading to the other uncertainties, a dedicated study would be desired in the future.

\begin{figure}
\includegraphics[width=0.5\textwidth]{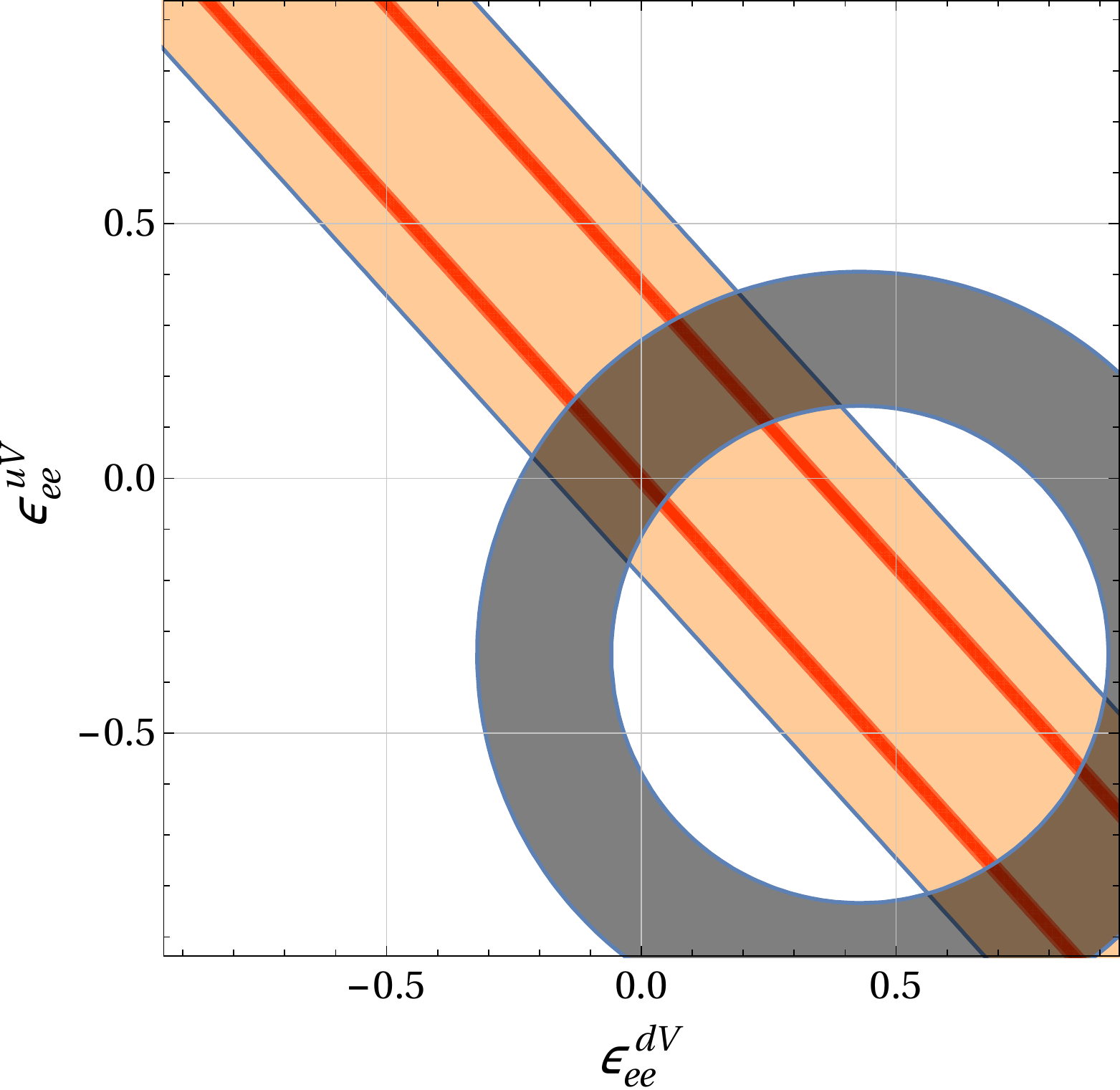}
\centering
\caption{The regions allowed at 90\% C.L. for electron neutrino NSI with vector up and down quark currents \eqref{eq:LNSI}. The black, orange and red regions correspond to CHARM, COHERENT and expected NaI 2T constraints, respectively, see text for details.}
         \label{fig:plotNSI}
\end{figure}

Our prediction for the SM rate in the COHERENT detector is  $N_{\rm th}\big( \hat {\cal C}_{a}^{(d)}|_{\rm SM} \big)=188\pm 53$ events. Comparison of this prediction with the COHERENT measurement gives the following 90\% C.L.  bounds on the NSI due to dimension 6 operators,
\begin{align}
-0.11 &<\varepsilon_{ee}^{uV} < 0.49\,, \quad &-0.10& <\varepsilon_{ee}^{dV} < 0.44\,,
\\
-0.06 &<\varepsilon_{\mu\mu}^{uV} < 0.12\,, \quad &-0.06& <\varepsilon_{\mu\mu}^{dV} < 0.11.
\end{align}
This is comparable to the sensitivity obtained in the global fits to the oscillation data, cf. Eqs. \eqref{eq:eps:uV:oscill}, \eqref{eq:eps:dV:oscill}. 
The limits on $\varepsilon_{\alpha\beta}^{uV}, \varepsilon_{\alpha\beta}^{dV}$ are corelated, see Fig. \ref{fig:plotNSI} for the case of $\nu_e$.
For NSI couplings to the strange quark we obtain a relatively weak bound, $|\varepsilon_{ee,\mu\mu}^{sV}| \lesssim 10^3$, because the sensitivity comes only from the ${\mathcal O}(q^2)$ term in the expansion of $F_1^{s/N}(q^2)$, see Eq. \eqref{eq:F1exp}. The axial couplings $\varepsilon_{ee}^{qA}$ and $\varepsilon_{\mu\mu}^{qA}$ are also poorly constrained, since they lead to spin-dependent interactions that are not coherently enhanced.

\begin{figure}[t]
  \includegraphics[height=0.33\textheight]{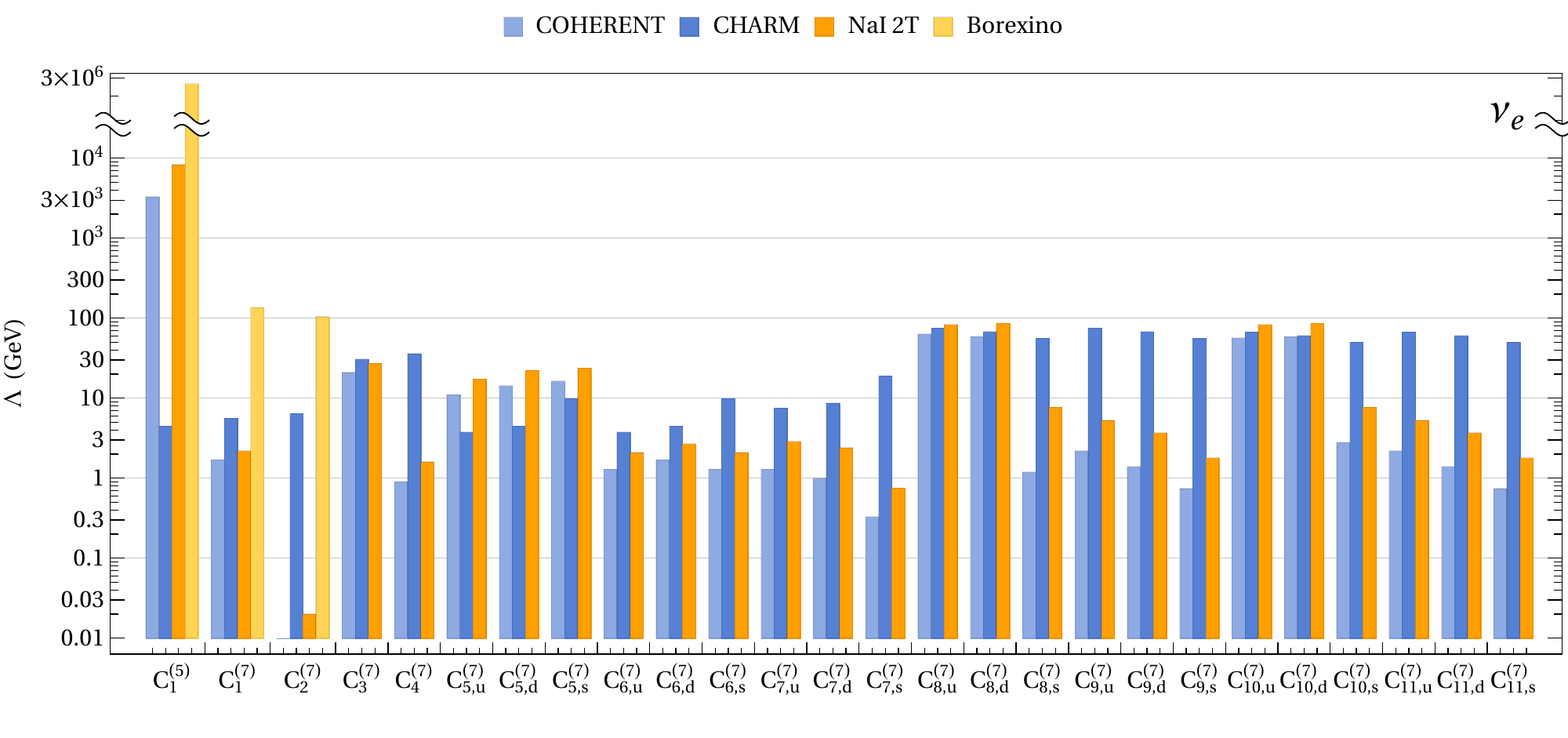}\\ \vspace{1cm}
  \includegraphics[height=0.33\textheight]{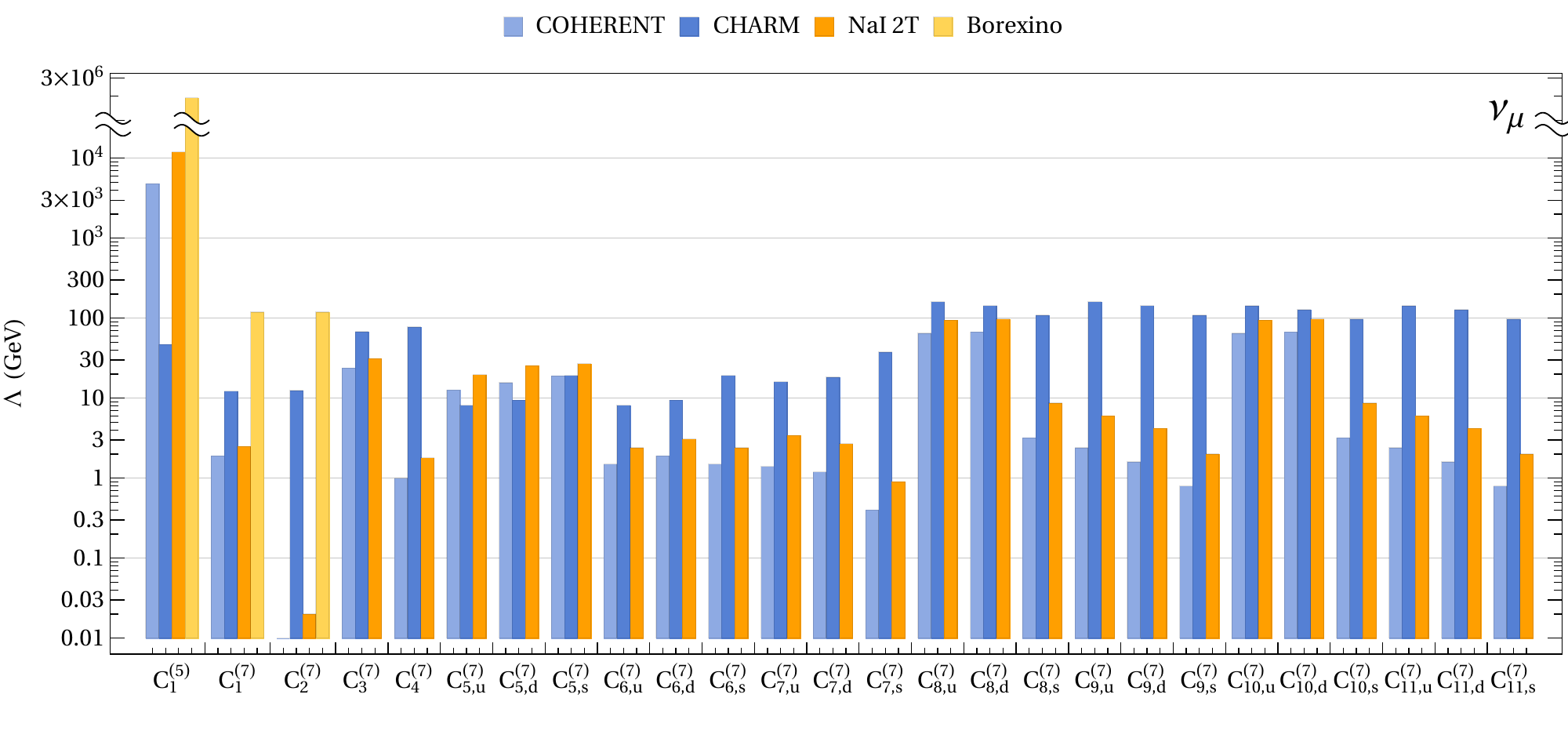}
\caption{Limits from COHERENT, CHARM, Borexino, and projected limits from a NaI 2T experiment on the scale $\Lambda$ of dimension 5 and dimension 7 NSI operators for electron neutrinos (top) and muon neutrinos (bottom). 
}
\label{fig:chart-all}
\end{figure}

\begin{figure}[t]
  \includegraphics[height=0.33\textheight]{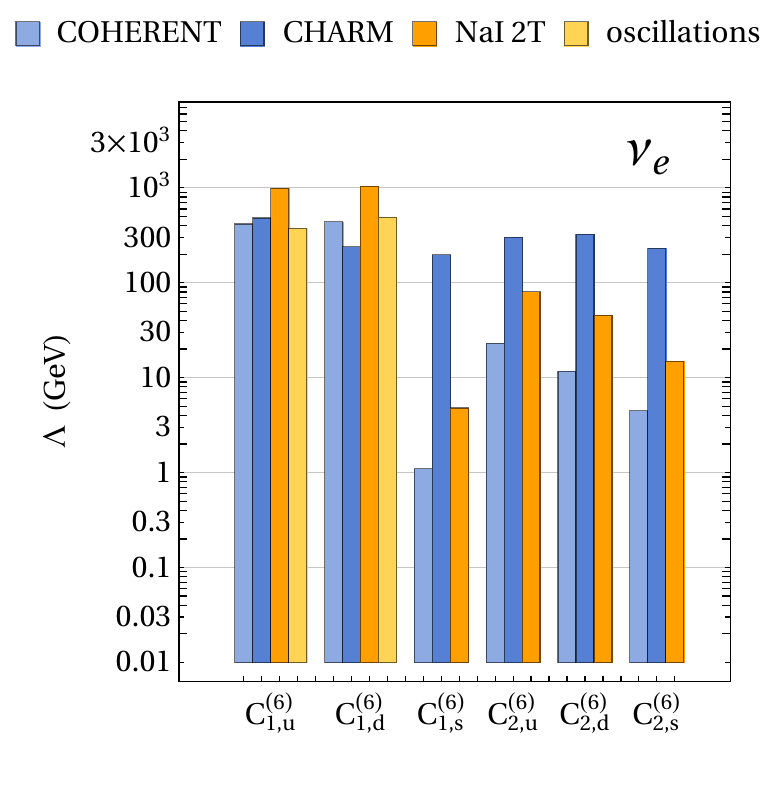} ~~~~~ \includegraphics[height=0.33\textheight]{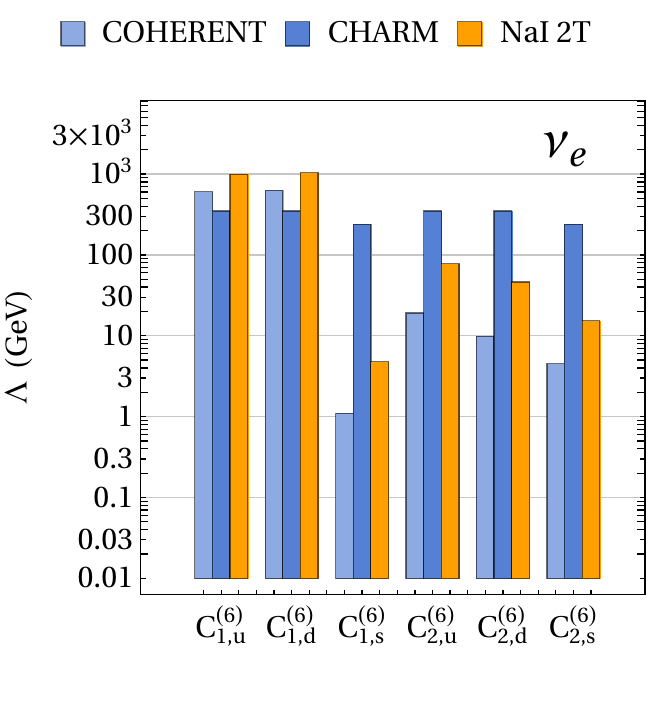}\\ \vspace{1cm}
  \includegraphics[height=0.33\textheight]{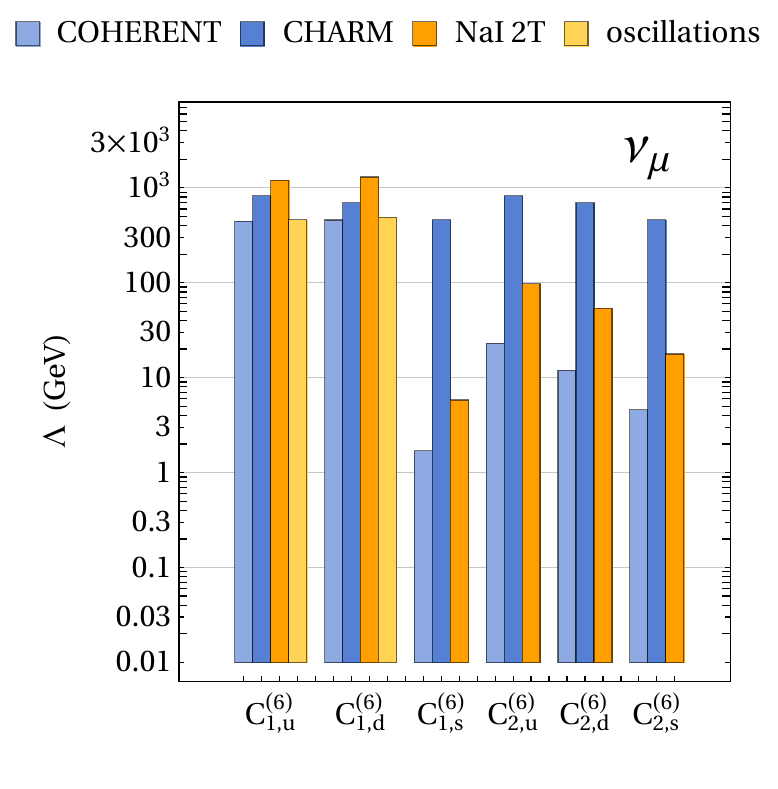} ~~~~~ \includegraphics[height=0.33\textheight]{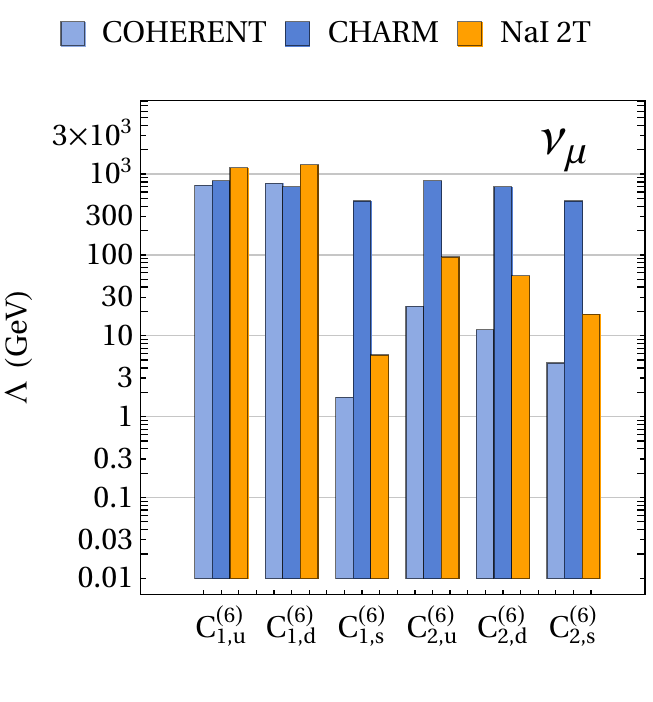}
\caption{Limits from COHERENT, CHARM, neutrino oscillations and projected limits from a NaI 2T experiment on the scale $\Lambda$ of dimension 6 operators for electron neutrinos (top) and muon neutrinos (bottom). The two panels on the left correspond to lepton flavor conserving operators ($\nu_i \to \nu_i$), the ones on the right to lepton flavor violating operators ($\nu_i \to \nu_j$, $i\neq j$). 
}
\label{fig:chart-all2}
\end{figure}

We collect the 90\% C.L. bounds on the NSI coefficients in Table \ref{table:Lambdanue} for $\nu_e\to \nu_X$ transitions and in Table \ref{table:Lambdanumu} for $\nu_\mu \to \nu_X$ transitions. The bounds on $\hat C_a^{(d)}=C_a^{(d)}/\Lambda^{d-4}$ are converted to lower bounds on the NP scale, $\Lambda$, setting the dimensionless Wilson coefficients to $C_a^{(d)}=1$. The two tables also contain bounds from deep inelastic scattering, Section \ref{sec:Charm}, and 
the expected reach from the proposed 2 ton NaI experiment \cite{Akimov:2018ghi}, cf.~Table \ref{table:propexp}. For easier comparison the bounds are also illustrated with barcharts in Fig. \ref{fig:chart-all} for dimension 5 and dimension 7 operators, and in Fig. \ref{fig:chart-all2} for dimension 6 operators. For dimension 6 operators with vector couplings to quarks there are two possibilities. For flavor diagonal transitions,  $\nu_\mu \to \nu_\mu$ and  $\nu_e \to \nu_e$, the NSI contributions interfere with the SM. The resulting bounds are shown in Fig. \ref{fig:chart-all2} (left).  If the transition is not flavor diagonal, the bounds are weaker, shown in the right panels in  Fig. \ref{fig:chart-all2}. For all the other operators the final neutrino,  $\nu_X$, can be of any neutrino flavor, including sterile neutrinos. 

The bound on $\hat C_{7,s}^{(7)}$ is controlled by the $s$-quark tensor form factor $F_{T,0}^{s/N}$, which is not well known.  At present it is even consistent with zero, which  
 implies that there is no reliable bound on the $\hat\C_{7,s}^{(7)}$. In Tables  \ref{table:Lambdanue}, \ref{table:Lambdanumu}  we use the central value $F_{T,0}^{s/N}/m_s=3.2\cdot 10^{-4}$ from \cite{Bishara:2017pfq}, with $\sigma_{F_i}=0$, to gauge the rough potential reach of COHERENT, once lattice determinations of $F_{T,0}^{s/N}$ become precise enough. 
We use a similar approach for the Rayleigh operators, $\hat C_{1,2}^{(7)}$ where we (i) use the phenomenological form factor for two body currents for $\hat C_{1}^{(7)}$, Eq. \eqref{eq:twobody:form}, (ii) neglect the two-body current contributions for $\hat C_{2}^{(7)}$, and (iii) and use the NDA estimates \eqref{eq:NDA:Rayleigh} for the non perturbative single nucleon matrix elements, and do not assign any associated errors to these approximations. The bounds shown are thus just giving a rough potential reach of \CE experiments once theoretical errors will be under control (with probably a better guesstimate for  $\hat C_{1}^{(7)}$ than  $\hat C_{2}^{(7)}$). 

Tables  \ref{table:Lambdanue}, \ref{table:Lambdanumu} show in the case of dimension 6 operators  the bounds for flavor non-diagonal processes $\nu_{e}\to\nu_X$ with $X\ne e$, and $\nu_{\mu}\to\nu_X$ with $X\ne \mu$, respectively. For flavor diagonal transitions, $\nu_e\to \nu_e$ or $\nu_\mu\to \nu_\mu$,  the bounds on NSI from the COHERENT measurement are instead
\begin{align}
{\cal C}_{1,u(d,s)}^{(6)}:&\qquad  \Lambda > 417(440,1.1)~\text{GeV}&(\nu_e),&\qquad &\Lambda > &443(458,1.7)~\text{GeV} &(\nu_\mu),& 
\\
{\cal C}_{2,u(d,s)}^{(6)}:&\qquad   \Lambda > 22.9(11.7,4.5)~\text{GeV}&(\nu_e),&\qquad  &\Lambda >& 23.1(11.9,4.6)~\text{GeV} &(\nu_\mu),&
\end{align} 

\begin{table}
\begin{center} 
\begin{tabular}{cccccccc }
 \hline\hline
\multicolumn{8}{c}{~~~~~~Lower bounds on $\Lambda$ in GeV, $\nu_e\to \nu_X$ transitions~~~~~~}  \\
~~$ \hat {\cal C}_{i,q}^{(d)}$~~ &~COHERENT~&~~CHARM~~&~~NaI 2T~~&~~$ \hat {\cal C}_{i,q}^{(d)}$~~ &~COHERENT~&~~CHARM~~&~~NaI 2T~~\\
\hline
$ \hat {\cal C}_{1}^{(5)}$    & $3.3 \cdot 10^3$    &   4.5 & $8.3 \cdot 10^3$ & $ \hat {\cal C}_{7,u}^{(7)}$   & $1.3 \cdot\left( \tfrac{2.3\, \rm MeV}{m_u} \right)^{1/3}$   &   7.5 & $2.9\cdot\left(\tfrac{2.3\, \rm MeV}{m_u} \right)^{1/3}$ \\
$ \hat {\cal C}_{1,u}^{(6)}$     & $603$    &   $349$ & $993$ &$ \hat {\cal C}_{7,d}^{(7)}$  & $1 \cdot \left(\tfrac{4.8\, \rm MeV}{m_d} \right)^{1/3}$   &   8.7 & $2.4\cdot \left(\tfrac{4.8\, \rm MeV}{m_d} \right)^{1/3}$\\
 $ \hat {\cal C}_{1,d}^{(6)}$    & $632$    &   $349$ & $1.04\cdot 10^3$ &$ \hat {\cal C}_{7,s}^{(7)\dagger}$   & $0.33  \cdot \left(\tfrac{95\, \rm MeV}{m_s} \right)^{1/3}$   &   19  & $0.75 \cdot \left(\tfrac{95\, \rm MeV}{m_s} \right)^{1/3}$\\
$ \hat {\cal C}_{1,s}^{(6)}$  & 1.1  &   $239$ & 4.8 &$ \hat {\cal C}_{8,u}^{(7)}$   & $64$   &  75 & 83.4\\
 $ \hat {\cal C}_{2,u}^{(6)}$     & $19.1$    &   $349$ & 78.2 &$ \hat {\cal C}_{8,d}^{(7)}$  & $59$   &  68 & 86.2\\
 $ \hat {\cal C}_{2,d}^{(6)}$    & $9.8$    &   $349$ & 46.4 & $ \hat {\cal C}_{8,s}^{(7)}$   & $1.2$    & 56 & 7.7\\
$ \hat {\cal C}_{2,s}^{(6)}$  & $4.5$    &   $239$ & 15.4& $ \hat {\cal C}_{9,u}^{(7)}$   & $2.2$   &   75 & 5.3 \\
$ \hat {\cal C}_{1}^{(7)\dagger}$    & 1.7  &  5.6 &  2.2 &$ \hat {\cal C}_{9,d}^{(7)}$  & $1.4$   &   68 & 3.7\\
$ \hat {\cal C}_{2}^{(7)\dagger}$   & 0.01 &   6.4 & 0.02  & $ \hat {\cal C}_{9,s}^{(7)}$   & $0.74$    & 56 & 1.8\\
$ \hat {\cal C}_{3}^{(7)}$    & $21$    &   31 & 27.3  & $ \hat {\cal C}_{10,u}^{(7)}$   & $57$   &   67 & 83.4\\
$ \hat {\cal C}_{4}^{(7)}$   & $0.9$    &   36&  1.6 &$ \hat {\cal C}_{10,d}^{(7)}$  & $59$   &   61 & 86.2\\
$ \hat {\cal C}_{5,u}^{(7)}$   & $11$  &   3.8 & 17.3 &$ \hat {\cal C}_{10,s}^{(7)}$   & $2.8$    & 50 & 7.7\\
$ \hat {\cal C}_{5,d}^{(7)}$  & $14.4$  &   4.5 & 22.3 &$ \hat {\cal C}_{11,u}^{(7)}$   & $2.2$   &   67 & 5.3\\
$ \hat {\cal C}_{5,s}^{(7)}$   & $16.4$   &  9.9 & 23.7 &$ \hat {\cal C}_{11,d}^{(7)}$  & $1.4$   &   61 & 3.7\\
$ \hat {\cal C}_{6,u}^{(7)}$   & $1.3$    &   3.8& 2.1&$ \hat {\cal C}_{11,s}^{(7)}$   & $0.74$    & 50 & 1.8\\
$ \hat {\cal C}_{6,d}^{(7)}$   & $1.7$    &   4.5& 2.7 &---&---&---&---\\
$ \hat {\cal C}_{6,s}^{(7)}$  & $1.3$    &   9.9&  2.1  &---&---&---&---\\
\hline\hline
\end{tabular}
\caption{The 90 \% C.L. lower bounds on $\Lambda$ from COHERENT, CHARM, and NaI 2T for $\nu_e \to \nu_X$, ($X\ne e$, for $X=e$ see main text) NSI Wilson coefficients, $\hat \C_a^{(d)}$, Eq.~\eqref{eq:lightDM:Lnf5}, setting  $\C_a^{(d)}=1$, and assuming only one such NSI Wilson coefficient is nonzero. 
For $\hat C_{7,s}^{(7)}$ ($\hat C_{1,2}^{(7)}$) we use only the central value of the form factor (NDA estimates) so the bounds are merely indicative, see main text for details. } 
\label{table:Lambdanue}
\end{center}
\end{table}

\begin{table}
\begin{center} 
\begin{tabular}{cccccccc }
 \hline\hline
\multicolumn{8}{c}{~~~~~~Lower bounds on $\Lambda$ in GeV, $\nu_\mu\to \nu_X$ transitions~~~~~~}  \\
~~$ \hat {\cal C}_{i,q}^{(d)}$~~ &~COHERENT~&~~CHARM~~&~~NaI 2T~~&~~$ \hat {\cal C}_{i,q}^{(d)}$~~ &~COHERENT~&~~CHARM~~&~~NaI 2T~~\\
\hline
$ \hat {\cal C}_{1}^{(5)}$    & $4.8\cdot10^3$    &   47.1 & $1.2\cdot10^4$ & $ \hat {\cal C}_{7,u}^{(7)}$   & $1.4 \cdot \left(\tfrac{2.3\, \rm MeV}{m_u} \right)^{1/3}$   &   16.1  & $3.4 \cdot \left(\tfrac{2.3\, \rm MeV}{m_u} \right)^{1/3}$ \\
$ \hat {\cal C}_{1,u}^{(6)}$    & $726$    &   826 & $1.2\cdot 10^3$& $ \hat {\cal C}_{7,d}^{(7)}$  & $1.2 \cdot \left(\tfrac{4.8\, \rm MeV}{m_d}\right)^{1/3}$   &   18.3 & $2.7 \cdot \left(\tfrac{4.8\, \rm MeV}{m_d}\right)^{1/3}$  \\
 $ \hat {\cal C}_{1,d}^{(6)}$    & $767$    &   697 &  $1.3\cdot 10^3$& $ \hat {\cal C}_{7,s}^{(7)\dagger}$   & $0.4 \cdot \left(\tfrac{95\, \rm MeV}{m_s} \right)^{1/3}$   &   37.7 & $0.9 \cdot \left(\tfrac{95\, \rm MeV}{m_s} \right)^{1/3}$ \\
$ \hat {\cal C}_{1,s}^{(6)}$  & 1.74 &   463 & 5.8 &$ \hat {\cal C}_{8,u}^{(7)}$   & $64.7$   &   160.6 & 94.5\\
 $ \hat {\cal C}_{2,u}^{(6)}$    & $23.1$    &   826 & 94.3 & $ \hat {\cal C}_{8,d}^{(7)}$  & $67.5$   &   143.4 & 97.7\\
 $ \hat {\cal C}_{2,d}^{(6)}$    & $11.9$    &   697 & 55.5 & $ \hat {\cal C}_{8,s}^{(7)}$   & $3.2$    & 109.1 & 8.7 \\
$ \hat {\cal C}_{2,s}^{(6)}$  & $4.6$    &  463 & 18.5 & $ \hat {\cal C}_{9,u}^{(7)}$   & $2.4$   &   160.6 & 6\\
$ \hat {\cal C}_{1}^{(7)\dagger}$    & 1.9  &   12.3 & 2.5 &$ \hat {\cal C}_{9,d}^{(7)}$  & $1.6$   &   143.4 & 4.2\\
$ \hat {\cal C}_{2}^{(7)\dagger}$   & 0.01 &   12.5 & 0.02 & $ \hat {\cal C}_{9,s}^{(7)}$   & $0.8$    & 109.1 & 2 \\
$ \hat {\cal C}_{3}^{(7)}$    & $23.7$    &   67.9 & 31.2 & $ \hat {\cal C}_{10,u}^{(7)}$   & $64.7$   &  143.1 & 94.5\\
$ \hat {\cal C}_{4}^{(7)}$   & $1$    &   77.8 & 1.8 &$ \hat {\cal C}_{10,d}^{(7)}$  & $67.5$   &   127.7 & 97.7\\
$ \hat {\cal C}_{5,u}^{(7)}$   & $12.6$  &   8.2 & 19.7 &$ \hat {\cal C}_{10,s}^{(7)}$   & $3.2$    & 97.2 & 8.7\\
$ \hat {\cal C}_{5,d}^{(7)}$  & $15.6$  &   9.4 & 25.5 & $ \hat {\cal C}_{11,u}^{(7)}$   & $2.4$   &  143.1 & 6\\
$ \hat {\cal C}_{5,s}^{(7)}$   & $19$   &   19.3 & 27.1 &$ \hat {\cal C}_{11,d}^{(7)}$  & $1.6$   &   127.7 & 4.2\\
$ \hat {\cal C}_{6,u}^{(7)}$   & $1.5$    &   8.2 & 2.4 & $ \hat {\cal C}_{11,s}^{(7)}$   & $0.8$    & 97.2 & 2\\
$ \hat {\cal C}_{6,d}^{(7)}$   & $1.9$    &   9.4 & 3.1  &---&---&--- &---\\
$ \hat {\cal C}_{6,s}^{(7)}$  & $1.5$    &   19.3 & 2.4 & ---&---&---&---\\
\hline\hline

\end{tabular}
\caption{The 90 \% C.L. lower bounds on $\Lambda$ from COHERENT, CHARM and NaI 2T for $\nu_\mu \to \nu_X$ ($X\ne \mu$, for $X=\mu$ see main text) NSI Wilson coefficients, $\hat \C_a^{(d)}$, Eq.~\eqref{eq:lightDM:Lnf5}, setting  $\C_a^{(d)}=1$, and assuming only one such NSI Wilson coefficient is nonzero. 
For $\hat C_{7,s}^{(7)}$ ($\hat C_{1,2}^{(7)}$) we use only the central value of the form factor (NDA estimates) so the bounds are merely indicative, see main text for details. 
} 
\label{table:Lambdanumu}
\end{center}
\end{table}

The relative sizes of bounds in Tables \ref{table:Lambdanue}, \ref{table:Lambdanumu} are easy to understand using  approximate scalings in Eqs.~\eqref{eq:cohNDA1}-\eqref{eq:cohNDA3}.  The ${\cal O}(A^2)$ enhancements of operators ${\cal Q}_1^{(5)}$, ${\cal Q}_{1,q}^{(6)}$, ${\cal Q}_3^{(7)}$, ${\cal Q}_{5,q}^{(7)}$ and ${\cal Q}_{8,q}^{(7)}$, ${\cal Q}_{10,q}^{(7)}$, Eq. \eqref{eq:cohNDA2}, translate into more stringent bounds for $\Lambda$. The bounds are significantly weaker for the remaining non-enhanced operators. The bounds on operators with strange quarks are also weaker, since the corresponding form factors are smaller.

The bounds on NSI from \CE experiments are set to improve in the future with a number of new detectors either already taking data or being planned. The COHERENT collaboration is operating a 10kg Ge detector, a 22kg single-phase liquid Ar detector, and a 185 kg NaI[Tl] scintillating crystal detector \cite{Akimov:2018ghi}. The liquid Ar may increase to 1ton, and NaI[Tl] to 2 tons, in the future \cite{Berryman:2018jxt}, cf.~Table \ref{table:propexp}.
To take full advantage of these experimental progress an increased precision in the predictions of nuclear response functions and nuclear form factors will be called for. 

In Tables \ref{table:Lambdanue} and \ref{table:Lambdanumu} we show the expected improvements in the sensitivity to NSI due to the 2 ton NaI detector proposed by the COHERENT Collaboration \cite{Akimov:2018ghi}, with the same neutrino source but with a baseline of $28$m and a lower energy threshold of $\sim13$ keV. Furthermore, in the projections we assume that the total theoretical uncertainty is reduced 10-fold compared to the present ones, quoted in Table \ref{table:sigmaalpha}. This would  give the projected  total theoretical uncertainties $\sigma_{F_i}\sim 3-5 \%$. This will require more precise determinations of the neutrino flux, which is already planned, as well as a much better knowledge of the quenching factors, and major advances in the purely theoretical inputs -- the form factors and nuclear response functions entering the SM prediction. While such a decrease of uncertainties may be aggressive, they do give us a useful gauge of the potential reach of \CE experiments. 

In NaI detector the neutrino recoils on both the iodine and sodium nuclei. For the SM neutrino interactions the scattering on iodine completely dominates. The coherently enhanced cross section  is $\sim 40$ times larger for neutrino scattering on iodine as it is for sodium. This is the case also for coherently enhanced NSI interactions, where scattering on iodine similarly dominates.
Spin-dependent interactions, on the other hand, can be comparable, depending on the operator. 
We find that scattering on iodine dominates except for the operators $\Q_{2,q}^{(6)}$, $\Q_{7,q}^{(7)}$ and $\Q_{9,q}^{(7)}$, for which the main contribution to the scattering rate is from sodium, while the two cross sections are of the same order for $\Q_{1}^{(5)}$. The expected bounds from $\nu_{e, \mu}\to \nu_X$ scattering are shown in Tables \ref{table:Lambdanue} and \ref{table:Lambdanumu}. 

For flavor diagonal transitions, $\nu_{e, \mu}\to \nu_{e,\mu}$, the expected bounds on dimension 6 operators from the NaI 2T detector are, 
\begin{align}
{\cal C}_{1,u(d,s)}^{(6)}:&\qquad  \Lambda > 993(1040,4.8)~\text{GeV} &(\nu_e),&\qquad &\Lambda >& 1200(1300,5.8)~\text{GeV} &(\nu_\mu),& 
\\
{\cal C}_{2,u(d,s)}^{(6)}:&\qquad   \Lambda > 80.3(45.2,14.9)~\text{GeV} &(\nu_e),&\qquad  &\Lambda >& 97.6(53.8,17.8)~\text{GeV} &(\nu_\mu),&
\end{align} 

\begin{table}
\begin{center} 
\begin{tabular}{cccccc }
 \hline\hline
~~~~ &~$T_{\rm th}$~&~~Baseline (m)~~&~~Target~~&~~Mass (kg)~~&~~~Source~~~\\
\hline
 NaI 2T  COH\cite{Akimov:2018ghi}    & 13 keV    &   28 & NaI & 2000 & SPD  \\
Ge COH\cite{Akimov:2018ghi}    & 5 keV    &   22 & Ge & 10 & SPD  \\
LAr COH \cite{Akimov:2018ghi}    & 20 keV    &   29 & Ar & 22 & SPD  \\
\hline
RED100 \cite{Akimov:2012aya}    & 500 eV    &   19 & Xe & 100 & 3 GW reactor \\
MINER \cite{Agnolet:2016zir}    & 10 eV    &   1 & $^{72}$Ge+$^{28}$Si & 30 & 1 MW reactor  \\
CONNIE \cite{Aguilar-Arevalo:2016khx}    & 28 eV    &   30 & Si & 1 & 3.8 GW reactor  \\
RICOCHET \cite{Billard:2016giu}    & 50-100 eV    &   $<$10 & Ge/Zn & 10 & 8.54 GW reactor  \\
NU-CLEUS \cite{Strauss:2017cuu}    & 20 eV    &   $<$10 & CaWO$_4$,Al$_2$O$_3$ & 0.001 & 8.54 GW reactor  \\
$\nu$GEN \cite{Belov:2015ufh}    & 350 eV    &   10 & Ge & 4$\times$0.4 & 3 GW reactor  \\
CONUS \cite{2017:Lindner:Conus}    & $<$300 eV    & 17 & Ge & 4 & 3.9 GW reactor  \\
TEXONO \cite{Wong:2010zzc}    & 150-200 eV    & 28 & Ge & 1 & 2$\times$2.9 GW reactors  \\
\hline\hline
\end{tabular}
\caption{A list of proposed experiments to detect \CE using (anti)neutrinos from Stopped Pion Decay (SPD) or from reactors, with recoil energy threshold, $T_{\rm th}$, the distance from the source, the target material and its mass given in 2nd to 4th columns. 
} 
\label{table:propexp}
\end{center}
\end{table}

While COHERENT uses stopped pions as a source of neutrinos, there are also a number of  planned or already operating \CE experiments that use reactor antineutrinos, see Table \ref{table:propexp} as well as, e.g., Refs. \cite{Billard:2018jnl,Canas:2018rng,Qian:2018wid}. Reactors produce large quantities of low energy electronic antineutrinos. On average about $\sim$6 antineutrinos are produced per fission, for a total of $\sim 2\times10^{20}\bar\nu_e$ per second per GW of thermal reactor power \cite{Qian:2018wid,Huber:2011wv,Vogel:1989iv}, with a maximum energy of $\sim8$ MeV. 

As two representative examples of reactor \CE experiments we chose the proposed RED100 \cite{Akimov:2012aya} and MINER \cite{Agnolet:2016zir} experiments, and checked their respective sensitivities to different NSI, assuming in both cases  a total uncertainty of $\sigma_{F_i}\sim 10\%$. RED100 has a proposed 100kg target of liquid Xenon, with a baseline of $19$m from a 1GW reactor and an energy threshold of $500$ eV.  While this energy threshold is lower than for stopped pion decay experiments, it is the highest among the reactor experiments. MINER \cite{Agnolet:2016zir} has  a proposed 30kg detector composed of ${}^{72}$Ge and $^{28}$Si in 2:1 ratio,
with a baseline of $1$m from a 1MW detector and a very low energy threshold of $10$ eV.

Because of lower thresholds, both RED100 and MINER would have one to two orders of magnitude better sensitivity to the neutrino dipole moment, $\hat \C_1^{(5)}$, compared to NaI 2T. The reach for the vector NSI current operators, $\hat \C_{1,u(d)}^{(6)}$ could exceed the ones from global oscillation fits, while there would be also an appreciable improvement on the derivative couplings, $\hat \C_{8,q}^{(7)}, \hat \C_{10,q}^{(7)}$.  The operators inducing spin-dependent interactions, on the other hand,  cannot be probed better in these experiments, since xenon has a smaller nuclear spin than iodine, while ${}^{72}$Ge and ${}^{28}$Si have nuclear spin 0.

Future experiments can improve their sensitivity to specific NSI operators by changing experimental conditions. As just stated, lowering the energy threshold improves the sensitivity to the magnetic monopole operator ${\cal Q}_1^{(5)}$, whose contribution is enhanced by the $1/\vec q^{\,2}$ photon pole, cf. Eqs.  \eqref{eq:c56(0)}, \eqref{eq:cohNDA2}. For instance, the CONNIE collaboration proposes a 1kg Si detector with the energy threshold of $28$ eV that would be situated $30$m away from the reactor \cite{Aguilar-Arevalo:2016khx}. Taking all the other parameters as in our projection for the NaI 2T limit, leads to a projected limit for ${\cal Q}_1^{(5)}$ of $\Lambda\simeq 50$ TeV, to be compared with $8.3$ TeV at NaI 2T, Table~\ref{table:Lambdanue}. 

\begin{figure}
   \includegraphics[scale=0.5]{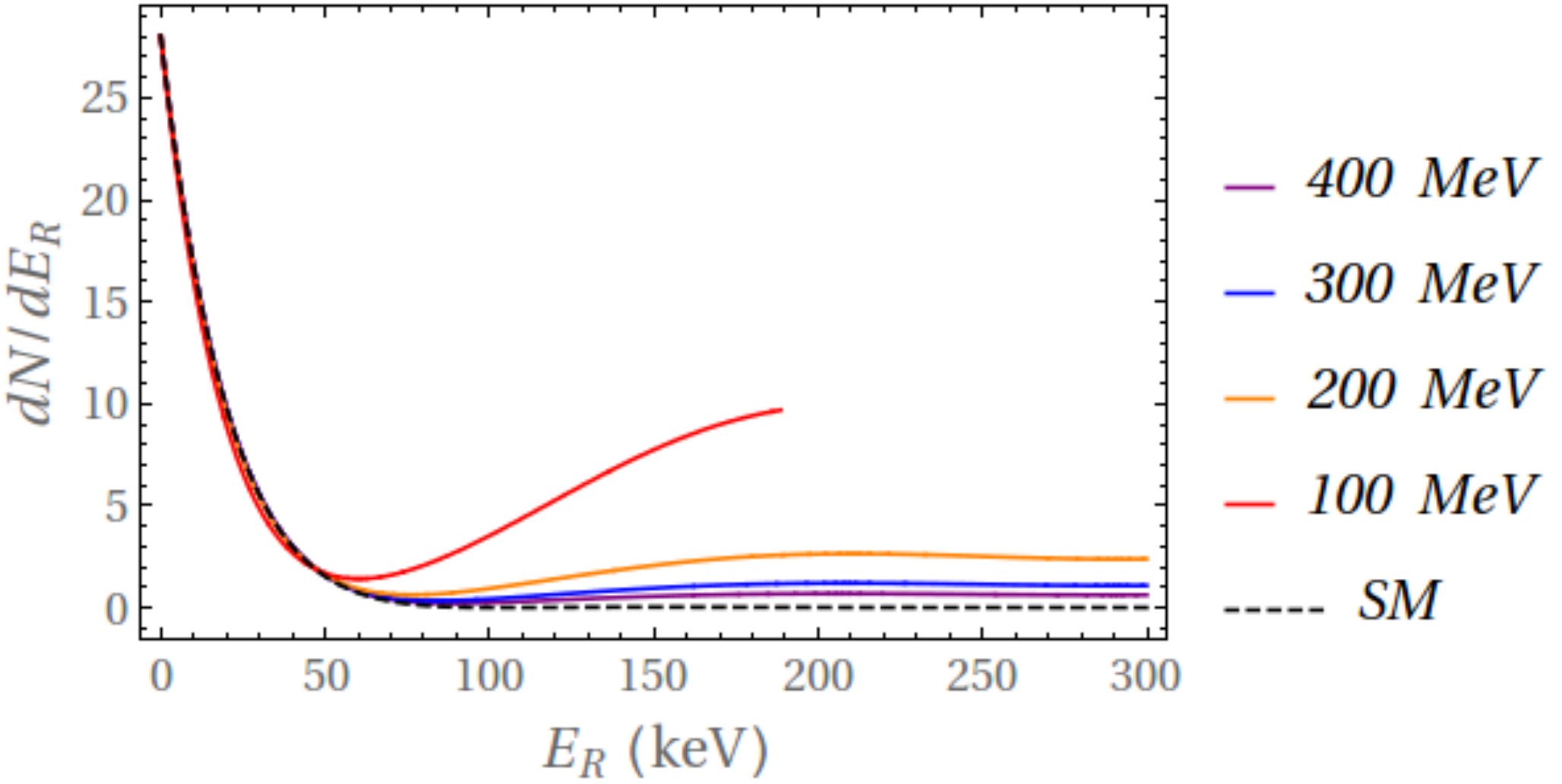}
\caption{Effect at high recoil energy of the operator ${\cal Q}_4^{(7)}$, assuming $\hat{\cal C}_4^{(7)} = 0.1$ GeV$^{-3}$, for four different neutrino energies, using monochromatic neutrino beams for illustration. The black dashed line is the SM rate.}
\label{fig:HEC47}
\end{figure}

Varying the neutrino energy, $E_\nu$, as well as increasing the range of nuclear recoil energies, $E_R$, can also be beneficial. In this way one could uncover $\vec q^{\,2}$ dependence of the neutrino scattering cross section that differs from the SM one. We illustrate this in Fig. \ref{fig:HEC47}, taking as the example the $\Q_4^{(7)}$ operator. The $\Q_4^{(7)}$ operator matches onto the NR operator ${\cal O}_{1,N}^{(1)}$, whose matrix element squared starts at ${\mathcal O}(\vec q^{\,4})$ and is independent of $E_\nu$, see Eq. \eqref{eq:RSpp}. In contrast, the SM matrix element squared grows with neutrino energy, $\propto E_\nu^2$.  For high enough recoil energies, $E_R\sim {\mathcal O}(100{\rm~keV})$, and neutrino energies $E_\nu\sim {\mathcal O}(100{\rm~MeV})$, there is clear distinction between the scattering rate with and without the presence of $\Q_4^{(7)}$ (the NP scale was taken very low in Fig. \ref{fig:HEC47} to exaggerate the effect).

\subsection{Constraints on NSI from deep inelastic scattering}
\label{sec:Charm}

The CHARM neutrino detector \cite{Diddens:1980xz} was composed of 78 plates of marble (CaCO$_3$) with a total fiducial mass of 87.4 tons. The target is to a very good approximation an isoscalar -- the correction to the cross section from the isotriplet component is ${\mathcal O}(0.2 \%)$ \cite{Allaby:1988bb}. Data were recorderd exposing the detector to neutrinos and antineutrinos from the CERN 400~GeV SPS proton beam dump. 

In our analysis we focus on the ratio of NC and CC total cross sections for electron neutrinos and antineutrinos that has been measured to be \cite{Dorenbosch:1986tb}
\beq \label{eq:ReCHARM}
R_e = \frac{\sigma(\nu_e N \rightarrow \nu_e X) + \sigma(\bar\nu_e N \rightarrow \bar\nu_e X)}{\sigma(\nu_e N \rightarrow e^- X) + \sigma(\bar\nu_e N \rightarrow e^+ X)} = 0.406 \pm 0.140,
\eeq
where an equal flux of $\nu_e$ and $\bar \nu_e$ has been assumed, while similarly for muon neutrinos \cite{Allaby:1987vr}
\beq \label{eq:ReCHARM1}
R_{\nu_\mu} = \frac{\sigma(\nu_\mu N \rightarrow \nu_\mu X)}{\sigma(\nu_\mu N \rightarrow \mu^- X)} = 0.3093 \pm 0.0031,
\eeq
The ratios $R_e$ and $R_{\nu_\mu} $ are predicted in the SM to be \cite{Erler:2013xha,Falkowski:2017pss} 
\beq \label{eq:ReSM}
R_{e}^{\rm SM} = 0.3221 \pm 0.0006, \qquad R_{\nu_\mu}^{\rm SM} = 0.3156 \pm 0.0006.
\eeq
The dominant theoretical uncertainty in the two predictions is due to the approximation that the target was taken to be an isospin singlet, which is correct within ${\mathcal O}(0.2 \%)$.

Since the neutrino CC cross section is strongly constrained, we can assume that NSI only affect NC transitions. The ratio $R_e$ in Eq. \eqref{eq:ReSM} then receives the NSI correction as
\beq \label{eq:ReNSI}
R_{e}^{\rm NSI} = R_{e}^{\rm SM} + \frac{\Delta \sigma_{\rm NSI}}{\sigma_{\rm CC}},
\eeq
where $\sigma_{\rm CC}$ is the total neutrino and antineutrino CC cross section and $\Delta \sigma_{\rm NSI}$ is the NSI contribution to the NC cross section. In the analysis we take only a single NSI operator at a time to be nonzero. Similarly, $R_{\nu_\mu}$ is modified to 
\beq \label{eq:RnumuNSI}
R_{\nu_\mu}^{\rm NSI} = R_{\nu_\mu}^{\rm SM} + \frac{\Delta \sigma_{\rm NSI}^{\nu_\mu}}{\sigma_{\rm CC}^{\nu_\mu}},
\eeq
where the cross sections in the ratios now refer to the muonic neutrino only.
Comparing  Eqs.~\eqref{eq:ReNSI} and~\eqref{eq:RnumuNSI} with the experimental results in Eqs.~\eqref{eq:ReCHARM} and~\eqref{eq:ReCHARM1}, gives the upper bound on the allowed sizes of NSI Wilson coefficients, $\hat \C_a^{(d)}$, Eq. \eqref{eq:lightDM:Lnf5}. Taking the dimensionless Wilson coefficients to be $\C_a^{(d)}=1$, this translates to a 90\% CL lower bound on NP scale $\Lambda$ for each NSI operator, given in Tables  
\ref{table:Lambdanue} and \ref{table:Lambdanumu}. These bounds are also shown as dark blue bars in Figs. \ref{fig:chart-all} and \ref{fig:chart-all2}.

The relative values of bounds are easily understood from the matrix elements in Eqs.~\eqref{eq:c15matr}-\eqref{eq:c7matr}. In particular, the two scalar operators ${\cal Q}_{5,q}^{(7)}$ and ${\cal Q}_{6,q}^{(7)}$ have exactly the same matrix elements and thus the same bounds for given flavor $q$. The difference between the bounds on $\Lambda$ for three light quark flavors comes predominantly from the factor $m_q$ that is part of the definition of the operators, leading to $\sim (m_s/m_{u,d})^{1/3}$ larger  $\Lambda$ exclusion for the strange quark. 
The matrix element of the tensor operator, ${\cal Q}_{7,q}^{(7)}$, is bigger, cf. Eq. \eqref{eq:c7matr}, which translates to roughly factor of 2 more stringent bounds on $\Lambda$.  

In the case of flavor diagonal transitions, the bounds on dimension 6 operators are
\begin{align}
{\cal C}_{1,u(d,s)}^{(6)}:\qquad  \Lambda > 480(239,198)~\text{GeV}\quad (\nu_e),\qquad \Lambda > 826(697,463)~\text{GeV}\quad (\nu_\mu), 
\\
{\cal C}_{2,u(d,s)}^{(6)}:\qquad   \Lambda > 301(324,231)~\text{GeV}\quad (\nu_e),\qquad  \Lambda > 826(697,463)~\text{GeV}\quad (\nu_\mu),
\end{align} 
For the reader's convenience we translate the CHARM bounds on dimension 6 operators also into the bounds on $\varepsilon_i$ parameters, Eq. \eqref{eq:LNSI}. For the electron neutrino, one obtains
\begin{align}
-0.11 <&\,\varepsilon_{ee}^{uV} < 0.27\,, &\quad &-0.38 <\varepsilon_{ee}^{uA} < 0.69\,, \\
-0.24 <&\,\varepsilon_{ee}^{dV} < 0.08\,, &\quad &-0.59 <\varepsilon_{ee}^{dA} < 0.44\,, \\
-0.74 <&\,\varepsilon_{ee}^{sV} < 1.60\,, &\quad &-1.17 <\varepsilon_{ee}^{sA} < 1.01\,,
\end{align}
while for the muon neutrino 
\begin{align}
-0.03 <&\,\varepsilon_{\mu\mu}^{uV} < 0.06\,, &\quad &-0.11 <\varepsilon_{\mu\mu}^{uA} < 0.08\,, \\
-0.05 <&\,\varepsilon_{\mu\mu}^{dV} < 0.02\,, &\quad &-0.31 <\varepsilon_{\mu\mu}^{dA} < 0.15\,, \\
-0.20 <&\,\varepsilon_{\mu\mu}^{sV} < 0.11\,, &\quad &-0.54 <\varepsilon_{\mu\mu}^{sA} < 0.40\,.
\end{align}
For the case of the electron neutrino these bounds are comparable, yet somewhat stronger, than the bounds from oscillations for vector currents involving $u$ and $d$ quarks, Eqs.~\eqref{eq:eps:uV:oscill} and~\eqref{eq:eps:dV:oscill}, and are significantly stronger for the case of the muon neutrino.

Note that for a number of operators the bounds on $\Lambda$ from the CHARM experiment are comparable to the momentum exchange $q\sim {\mathcal O}(10{\rm~GeV})$. This means that the EFT analysis may be applicable only for strongly coupled mediators, with couplings larger than ${\mathcal O}(1)$.  Another general comment regarding CHARM constraints on NSI is that for light mediators these are comparatively less effective than \CE constraints where the momentum exchange is smaller. 

\subsection{Other constraints}
\label{sec:Others}

The contribution to the neutrino scattering rates due to the neutrino magnetic moment has a pole at $\vec q^2=0$, cf. Fig. \ref{fig:C15rates} in Appendix \ref{app:plots}. This means that experiments with lower $E_R$ thresholds will have better sensitivity to the magnetic moment. Furthermore, scattering on electrons will in general lead to lower $\vec q^2$. 
Measurements of solar neutrinos scattering on electrons in  Borexino \cite{Borexino:2017fbd} give the current most stringent limits on the magnetic moment $\mu_{\nu_\alpha}$, or, equivalently,  on the Wilson coefficient ${\cal C}_1^{(5)}$,
\beq
{\cal C}_1^{(5)}:\qquad \Lambda > 2.7 \cdot 10^6\, \rm GeV \quad (\nu_e); \qquad\quad
\Lambda > 1.8 \cdot 10^6 \, \rm GeV\quad (\nu_\mu).
\eeq
The measured neutrino scattering rates in Borexino can also be translated in a bound on Rayleigh operators ${\cal C}_{1,2}^{(7)}$. The neutrino interactions mediated by the two Rayleigh operators result either in $\nu A\to \nu A$ scattering through 1-loop matrix elements or in $\nu A\to \nu A \gamma$, i.e., with an emission of an extra photon. In the Borexino experiment both processes lead to the same signal. Using the results from Section \ref{sec:Rayleigh:scatt} for the 1-loop contribution, with NDA estimates for the single nuclear matrix elements and neglecting two-body currents for $\Q_{2}^{(7)}$ operator, and adding the cross section for the process with a photon emission, give the total NSI scattering rate. 
Comparing it with the Borexino measurement \cite{Borexino:2017fbd} of the solar neutrino flux gives
\begin{align}
{\cal C}_{1}^{(7)}:\qquad  \Lambda > 136~\text{GeV}\quad (\nu_e),\qquad \Lambda > 119~\text{GeV}\quad (\nu_\mu), 
\\
{\cal C}_{2}^{(7)}:\qquad   \Lambda >   104~\text{GeV}\quad (\nu_e),\qquad  \Lambda >119~\text{GeV}\quad (\nu_\mu).
\end{align}
A different set of bounds on NSI operators \eqref{eq:dim6EW:Q1Q2:light}-\eqref{eq:dim5:Q10Q11:light} comes from collider experiments.  For a proper analysis we need to extend the EFT analysis to above the electroweak symmetry breaking scale, which we do in the next section. 
However, some of the bounds directly apply to the operators \eqref{eq:dim6EW:Q1Q2:light}-\eqref{eq:dim5:Q10Q11:light}. For instance, the searches for dark matter can be re-interpreted as bounds on NSI operators with two neutrinos replacing the two DM particles in the final state. 

The monojet searches at CMS and ATLAS~\cite{Agrawal:2013hya,Khachatryan:2014rra,Aad:2015zva} thus result in lower limits of $\Lambda \gtrsim 1$ TeV for (axial-)vector current operators, Eq.~\eqref{eq:dim6EW:Q1Q2:light}, $\Lambda \gtrsim 200$ GeV for gluon-gluon operators, Eq.~\eqref{eq:dim7:Q3Q4:light}, $\Lambda \gtrsim 40$ GeV for scalar operators, Eq.~\eqref{eq:dim7:Q5Q6:light}, and $\Lambda \gtrsim 20$ GeV for tensor current operators, Eq.~\eqref{eq:dim5:Q7:light}, after converting to our normalization of the operators. The monophoton searches bound di-photon operators, Eq. \eqref{eq:dim7:Q1Q2:light}, giving $\Lambda\gtrsim 30$ GeV \cite{Nelson:2013pqa}. These bounds, apart from (axial-)vector current operators, are quite weak, with values of $\Lambda$ allowed that are even below the kinematical cuts on $p_T$ and/or MET. One may thus question the applicability of the EFT with the actual bounds from colliders in reality even weaker, unless the couplings are large. We do not attempt to correct for these effects since this is beyond the scope of present manuscript, see, however, \cite{Pobbe:2017wrj} on how to properly obtain EFT bounds from LHC searches.

Once the EFT is uplifted above the electroweak scale, the bounds from collider searches can become more severe. 
For instance, searches for charged fermion contact interactions at LEP \cite{Schael:2013ita} give constraints on the $SU(2)_L\times U(1)_Y$ symmetric operators, defined in Eqs. \eqref{eq:4fops}-\eqref{eq:4fopsd} below.  These operators result in dimension 6 NSI operators \eqref{eq:dim6EW:Q1Q2:light} below the electroweak scale, but with bounds on $\Lambda$ from LEP of about ${\cal O}(1$~TeV) even when coupling only to leptons (see Section \ref{sec:ew:matching} for details). 

\section{NSI above the electroweak scale}
\label{sec:ew:matching}
In this Section we turn to the question of how the NSI interactions \eqref{eq:lightDM:Lnf5} are generated. Since the bounds on many of the operators are relatively mild, cf. Tables \ref{table:Lambdanue} and \ref{table:Lambdanumu} , it is possible that light NP could be responsible for their generation. This interesting direction was pursued, e.g., in Refs. \cite{Pospelov:2013rha,Pospelov:2012gm,Bertuzzo:2018ftf,Bertuzzo:2018itn}. 

The other option is that the NP responsible for NSI is heavy, heavier than the electroweak scale. If this is the case, the NP states can be integrated out leading to an EFT that is valid between the scale of NP, $\Lambda$, and the electroweak scale, $v_{\rm EW}$, with the effective Lagrangian (see also \cite{Gavela:2008ra,Brivio:2017vri,Grzadkowski:2010es,Buchmuller:1985jz})
\beq
{\cal L}_{\rm EW}=\sum_{a,d} \frac{ C_a^{(d)}}{\Lambda^{d-4}}Q_a^d,
\eeq
where $Q_a^d$ are $SU(2)_L\times U(1)_Y$ invariant operators.
In constructing the electroweak (EW) EFT operators we add to the SM field content the right-handed neutrino, $\nu_R$, which is a SM singlet. This allows for SM neutrinos to be either Dirac or Majorana. In the rest of this Section we discuss the bounds on operators $Q_a^d$ and their matchings onto the low energy EFT for NSI, Eq. \eqref{eq:lightDM:Lnf5}. We only consider operators that do not violate lepton number, since the lepton number violating operators are severely constrained by bounds on neutrinoless double beta decay.

The two dimension 6 operators in EW EFT that lead to an neutrino magnetic dipole moment are (throughout this section we do not display generational indices on leptonic fields, and assume flavor conservation for quark currents)
\beq
Q_{1,B}^{(6)} = \frac{g_1}{8\pi^2}\big(\bar \nu_R \sigma_{\mu\nu} \tilde H^\dagger L_L \big) B^{\mu\nu}\,, \qquad Q_{1,W}^{(6)} =\frac{g_2}{8\pi^2}\big(\bar \nu_R \sigma_{\mu\nu}  \tilde H^\dagger \tau^a   L_L\big) \, W^{a,\mu\nu}\,,
\eeq
where $\tilde H=i \sigma_2 H^*$, 
with $H$ the Higgs doublet, $L_L$ the lepton doublet, $B_{\mu\nu}$ and $W^a_{\mu\nu}$ the hypercharge and $SU(2)_L$ field strengths, and $\tau^a$ the $SU(2)_L$ generator in the fundamental representation. Both of the above operators contribute below the EW scale to the dimension five magnetic dipole operator, Eq. \eqref{eq:dim5:nf5:Q1:light}, 
\beq
\hat{\cal C}_1^{(5)} = \frac{v_{\rm EW}}{\sqrt{2}\Lambda^2} \Big(  C_{1,B}^{(6)} + \frac{1}{2}  C_{1,W}^{(6)} \Big),
\eeq
where $v_\text{EW} \simeq 246$~GeV is the Higgs vev.  If either $Q_{1,B}^{(6)}$ or $Q_{1,W}^{(6)}$ are generated in the UV, the neutrinos will have magnetic moments. The exception is, if the two contributions cancel against each other, i.e., for $2 \hat C_{1,B}^{(6)} =- \hat C_{1,W}^{(6)}$, in which case the neutrino magnetic moment vanishes. 

The dimension six operators $\Q_{1(2),f}^{(6)}$, Eq.  \eqref{eq:dim6EW:Q1Q2:light}, can arise from the following  four-fermion operators in the EW EFT (for $i=j$ the $Q_{4F,ii}^{(6)}$ operator is equivalent to $Q_{3F,ii}^{(6)}$ and should be dropped), 
\begin{align} \label{eq:4fops}
Q_{1F,ij}^{(6)} &= \left(\bar L_L^i \gamma_\mu L_L^i\right)\left(\bar Q_L^j \gamma^\mu Q_L^j\right)\,, \quad 
&Q_{2F,ij}^{(6)} &= \left(\bar L_L^i \gamma_\mu \tau^a L_L^i\right)\left(\bar Q_L^j \gamma^\mu \tau^a Q_L^j\right)\,,\\
Q_{3F,ij}^{(6)} &= \left(\bar L_L^i \gamma_\mu L_L^i\right)\left(\bar L_L^j \gamma^\mu L_L^j\right)\,, \quad &Q_{4F,ij}^{(6)} &= \left(\bar L_L^i \gamma_\mu \tau^a L_L^i\right)\left(\bar L_L^j \gamma^\mu \tau^a L_L^j\right)\,,\\
Q_{5F,ij}^{(6)} &= \left(\bar L_L^i \gamma_\mu L_L^i\right)\left(\bar e_R^j \gamma^\mu e_R^j\right)\,, \quad &Q_{6F,ij}^{(6)}& = \left(\bar L_L^i \gamma_\mu L_L^i\right)\left(\bar u_R^j \gamma^\mu u_R^j\right)\,, \\
Q_{7F,ij}^{(6)} &= \left(\bar L_L^i \gamma_\mu L_L^i\right)\left(\bar d_R^j \gamma^\mu d_R^j \right)\,, \label{eq:4fopsd}
\end{align}
which will then give for the Wilson coefficients of dimension 6 operators below electroweak scale  ($u_1=u$, $d_1 = d, d_2 = s$ and $\alpha$ the lepton flavor, not displayed on l.h.s.)
\begin{align}
\hat{\cal C}_{1(2),u_1}^{(6)} &= \frac{1}{2\Lambda^2} \left[ \pm\Big(C_{1F,\alpha 1}^{(6)} + \frac{1}{4} C_{2F,\alpha 1}^{(6)}\Big) + C_{6F,\alpha 1}^{(6)}  \right], 
\\
\hat{\cal C}_{1(2),d_i}^{(6)} &= \frac{1}{2\Lambda^2} \left[ \pm\Big(C_{1F,\alpha i}^{(6)} + \frac{1}{4} C_{2F,\alpha i}^{(6)}\Big) + C_{7F,\alpha i}^{(6)}  \right],
\\
\hat{\cal C}_{1(2),e}^{(6)} &= \frac{1}{2\Lambda^2} \left[ \pm\Big(C_{3F,\alpha 1}^{(6)} + \frac{1}{4} C_{4F,\alpha 1}^{(6)}\Big)+ C_{5F,\alpha 1}^{(6)}  \right].
\end{align}
The problem with generating large contributions to $\hat{\cal C}_{1(2),f}^{(6)}$ Wilson coefficients in this way is that the dimension 6 operators in Eq.~\eqref{eq:4fops} are extremely well bounded. Translating the results from \cite{Falkowski:2015krw,Falkowski:2017pss} the bounds for $Q_{nF,ij}^{(d)}$ for electrons coupling to first generations quarks, $i=1,j=1$  are $\Lambda> 4.7;4.8; 3.4;4.4; 3.3; 3.5$~TeV, while for muons, $i=2,j=1$,  $\Lambda> 1.6;3.1; 3.6;3.6; 1.0; 0.4$~TeV, where respectively $n=1,2,3,5,6,7$,  and we set $C_{nF,ij}^{(6)}=1$ (for $Q_{4F,21}^{(d)}$ the bound is of the same order as for $Q_{3F,21}^{(d)}$, but a precise determination would require a correlation matrix to properly account for the change of basis).  For electron-quark couplings the most stringent bounds come from measurements of $d\to u e\nu$ transitions and atomic parity violation, and are much more severe than the bounds from neutrino scattering. For muon-quark couplings the most stringent bound for $Q_{2F,21}^{(6)}$ is from $d\to u \mu\nu$ transitions, while for the other operators, $Q_{1F,2 i}^{(6)}$, $Q_{6F,2i}^{(6)}$, $Q_{7F,2i}^{(6)}$, it is mainly from neutrino scattering. The severe bounds from transitions involving charged leptons can be avoided in the special case, where $C_{1F,i}^{(6)} =\frac{1}{4} C_{2F,i}^{(6)}$, with all the other Wilson coefficients zero, since then the quarks only couple to neutrinos. 

Another option is that the leading contributions arise from dimension eight operators with two Higgs insertions (see also, e.g., \cite{Davidson:2003ha}),
\begin{align}
Q_{1F',i}^{(8)} &= \big(\bar L_L \tilde H  \gamma_\mu \tilde H^\dagger L_L \big)\left(\bar Q_L^i \gamma^\mu Q_L^i\right) \, ,  &Q_{2F'}^{(8)} &= \big(\bar L_L \tilde H  \gamma_\mu \tilde H^\dagger L_L \big) \left(\bar L_L \gamma^\mu L_L\right) \, , \\
Q_{3F'}^{(8)} &=  \big(\bar L_L \tilde H  \gamma_\mu \tilde H^\dagger L_L  \big)\left(\bar e_R \gamma^\mu e_R\right)\,,  
&Q_{4F'}^{(8)} &=  \big(\bar L_L \tilde H  \gamma_\mu \tilde H^\dagger L_L  \big)\left(\bar u_R \gamma^\mu u_R\right) \,, \\
Q_{5F',i}^{(8)} &=  \big(\bar L_L \tilde H  \gamma_\mu \tilde H^\dagger L_L \big) \left(\bar d_R^i \gamma^\mu d_R^i \right) \,.
\end{align}
After the Higgs obtains a vev only the neutrino is projected out of the lepton doublet, $\tilde H^\dagger L_L\to \nu_L v_{\rm EW}/\sqrt2$.
The Wilson coefficients for low energy dimension 6 operators are in this case,
\begin{align}
\hat{\cal C}_{1(2),u_1}^{(6)} &= \frac{v_{\rm EW}^2}{2\Lambda^4} \left( \pm C_{1F',1}^{(8)} + C_{4F'}^{(8)}  \right)\,, 
\qquad
&\hat{\cal C}_{1(2),d_i}^{(6)} &= \frac{v_{\rm EW}^2}{2\Lambda^4} \left( \pm C_{1F',i}^{(8)} + C_{5F'}^{(8)}  \right),
\\
\hat{\cal C}_{1(2),e}^{(6)} &= \frac{v_{\rm EW}^2}{2\Lambda^4} \left[ \pm C_{2F'}^{(8)} + C_{3F'}^{(8)}  \right]. &&
\end{align}
 Since these operators only lead to couplings of quarks to the neutrinos and not to charged leptons, this relaxes some of the bounds. The remaining bounds are ``inevitable'', as they come from processes that involve neutrinos -- these are the bounds discussed at the end of Section \ref{sec:Others}.

The gauge-gauge operators in \eqref{eq:dim7:Q1Q2:light} and \eqref{eq:dim7:Q3Q4:light} can arise from dimension eight operators
\begin{align}
Q_{1B}^{(8)} &= \frac{\alpha_1}{12\pi}\big(\bar \nu_R \tilde H^\dagger L_L \big)  B^{\mu\nu}B_{\mu\nu}\,, \quad &Q_{2B}^{(8)}&=\frac{\alpha_1}{8\pi}\big(\bar \nu_R \tilde H^\dagger L_L\big) B^{\mu\nu}\tilde B_{\mu\nu}\,, 
\\
Q_{1W}^{(8)} &= \frac{\alpha_2}{12\pi}\big(\bar \nu_R \tilde H^\dagger L_L \big) W^{a,\mu\nu}W_{\mu\nu}^a\,, \quad &Q_{2W}^{(8)} &=\frac{\alpha_2}{8\pi}\big(\bar \nu_R \tilde H^\dagger L_L\big) W^{a,\mu\nu} \tilde W_{\mu\nu}^a\,, 
\\
Q_{3W}^{(8)} &= \frac{\alpha_{12}}{12\pi}\big(\bar \nu_R \tilde H^\dagger \tau^a L_L \big) W^{a,\mu\nu}B_{\mu\nu}\,, \quad &Q_{4W}^{(8)} &=\frac{\alpha_{12}}{8\pi}\big(\bar \nu_R \tilde H^\dagger \tau^a  L_L\big) W^{a,\mu\nu} \tilde B_{\mu\nu}\,, 
\\
Q_{1G}^{(8)} &= \frac{\alpha_s}{12\pi}\big(\bar \nu_R \tilde H^\dagger L_L \big) G^{a,\mu\nu}G_{\mu\nu}^a\,, \quad &Q_{2G}^{(8)} &=\frac{\alpha_s}{8\pi}\big(\bar \nu_R \tilde H^\dagger L_L\big) G^{a,\mu\nu} \tilde G_{\mu\nu}^a\,,
\end{align}
giving the Wilson coefficients 
\begin{align}
&\hat{\cal C}_{1(2)}^{(7)} =  \frac{v_{\rm EW}}{\sqrt{2}\Lambda^4} \left(C_{1(2)B}^{(8)} + C_{1(2)W}^{(8)} + \frac{1}{2} C_{3(4)W}^{(8)}  \right)\,, \qquad \hat{\cal C}_{3(4)}^{(7)} =  \frac{v_{\rm EW}}{\sqrt{2}\Lambda^4} C_{1(2)G}^{(8)}. 
\end{align}
The reinterpretation of the 8 TeV ATLAS $W$+MET search, Ref.~\cite{Aad:2013oja}, bounds $Q_{1(2)W}^{(8)}$ to $\Lambda \gtrsim 0.7 (0.8)$ TeV \cite{Lopez:2014qja}, while the 7 TeV ATLAS $Z$+MET search, Ref.~\cite{Aad:2012awa}, gives  bounds on $Q_{1B}^{(8)},\ldots, Q_{4W}^{(8)}$ operators of roughly comparable strength, but with the details dependening on the relative sizes of photon and $Z$ exchange contributions \cite{Carpenter:2012rg}. 

The operators \eqref{eq:dim7:Q5Q6:light} and \eqref{eq:dim5:Q7:light} can arise from  the following dimension six operators 
\begin{align}
\label{eq:Q1R:Q2R}
Q_{1R,i}^{(6)} &= \left(\bar \nu_R L_L\right) \left(\bar Q_L^i u_R^i \right) \,, &\quad &Q_{2R,i}^{(6)} = \left(\bar \nu_R \sigma_{\mu\nu} L_L \right) \left(\bar Q_L^i \sigma_{\mu\nu}u_R^i \right) \,, 
\\
Q_{3R,i}^{(6)} &= \left(\bar \nu_R L_L\right) \left(\bar d_R^i Q_L^i  \right)\,,&\quad &Q_{4R,i}^{(6)} = \left(\bar \nu_R \sigma_{\mu\nu} L_L \right) \left(\bar d_R^i \sigma_{\mu\nu} Q_L^i \right)\,, 
\\
\label{eq:Q5R:Q6R}
Q_{5R}^{(6)} &= \left(\bar \nu_R L_L\right) \left(\bar e_R L_L \right)\,,&\quad &Q_{6R}^{(6)} = \left(\bar \nu_R \sigma_{\mu\nu} L_L \right) \left(\bar e_R \sigma_{\mu\nu} L_L \right) \,. 
\end{align}
The resulting Wilson coefficients are 
\begin{align}
\hat{\cal C}_{5,u_i(d_i)}^{(7)} &=  \frac{\Re1 C_{1(3)R,i}^{(6)}}{2m_{u_i(d_i)}\Lambda^2} \,, \quad&\hat{\cal C}_{6,u_i(d_i)}^{(7)}& =    \frac{\Im1  C_{1(3)R,i}^{(6)}}{2m_{u_i(d_i)}\Lambda^2}  \,,\quad &\hat{\cal C}_{7,u_i(d_i)}^{(7)}& =   \frac{C_{2(4)R,i}^{(6)}}{2m_{u_(d_i)}\Lambda^2}  \,, 
\\
\hat{\cal C}_{5,e}^{(7)}& =  \frac{ \Re1 C_{5R}^{(6)}}{2m_e\Lambda^2}  \,, \quad&\hat{\cal C}_{6,e}^{(7)}& = \frac{\Im1 C_{5R}^{(6)}}{2m_e\Lambda^2} \,,\quad &\hat{\cal C}_{7,e}^{(7)} &=   \frac{C_{6R}^{(6)} }{2m_e\Lambda^2} \,.
\end{align}
The bounds on these chirality flipping operators are quite stringent. For instance, from $pp\to \ell+\slashed E_T+X$ searches at the LHC we can expect bounds on the NP scale of the operators \eqref{eq:Q1R:Q2R}-\eqref{eq:Q5R:Q6R} at the order of $\Lambda\gtrsim{\mathcal O}(5 {\rm TeV})$  and at the similar level from semileptonic decays of light pseudoscalar mesons. (This estimate is based on the bounds  for tensor current SM-EFT operators with left-handed neutrinos obtained in \cite{Gonzalez-Alonso:2016etj}, which only give quadratic corrections to the SM rates, as do the operators \eqref{eq:Q1R:Q2R}-\eqref{eq:Q5R:Q6R}.  A dedicated analysis of operators with right-handed neutrinos is called for, which, however, is beyond the scope of our work.)

No such bounds exist for dimension eight operators, 
\begin{align}
Q_{1H,i}^{(8)} &= \big(\bar \nu_R \tilde H^\dagger L_L \big) \big(\bar Q_L^i \tilde H u_R^i  \big), 
\quad &Q_{2H,i}^{(8)}& = \big(\bar \nu_R \sigma_{\mu\nu}\tilde H^\dagger L_L \big) \big(\bar Q_L^i \tilde H \sigma_{\mu\nu} u_R^i  \big), 
\\
Q_{3H,i}^{(8)} &= \big(\bar \nu_R \tilde H^\dagger  L_L \big) \big(\bar Q_L^i  H d_R^i  \big),\quad 
&Q_{4H,i}^{(8)} &= \big(\bar \nu_R \sigma_{\mu\nu} \tilde H^\dagger L_L \big) \big(\bar Q_L^i  H \sigma_{\mu\nu} d_R^i  \big), 
\\
Q_{5H}^{(8)} &= \big(\bar \nu_R \tilde H^\dagger L_L \big) \big(\bar L_L H e_R  \big),\quad 
&Q_{6H}^{(8)} &= \big(\bar \nu_R \sigma_{\mu\nu} \tilde H^\dagger L_L \big) \big(\bar L_L H \sigma_{\mu\nu} e_R  \big). 
\\ 
Q_{7H,i}^{(8)} &= \big(\bar \nu_R \tilde H^\dagger \tau^a L_L  \big) \big(\bar Q_L^i \tau^a\tilde H u_R^i  \big) , 
\quad &Q_{8H,i}^{(8)} &= \big(\bar \nu_R \sigma_{\mu\nu} \tilde H^\dagger \tau^a L_L \big) \big(\bar Q_L^i \tau^a\tilde H \sigma_{\mu\nu} u_R^i\big), 
\\
Q_{9H,i}^{(8)} &= \big(\bar \nu_R \tilde H^\dagger \tau^a L_L \big) \big(\bar Q_L^i \tau^a H d_R^i  \big),\quad &Q_{10H,i}^{(8)}& = \big(\bar \nu_R \sigma_{\mu\nu} \tilde H^\dagger \tau^a L_L \big) \big(\bar Q_L^i \tau^aH \sigma_{\mu\nu} d_R^i  \big), \\
Q_{11H}^{(8)} &= \big(\bar \nu_R \tilde H^\dagger \tau^a L_L \big) \big(\bar L_L \tau^aH e_R  \big),\quad &Q_{12H}^{(8)} &= \big(\bar \nu_R \sigma_{\mu\nu} \tilde H^\dagger \tau^a L_L \big) \big(\bar L_L \tau^aH \sigma_{\mu\nu} e_R  \big),
\end{align}
which, after the Higgs obtains the vev, have the form $(\bar \nu_R \ldots \nu_L)(\bar f \ldots f)$. The relevant constraints on these operators thus come only from neutrino scattering experiments, discussed in the previous section. In principle there are also constraints from Higgs decaying to four body final states, $h\to 2j \nu\bar\nu$. However, these are at present much less constraining. 

The Wilson coefficients of the low energy EFT operators \eqref{eq:dim7:Q5Q6:light} and \eqref{eq:dim5:Q7:light}, generated from the above operators, are
\begin{align}
\hat{\cal C}_{5\{6\},u_i}^{(7)} &= \frac{1}{\Lambda^4}  \frac{v_{\rm EW}^2}{2m_{u_i}} \Re1 \{\Im1\} \Big( C_{1H,i}^{(8)} + \frac{1}{4}C_{7H,i}^{(8)} \Big), 
&\hat{\cal C}_{7,u_i}^{(7)} &= \frac{1}{\Lambda^4}  \frac{v_{\rm EW}^2}{2m_{u_i}} \Big( C_{2H,i}^{(8)} + \frac{1}{4}C_{8H,i}^{(8)} \Big),
\\
\hat{\cal C}_{5\{6\},d_i}^{(7)} &= \frac{1}{\Lambda^4}  \frac{v_{\rm EW}^2}{2m_{d_i}} \Re1 \{\Im1\} \Big( C_{3H,i}^{(8)} + \frac{1}{4}C_{9H,i}^{(8)} \Big), 
&\hat{\cal C}_{7,d_i}^{(7)} &= \frac{1}{\Lambda^4}  \frac{v_{\rm EW}^2}{2m_{d_i}} \Big( C_{4H,i}^{(8)} + \frac{1}{4}C_{10H,i}^{(8)} \Big),
\\
\hat{\cal C}_{5\{6\},e}^{(7)} &= \frac{1}{\Lambda^4}  \frac{v_{\rm EW}^2}{2m_{e}} \Re1 \{\Im1\} \Big( C_{5H}^{(8)} + \frac{1}{4}C_{11H}^{(8)} \Big), 
&\hat{\cal C}_{7,e}^{(7)} &= \frac{1}{\Lambda^4}  \frac{v_{\rm EW}^2}{2m_{e}} \Big( C_{6H}^{(8)} + \frac{1}{4}C_{12H}^{(8)} \Big).
\end{align}

Finally, the dimension 7 low energy EFT operators with derivatives on the neutrino current, Eqs.~\eqref{eq:dim7:Q8Q9:light} and \eqref{eq:dim5:Q10Q11:light}, can arise from the following electroweak EFT dimension 8 operators,
\begin{align}
Q_{1D,i}^{(8)} &= \partial_\mu\big(\bar \nu_R \sigma^{\mu\nu} \tilde H^\dagger L_L \big) \big(\bar Q_L^i \gamma_\nu Q_L^i\big), 
\quad &Q_{2D,i}^{(8)} &= \big(\bar \nu_R \ilrpartial_\mu \negmedspace \tilde H^\dagger L_L \big) \big(\bar Q_L^i \gamma^\mu Q_L^i \big), 
\\
Q_{3D}^{(8)} &= \partial_\mu\big(\bar \nu_R \sigma^{\mu\nu} \tilde H^\dagger L_L \big) \big(\bar L_L \gamma_\nu L_L  \big), 
\quad &Q_{4D}^{(8)} &= \big(\bar \nu_R \ilrpartial_\mu \negmedspace \tilde H^\dagger L_L \big) \big(\bar L_L \gamma^\mu L_L \big), 
\\
Q_{5D,i}^{(8)} &= \partial_\mu\big(\bar \nu_R \sigma^{\mu\nu} \tilde H^\dagger L_L \big) \left(\bar u_R^i \gamma_\nu u_R^i  \right), 
\quad &Q_{6D,i}^{(8)}& = \big(\bar \nu_R \ilrpartial_\mu \negmedspace\tilde H^\dagger L_L \big) \big(\bar u_R^i \gamma^\mu u_R^i \big), 
\\
Q_{7D,i}^{(8)} &= \partial_\mu\big(\bar \nu_R \sigma^{\mu\nu} \tilde H^\dagger L_L \big) \big(\bar d_R^i \gamma_\nu d_R^i  \big), 
\quad &Q_{8D,i}^{(8)} &= \big(\bar \nu_R \ilrpartial_\mu \negmedspace \tilde H^\dagger L_L \big) \big(\bar d_R^i \gamma^\mu d_R^i \big), 
\\
Q_{9D}^{(8)} &= \partial_\mu\big(\bar \nu_R \sigma^{\mu\nu} \tilde H^\dagger L_L \big) \big(\bar e_R \gamma_\nu e_R  \big), \quad &Q_{10D}^{(8)} &= \big(\bar \nu_R \ilrpartial_\mu \negmedspace \tilde H^\dagger L_L \big) \big(\bar e_R \gamma^\mu e_R \big).
\end{align}
After the Higgs obtains the vev, the operators have the form of (neutrino current)$\times$(charged fermion current). The most relevant bounds on these operators are, again, due to neutrino scattering experiments discussed in the previous Sections.  
The matching gives for the low energy Wilson coefficients 
\begin{align}
\hat{\cal C}_{8\{9\},u_i}^{(7)} &= \frac{v_{\rm EW}}{\Lambda^4} \Re1 \{\Im1\} \Big( C_{2D,i}^{(8)} + C_{6D,i}^{(8)} \Big), 
&\hat{\cal C}_{10\{11\},u_i}^{(7)} &= \frac{v_{\rm EW}}{\Lambda^4} \Re1 \{\Im1\} \Big( C_{1D,i}^{(8)} + C_{5D,i}^{(8)} \Big),
\\
\hat{\cal C}_{8\{9\},d_i}^{(7)} &= \frac{v_{\rm EW}}{\Lambda^4} \Re1 \{\Im1\} \Big( C_{2D,i}^{(8)} + C_{8D,i}^{(8)} \Big), 
&\hat{\cal C}_{10\{11\},d_i}^{(7)} &=\frac{v_{\rm EW}}{\Lambda^4} \Re1 \{\Im1\} \Big( C_{1D,i}^{(8)} + C_{7D,i}^{(8)} \Big),
\\
\hat{\cal C}_{8\{9\},e}^{(7)} &=\frac{v_{\rm EW}}{\Lambda^4}  \Re1 \{\Im1\} \Big( C_{4D}^{(8)} + C_{10D}^{(8)} \Big), 
&\hat{\cal C}_{10\{11\},e}^{(7)} &=\frac{v_{\rm EW}}{\Lambda^4} \Re1 \{\Im1\}\Big( C_{3D}^{(8)} + C_{9D}^{(8)} \Big).
\end{align}

\section{Conclusions}
\label{sec:conclusions}

In this manuscript we obtained predictions for \CE in the presence of nonstandard neutrino interactions described by an EFT at 2 GeV. Our analysis covers the complete basis of EFT operators up to and including dimension 7, and thus covers most of the viable models as long as the mediators are heavier than about 100MeV. We recast the recent measurement of \CE by the COHERENT collaboration using a CsI detector to obtain bounds on the EFT operators, assuming that only one NSI operator at the time contributes appreciably. The main results are collected in Figures \ref{fig:chart-all} and \ref{fig:chart-all2}, where they are compared with the bounds on NSI from neutrino oscillations, the solar neutrino flux measurements at Borexino, and from deep inelastic scattering. The obtained bounds apply to incoming electron or muon neutrinos scattering either through flavor diagonal interaction, or even, if the neutrino flavor changes (including scattering to sterile neutrinos). 

We see that already now the \CE measurements lead to the most stringent limits for some of the NSI operators, for instance for scalar currents. The NSI reach of \CE experiments is set to significantly improve in the future, with a number of new experiments either already running or being planned. In Figures \ref{fig:chart-all} and \ref{fig:chart-all2} we also show the projected limits for the NaI 2 ton detector planned by the COHERENT collaboration, also assuming that systematic errors can be decreased by an order of magnitude. The new experiments, as we show in the paper, can also increase their sensitivity to NSI by modifying running conditions, such as the incoming neutrino energy, the nuclear recoil energy thresholds, but also by trying to perform measurements at higher recoil energies. For this it will be important to investigate how well one can distinguish between NSI and the subleading corrections to our predictions (for instance from $q^2$ dependence of nuclear form factors away from zero recoil point, see, e.g., \cite{Cadeddu:2018dux}). 

When using our results it is important to note their validity. The DIS bounds assume that the mediators are heavier than a few 10s of GeV, the COHERENT bounds that they are heavier than about 100 MeV, while Borexino and oscillation bounds apply also to very light mediators (a typical momentum exchange in solar neutrinos scattering on electrons in Borexino is $q\sim$ few 100~keV$-$1~MeV, which sets the lower bound on mediator mass, such that use of EFT is justified for interpreting Borexino measurement). For light mediators, with masses below 10 GeV the bounds from DIS would get suppressed, and similarly for COHERENT bounds for mediators lighter than a few 10s MeV. It is easy to recast our bounds also in such cases, but one does need at least simplified models for the mediators in that case (some examples are, e.g., \cite{Bischer:2018zbd,Banerjee:2018eaf,Abdullah:2018ykz}). More challenging would be to extend our work to the intermediate range of $q^2\sim$ (few) GeV${}^2$, where neither ChPT methods that we used, nor the factorization used for DIS, apply. The benefit, on the other hand, is that many of the neutrino experiments are taking data precisely in this  theoretical difficult intermediate regime, so that any theoretical advances would be highly desired. 

{\bf Acknowledgments.} We thank Gil Paz for the discussions regarding Rayleigh operators, and Rex Tayloe and Kate Scholberg on the status of coherent neutrino scattering experiments. JZ and MT acknowledge support in part by the DOE grant de-sc0011784, and by DOE Neutrino Theory Network. MT thanks the Fermilab theory group for hospitality. The research of WA is supported by the National Science Foundation under Grant No. NSF PHY-1720252. WA thanks KITP for hospitality while part of this work was performed and acknowledges support by the National Science Foundation under Grant No. NSF PHY11-25915.

\begin{appendix}
\section{Nucleon form factors and nonrelativistic limits} \label{app:formfactors}
 
The nucleon form factors, $F_i$,  in Eqs. \eqref{eq:c1(0)}-\eqref{eq:c12(2)}, are defined as~\cite{Bishara:2017pfq}, 
\begin{align}
\langle N'|\bar q \gamma^\mu q|N\rangle&=\bar u_N'\Big[F_1^{q/N}(q^2)\gamma^\mu+\frac{i}{2m_N}F_2^{q/N}(q^2) \sigma^{\mu\nu}q_\nu\Big]u_N\,,
\\
\langle N'|\bar q \gamma^\mu \gamma_5 q|N\rangle&=\bar u_N'\Big[F_A^{q/N}(q^2)\gamma^\mu\gamma_5+\frac{1}{2m_N}F_{P'}^{q/N}(q^2) \gamma_5 q^\mu\Big]u_N\,,\label{eq:ff:A}
\\
\langle N'| m_q \bar q   q|N\rangle&= F_S^{q/N} (q^2)\, \bar u_N' u_N\,,
\\
\langle N'| m_q \bar q  i \gamma_5 q|N\rangle&= F_P^{q/N} (q^2)\, \bar u_N' i \gamma_5 u_N\,,
\\
\langle N'| \frac{\alpha_s}{12\pi} G^{a\mu\nu}G^a_{\mu\nu} |N\rangle&= F_G^{N} (q^2)\, \bar u_N' u_N\,,
\\
\langle N'| \frac{\alpha_s}{8\pi} G^{a\mu\nu}\tilde G^a_{\mu\nu}|N\rangle&= F_{\tilde G}^{N} (q^2)\, \bar u_N' i \gamma_5 u_N\,,
\\
\begin{split}
\langle N'|m_q \bar q \sigma^{\mu\nu} q |N\rangle&=  \bar u_N'\Big[F_{T,0}^{q/N} (q^2)\,  \sigma^{\mu\nu} +\frac{i}{2 m_N} \gamma^{[\mu}q^{\nu]} F_{T,1}^{q/N} (q^2) 
\\
&\qquad \qquad+ \frac{i}{m_N^2} q^{[\mu}k_{12}^{\nu]} F_{T,2}^{q/N} (q^2) \Big] u_N\,,
\end{split}
\end{align}
where we suppressed the dependence of nucleon states on their momenta, $\langle N'|\equiv\langle N(k_2)| $, $| N\rangle\equiv | N(k_1)\rangle $, and similarly, $\bar u_N'\equiv u_N(k_2)$, $u_N\equiv u_N(k_1)$, while $k_{12}^\mu=k_1^\mu+k_2^\mu$ and $q^\mu = k_2^\mu - k_1^\mu$, and $\gamma^{[\mu}q^{\nu]}=\gamma^{\mu}q^{\nu}-q^{\mu}\gamma^{\nu}$. 

The form factors, $F_i$, are functions of $q^2$ only. 

In the derivations of Eqs. \eqref{eq:c1(0)}-\eqref{eq:c12(2)} a non-relativistic reduction of nucleon currents is required. Counting
$v\cdot \partial\sim {\mathcal O}(q^2)$, with $q$ the typical soft three-momentum, the leading terms are  \cite{Bishara:2017pfq}
\begin{align}
\label{eq:HETlimit:scalar}
\bar N N&\to \bar N_v N_v+{\mathcal O}(q^2),
\\
\label{eq:HDMETlimit:pscalar}
\begin{split}
 \bar N i \gamma_5
N &\to \frac{1}{ m_N}\partial_\mu \big(\bar N_v S_N^\mu N_v \big) 
 +{\mathcal O}(q^3)\,,
\end{split}
\\
\label{eq:vecDM:expand}
\begin{split}
\bar N \gamma^\mu N &
\to  v^\mu \bar N_v N_v +{\mathcal O}(q),
\end{split}
\\
\label{eq:axialDM:expand}
\begin{split}
\bar N \gamma^\mu \gamma_5 N & 
\to  2 \bar N_v S_N^\mu N_v +\mathcal{O}(q^2),
\end{split}
\\
\label{eq:tensorDM:expand}
\begin{split}
\bar N \sigma^{\mu\nu} N& \to  \bar N_v \sigma_\perp^{\mu\nu}
N_v +  {\mathcal O}(q)\,,
  \end{split}
  \\
  \label{eq:axialtensorDM:expand}
  \begin{split}
  \bar
N \sigma^{\mu\nu} i\gamma_5 N&\to 
 2\bar
N_v S_N^{[\mu}v^{\nu]} N_v+{\mathcal O}(q)\,. 
\end{split}
\end{align}

\section{The NSI predictions for differential rates in COHERENT} 
\label{app:plots}
In this appendix we show the differential scattering rates in the presence of NSI, $dN/dE_R$, for the CsI[Tl]  detector of the COHERENT collaboration. In the numerical evaluations only one NSI operator, Eqs. \eqref{eq:dim5:nf5:Q1:light}-\eqref{eq:dim5:Q10Q11:light}, is taken to be nonzero. We set its Wilson coefficient to
\beq
\hat {\cal C}_{a}^{(d)} = \left(\frac{1}{\Lambda_{a;{\rm min}}}\right)^{d-4}\,,
\eeq
where $\Lambda_{a;{\rm min}}$ is the current lower limit for this operator, as obtained in Section \ref{subsec:cohlimit} from the COHERENT measurement, and listed in Tables \ref{table:Lambdanumu} and \ref{table:Lambdanue}. 

 \begin{figure}[t]
    \includegraphics[scale=0.46]{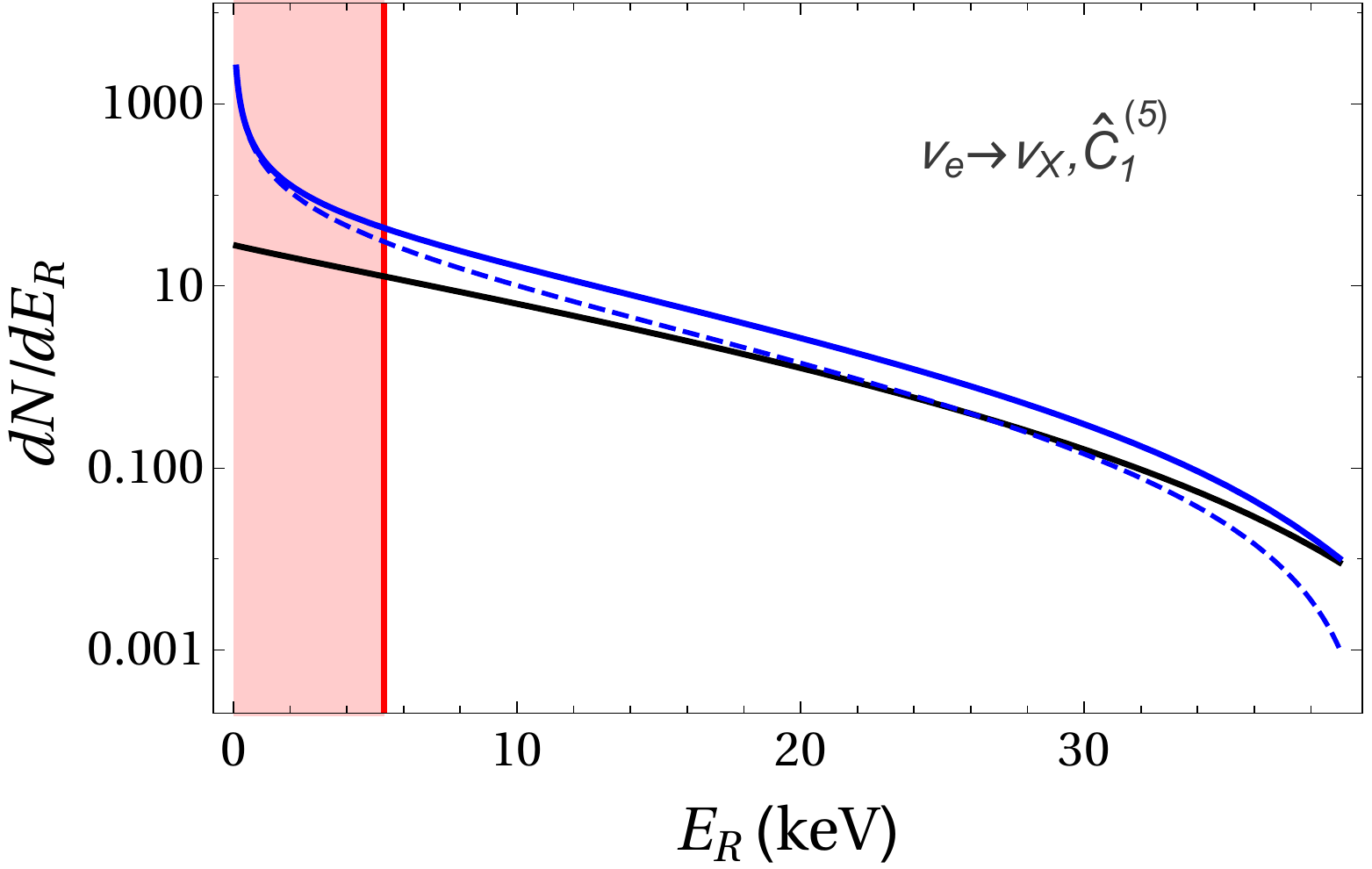} \hspace{1cm}
 \includegraphics[scale=0.46]{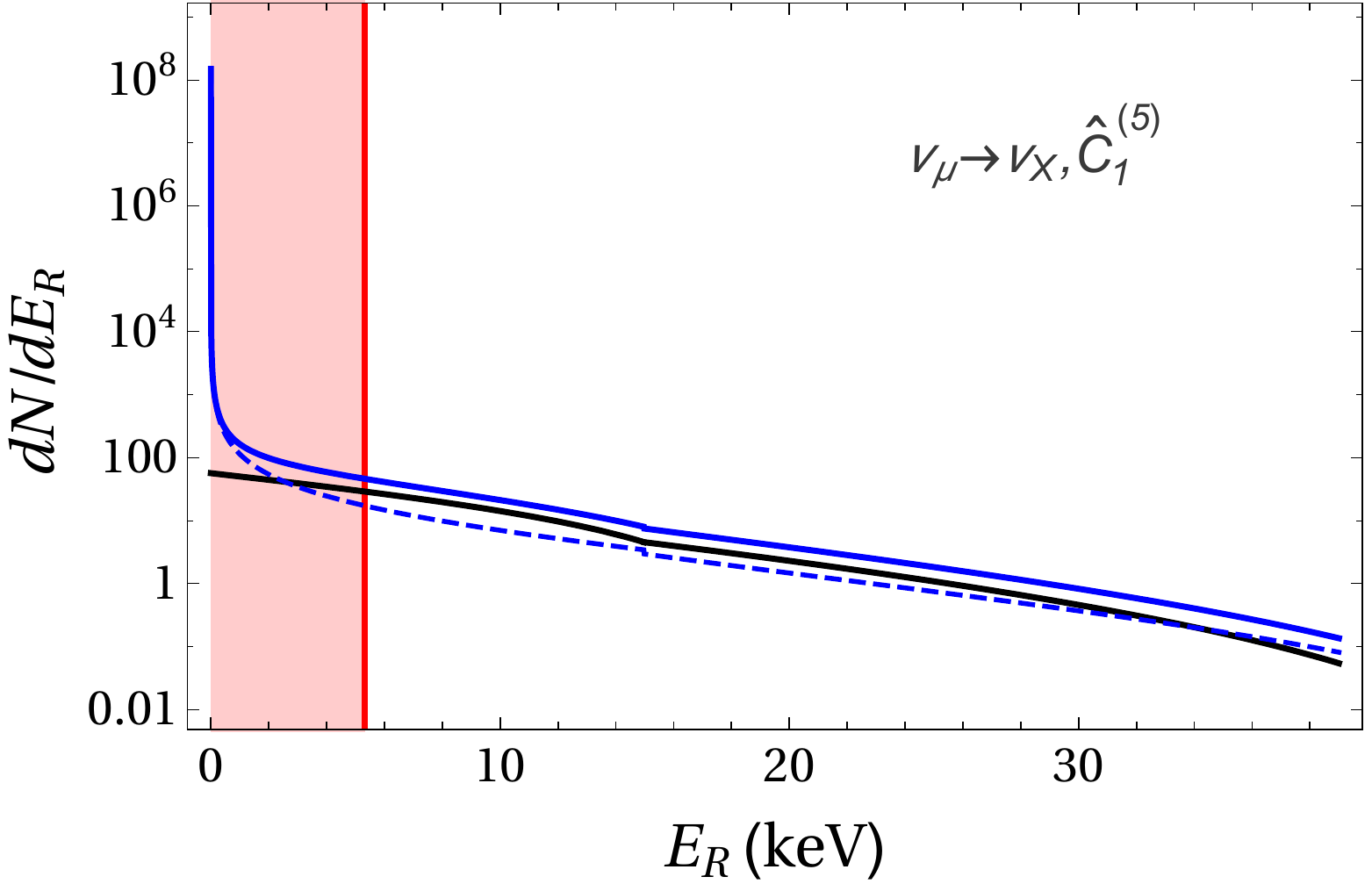}
\caption{The $\nu_e\to \nu_X$ (left) and $\nu_\mu\to \nu_X$ (right) rates in the COHERENT CsI detector including ${\cal Q}_{1}^{(5)}$ NSI operator contribution, for $\hat C_a^{(5)}=1/\Lambda_{a;{\rm min}}$, where $\Lambda_{a;{\rm min}}$ is taken to be the corresponding lower bound in Tables  \ref{table:Lambdanue} and \ref{table:Lambdanumu}, respectively. The SM rate is denoted by the solid black line, dashed blue line denotes the NSI only prediction, and solid blue line the sum of the two. The vertical red line denotes the energy threshold of the detector.}
\label{fig:C15rates}
\end{figure}
 
\begin{figure}[t]
    \includegraphics[scale=0.46]{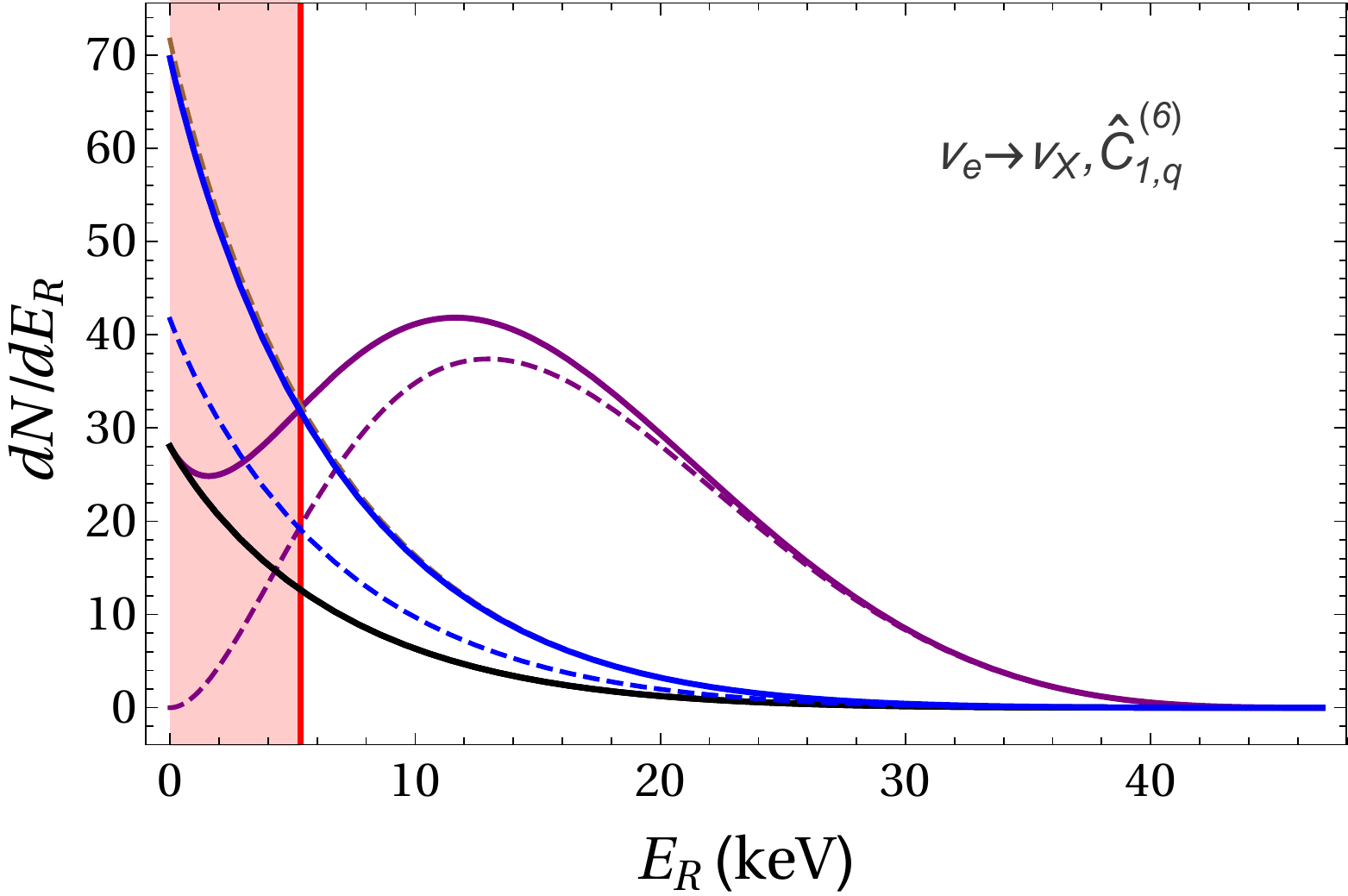}  \hspace{1cm}
    \includegraphics[scale=0.46]{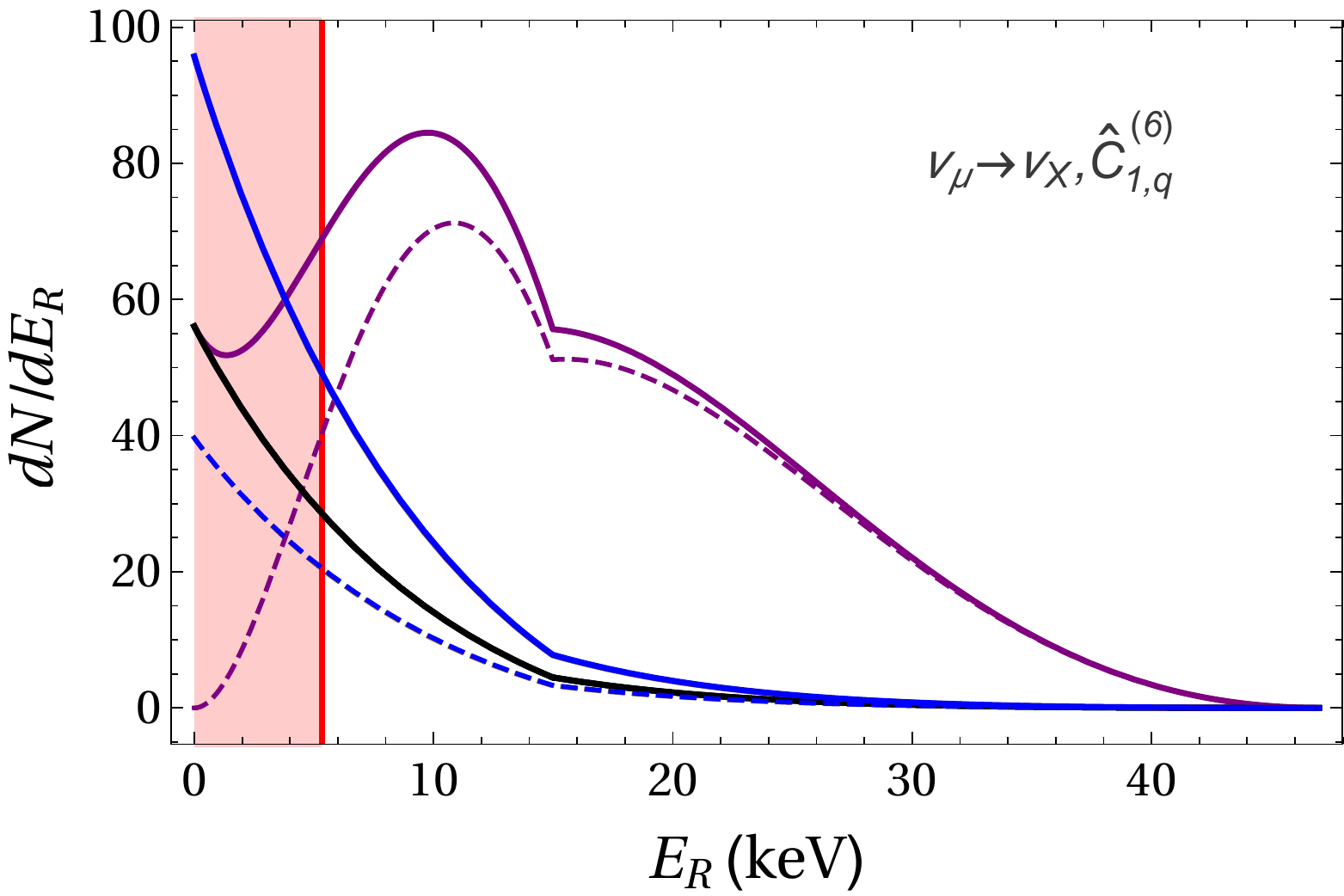}  \\[2mm]
       \includegraphics[scale=0.46]{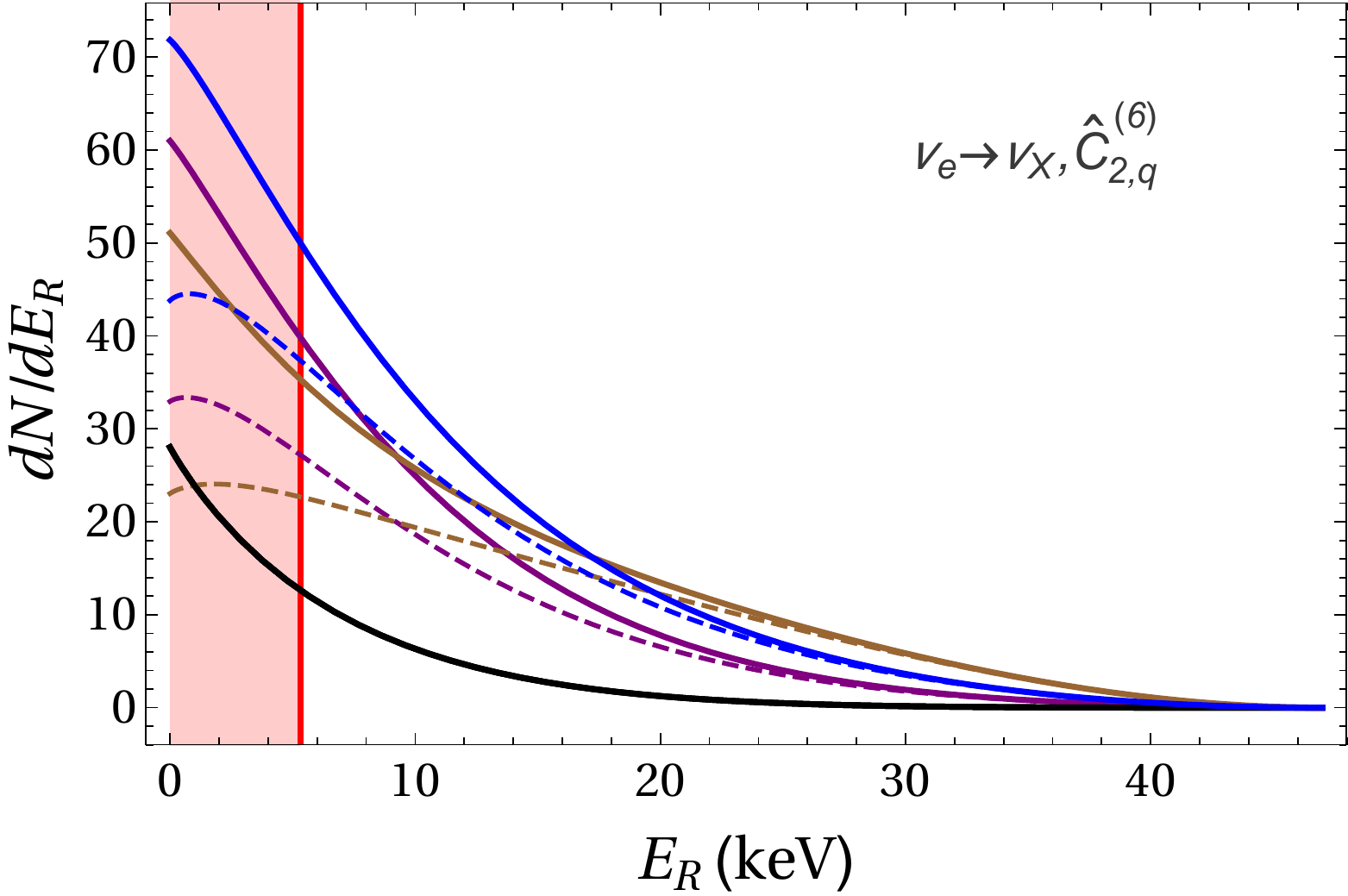}  \hspace{1cm}
    \includegraphics[scale=0.46]{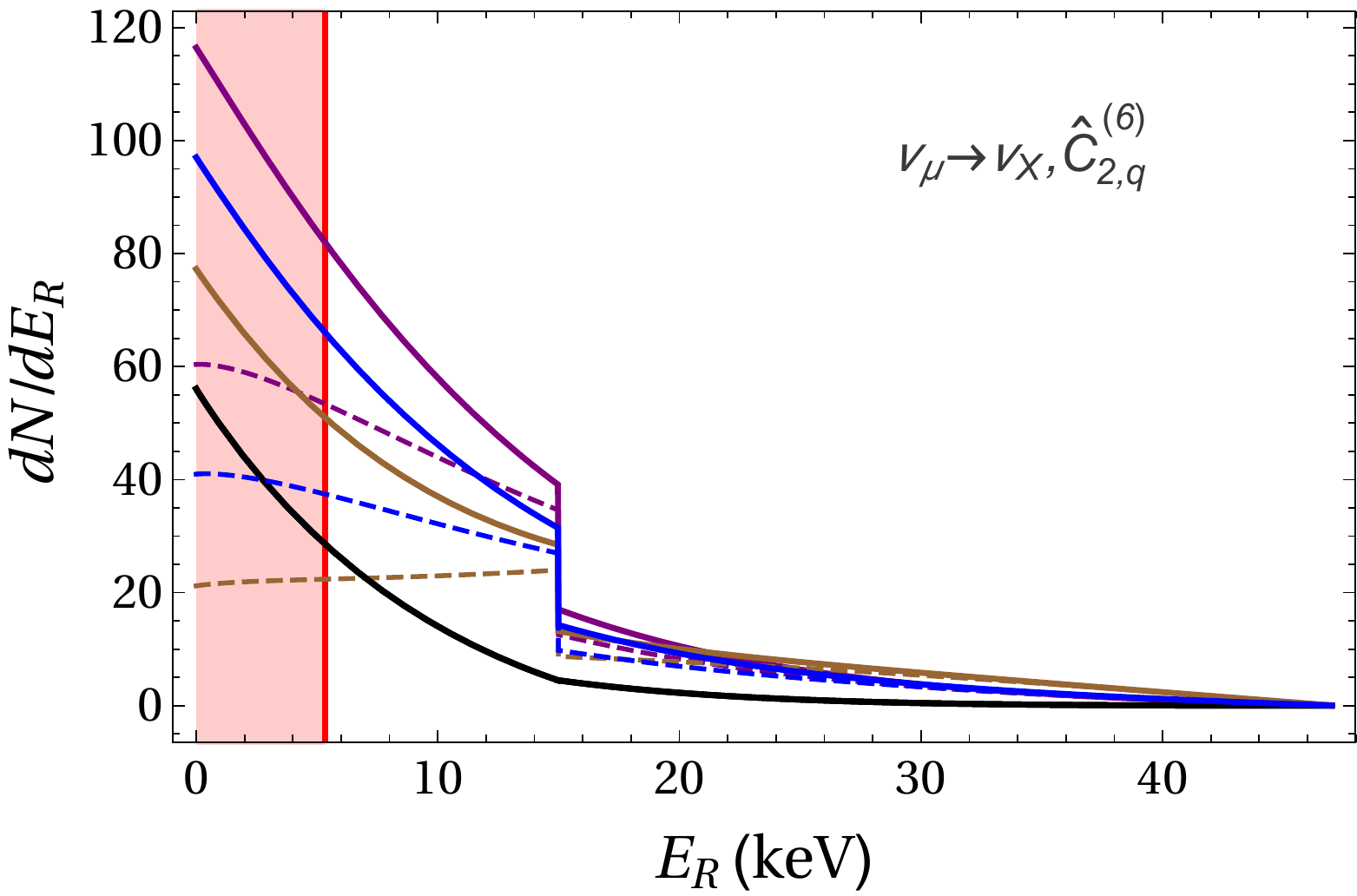}
\caption{The $\nu_e\to \nu_e$ (left) and $\nu_\mu\to \nu_\mu$ (right) scattering rates in the presence of  a single NSI operator, either ${\cal Q}_{1,q}^{(6)}$ (top) or ${\cal Q}_{2,q}^{(6)}$ (bottom), setting $\hat C_a^{(6)}=1/\Lambda_{a;{\rm min}}^2$, where $\Lambda_{a;{\rm min}}$ is taken to be the corresponding lower bound in Tables  \ref{table:Lambdanue} and \ref{table:Lambdanumu}, respectively.
The SM event rate is denoted by the black solid line, dashed lines denote NSI only and solid lines the sum of SM and NSI contributions (with blue, brown and purple representing couplings to up, down and strange quark currents, respectively). The energy threshold of the experiment is denoted by the vertical red line. 
}
\label{fig:C16rates}
\end{figure}

\begin{figure}[h!]
    \includegraphics[scale=0.46]{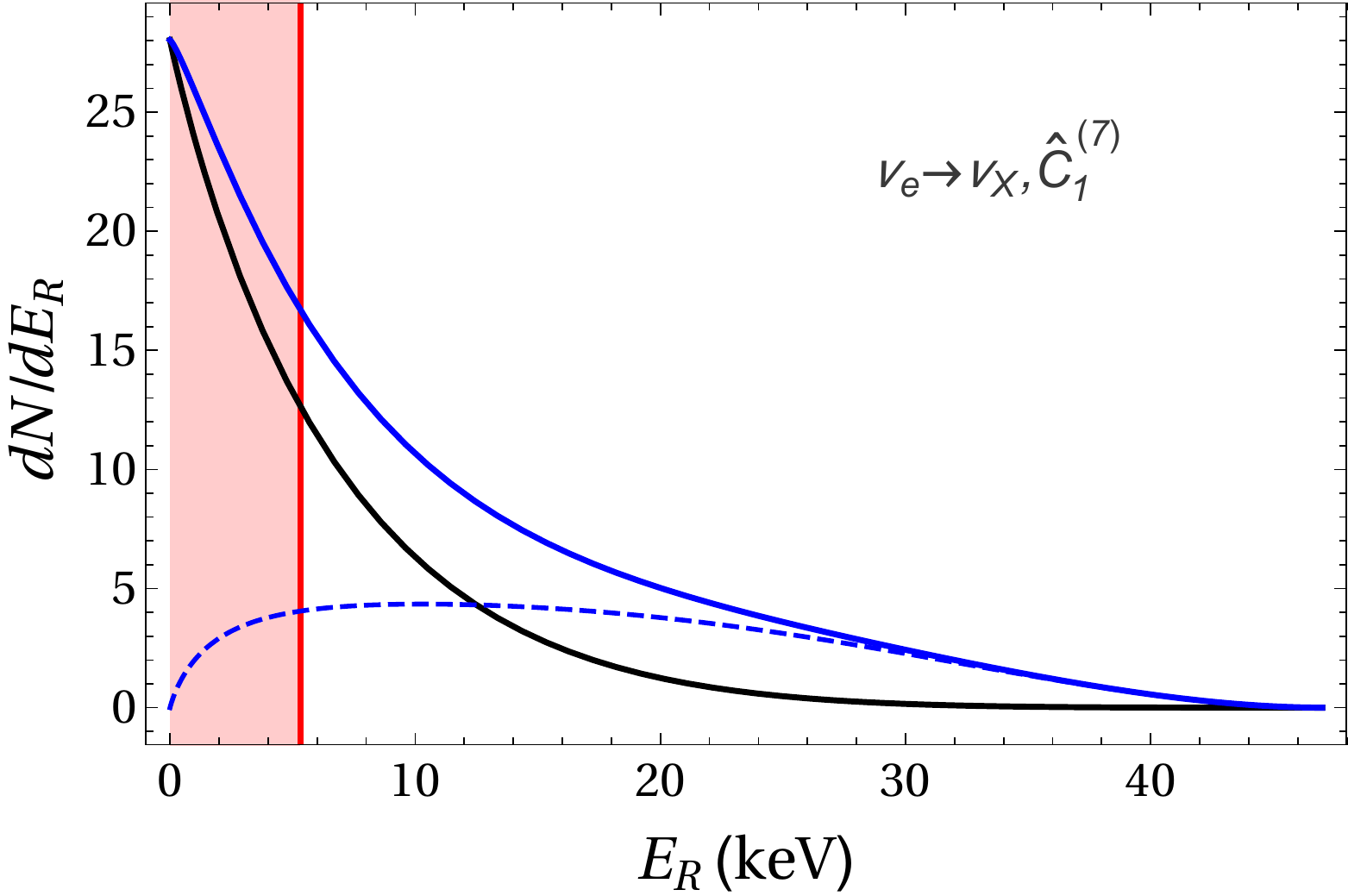} \hspace{1cm}
 \includegraphics[scale=0.46]{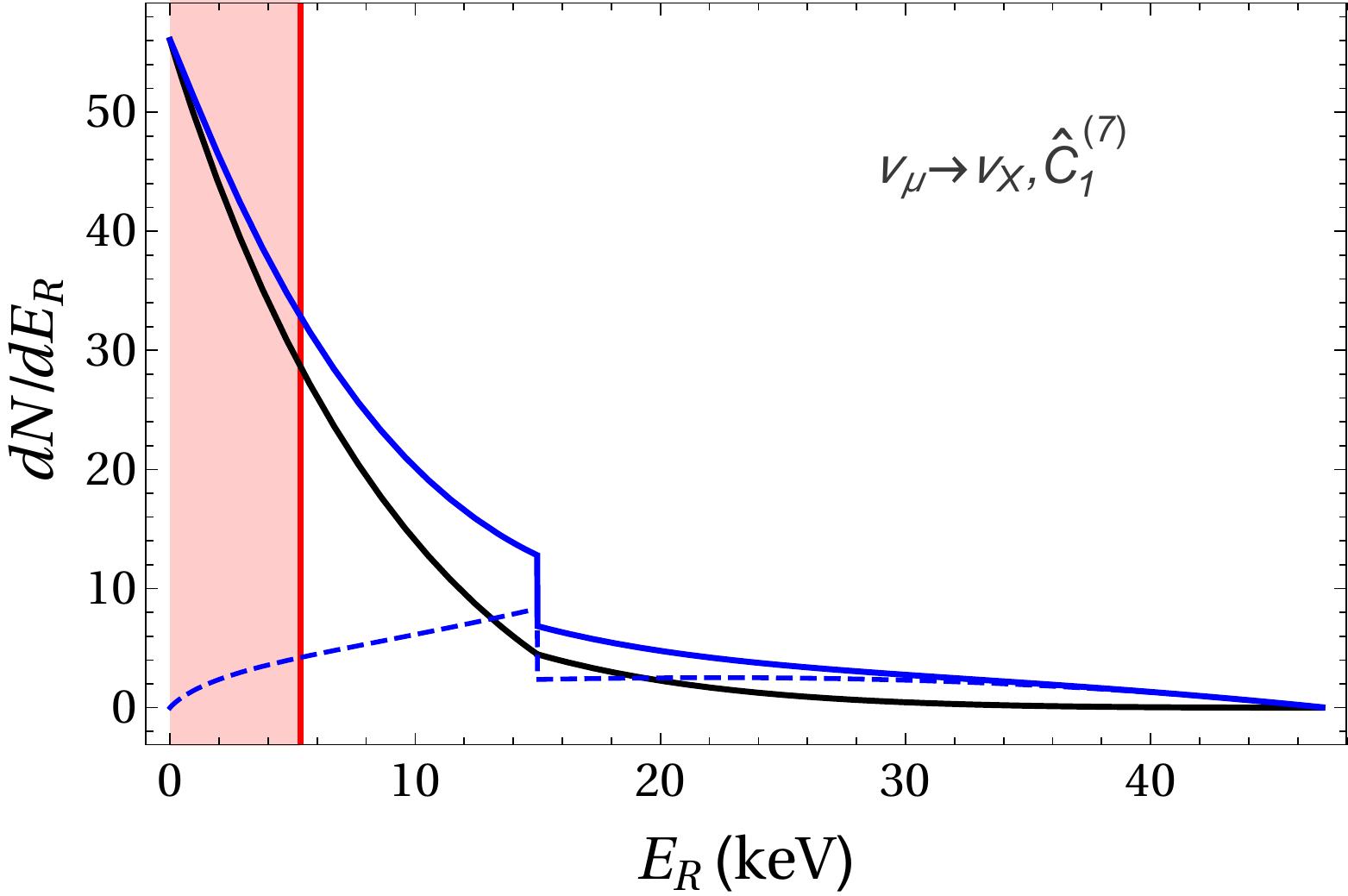}
 \\[2mm]
   \includegraphics[scale=0.46]{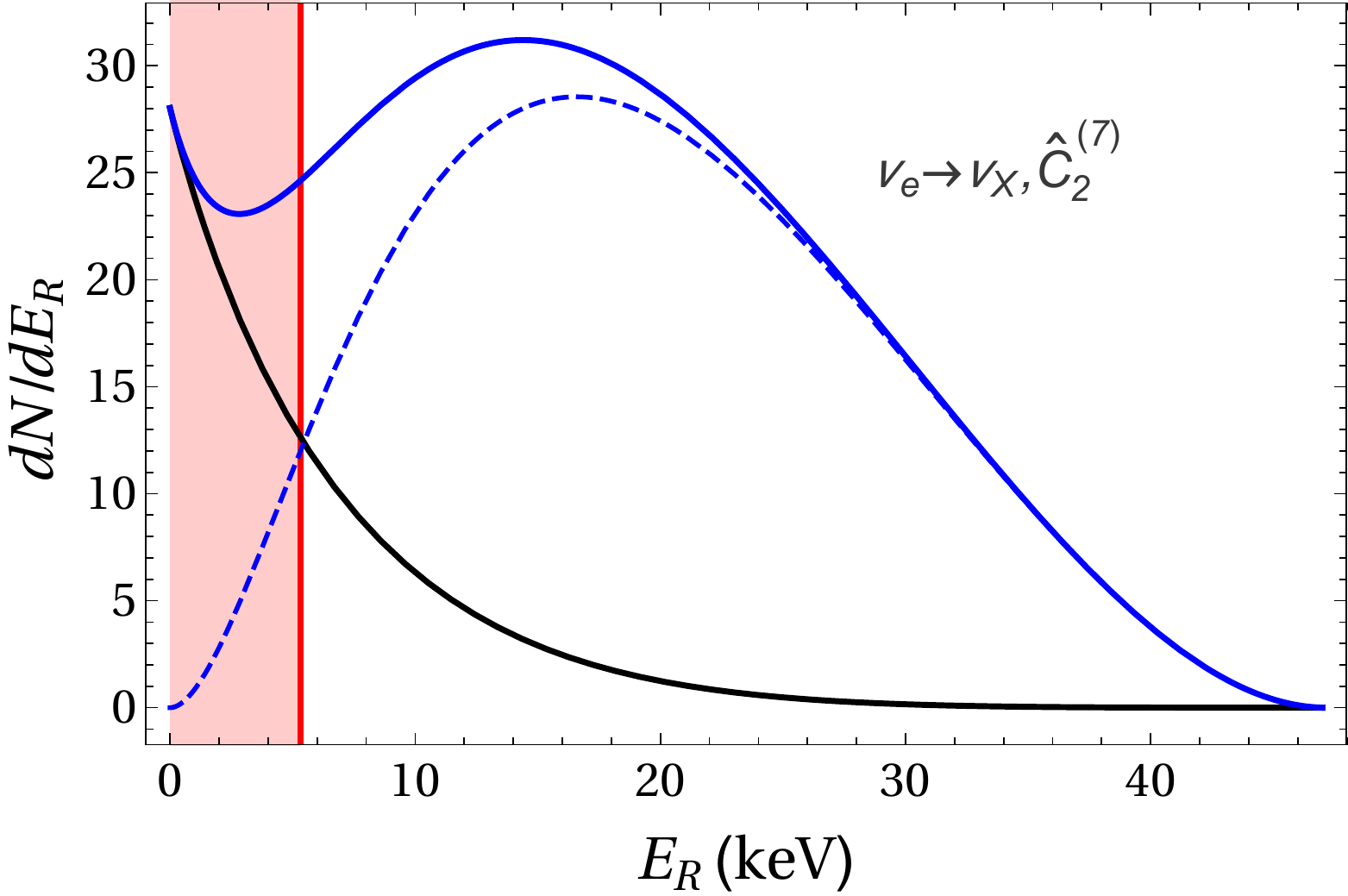} \hspace{1cm}
 \includegraphics[scale=0.46]{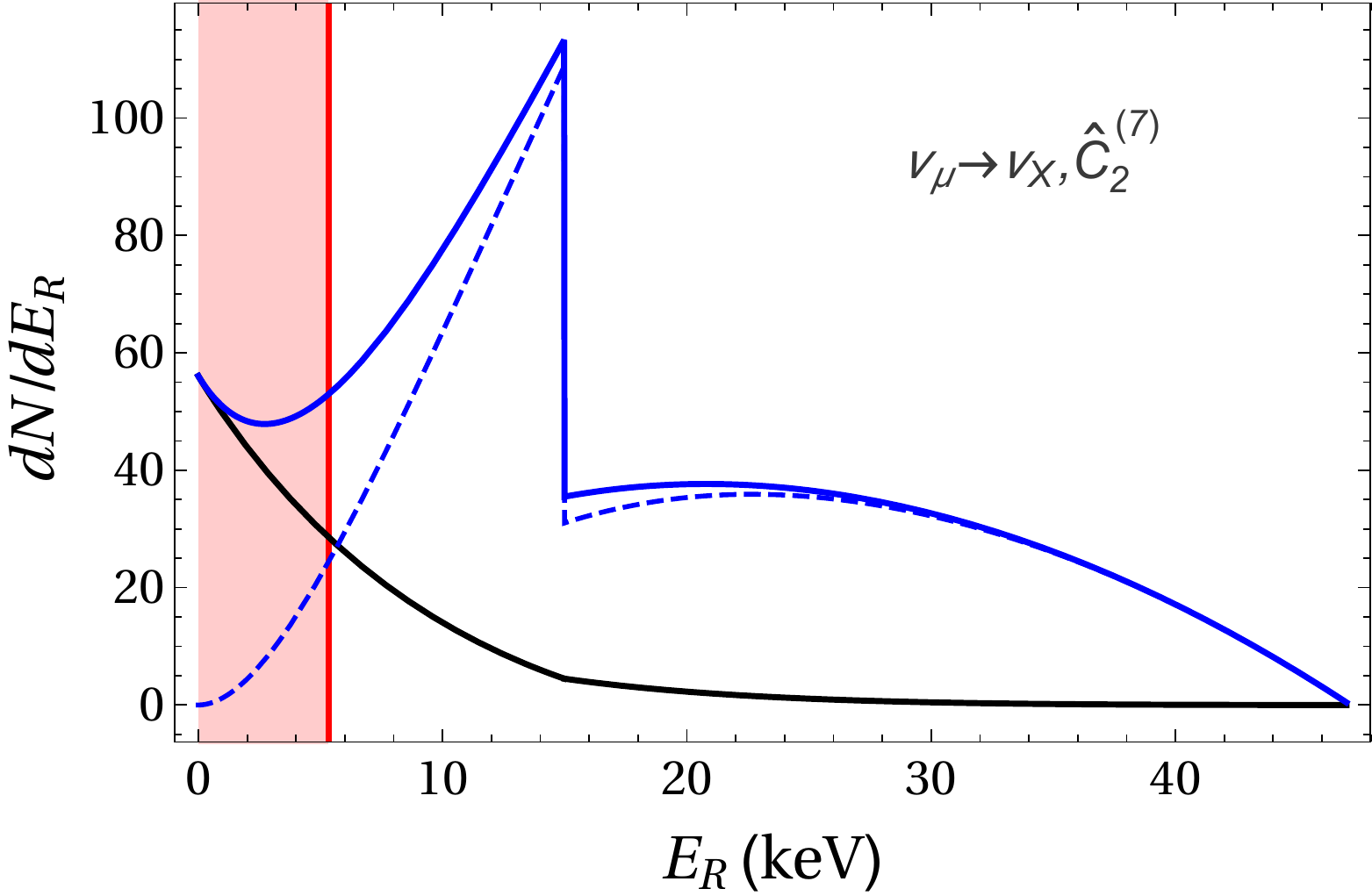}
 \\[2mm]
     \includegraphics[scale=0.46]{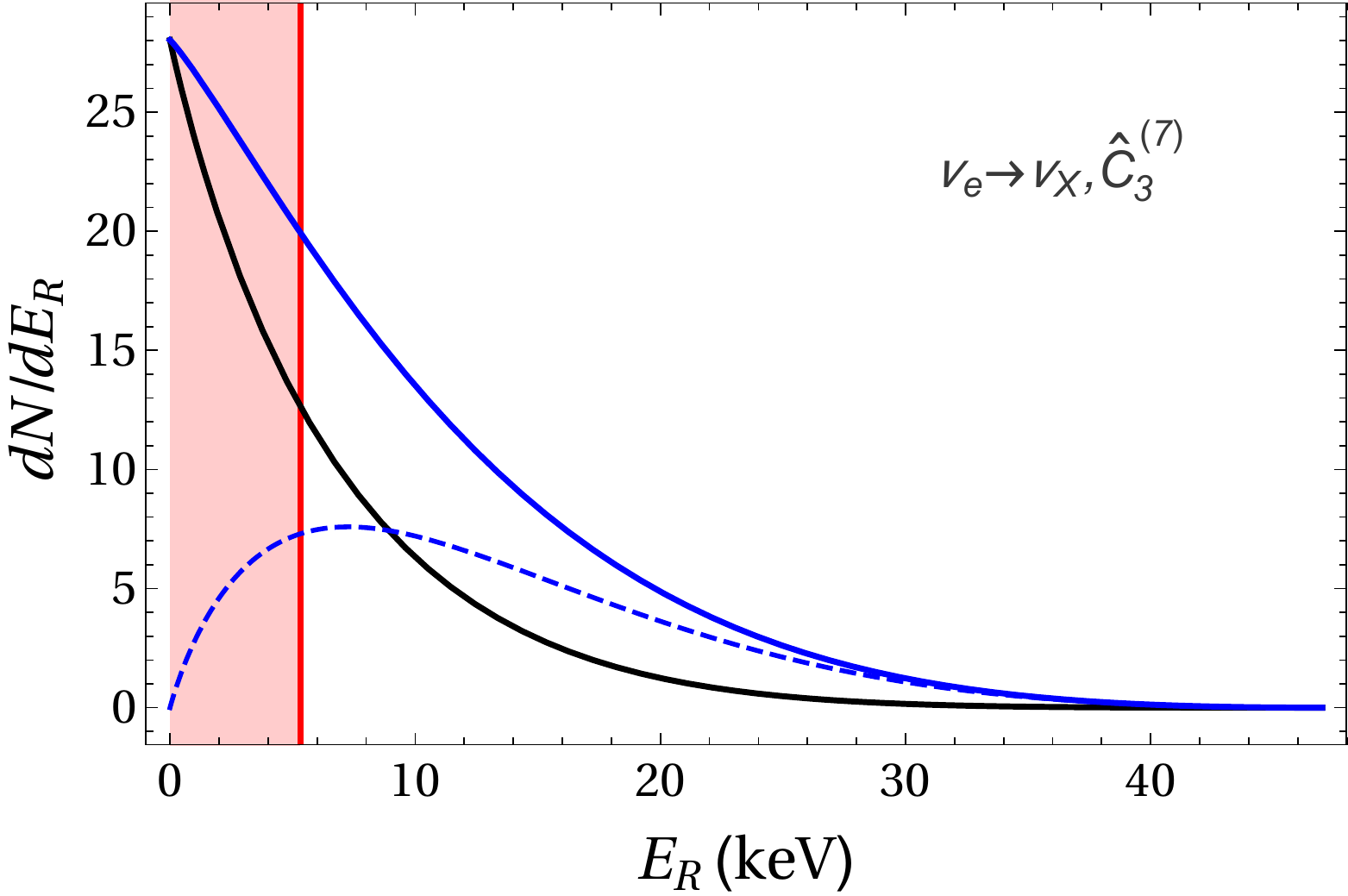} \hspace{1cm}
 \includegraphics[scale=0.46]{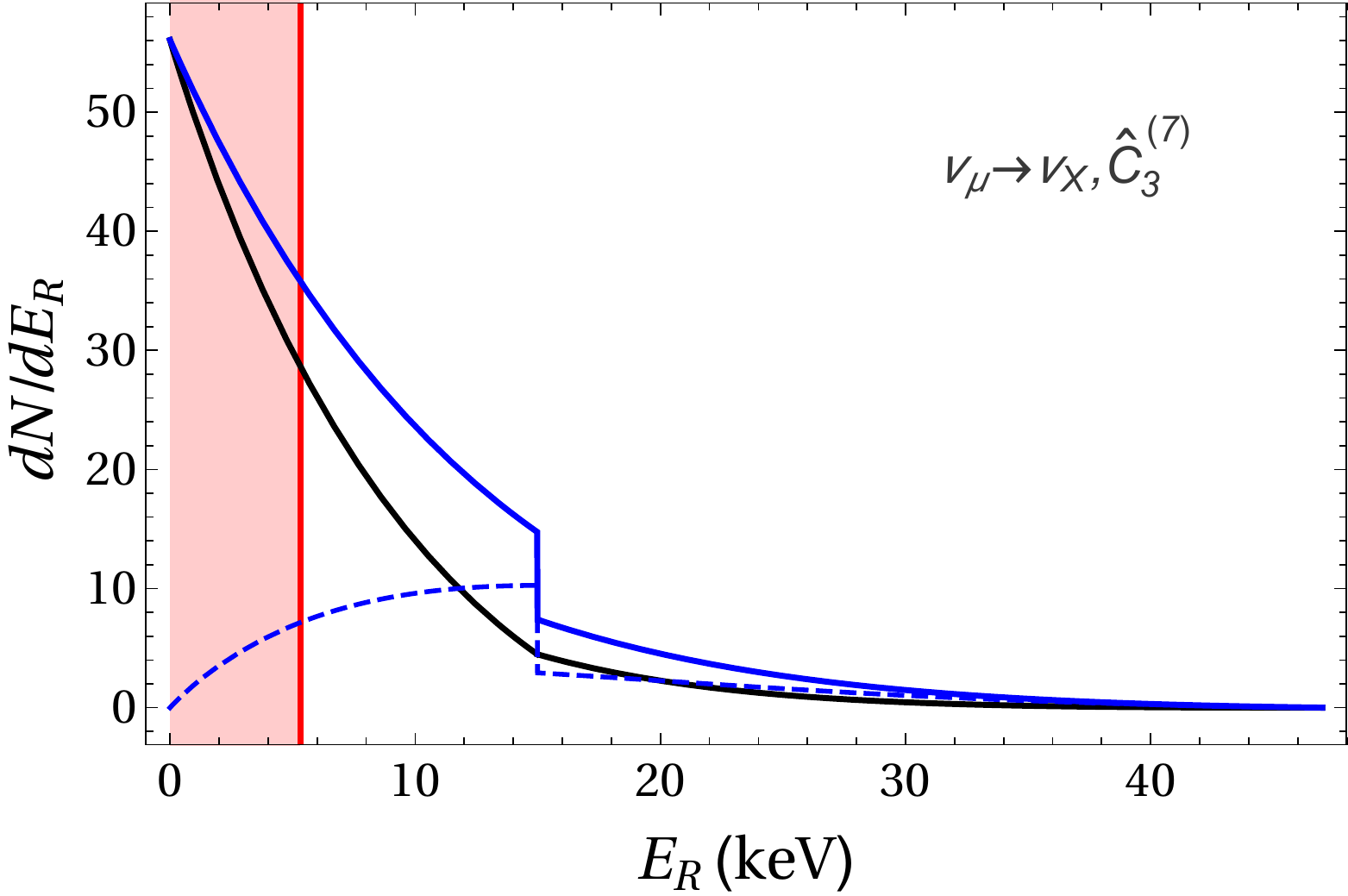}
 \\[2mm]
    \includegraphics[scale=0.46]{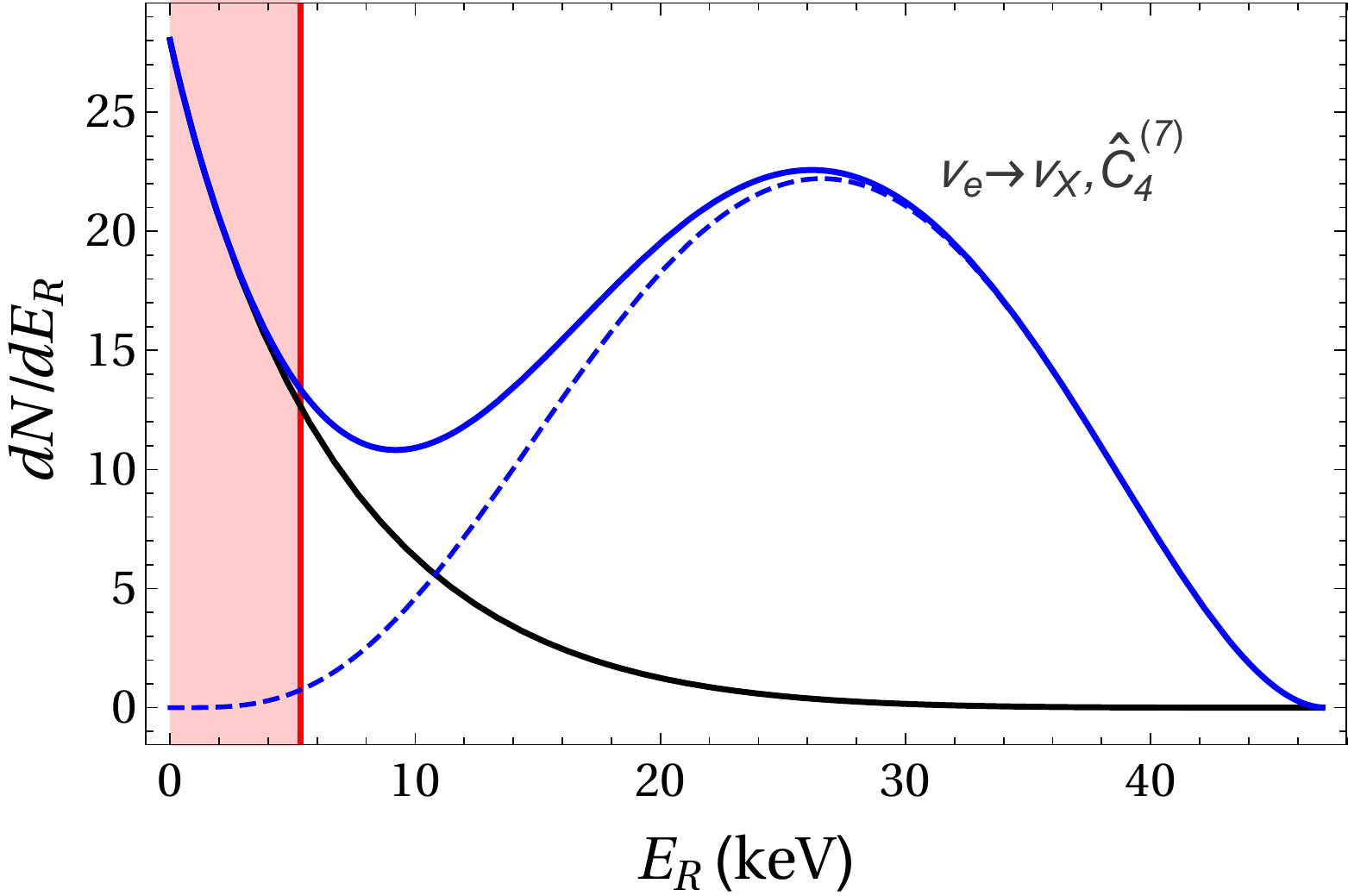} \hspace{1cm}
 \includegraphics[scale=0.46]{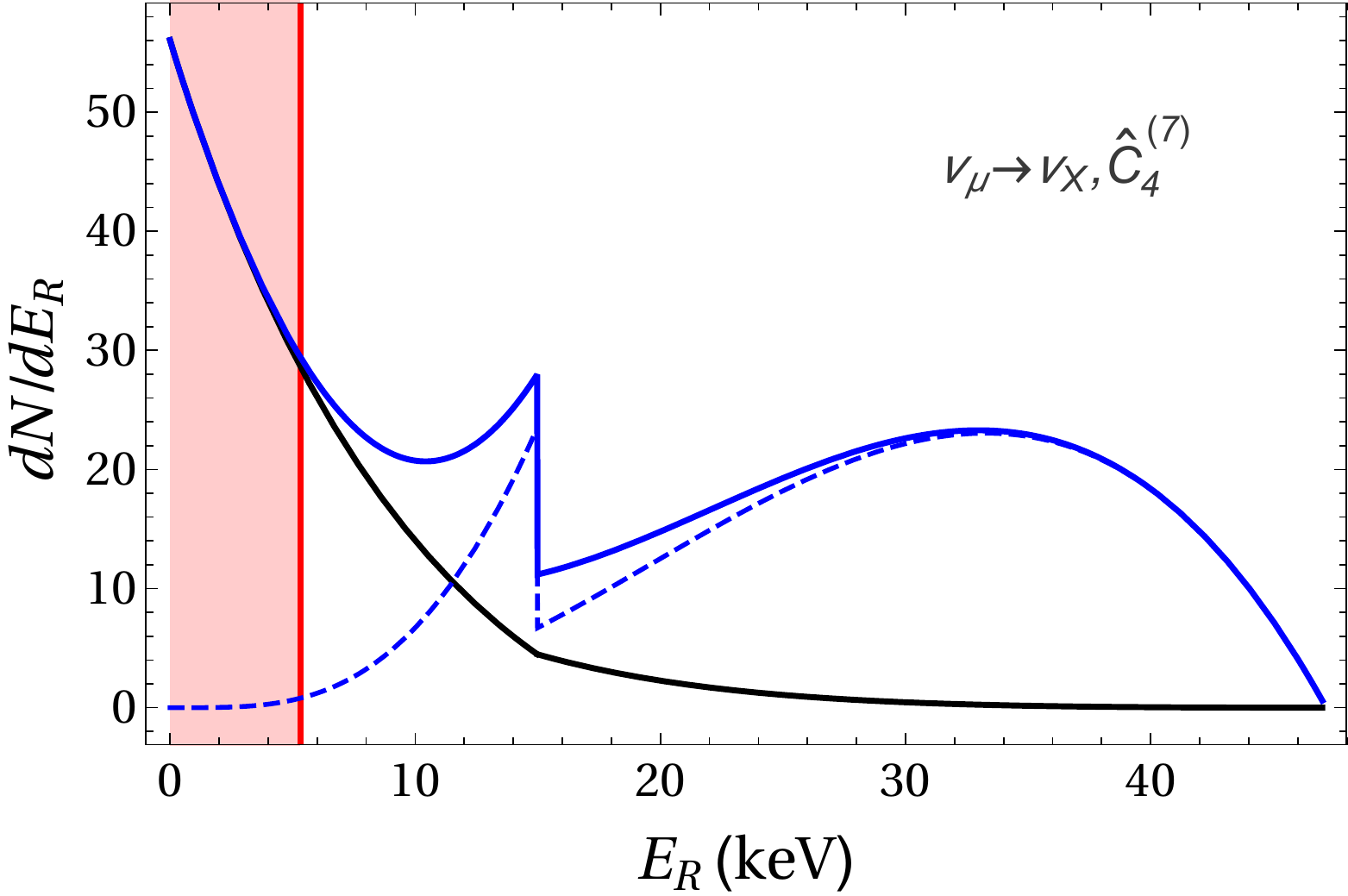}
\caption{The $\nu_e\to \nu_X$ (left) and $\nu_\mu\to \nu_X$ scattering rates in COHERENT CsI detector in the presence of NSI operators ${\cal Q}_{1}^{(7)}$, ${\cal Q}_{2}^{(7)}$, ${\cal Q}_{3}^{(7)}$, ${\cal Q}_{4}^{(7)}$ (top to bottom) setting $\hat C_a^{(7)}=1/\Lambda_{a;{\rm min}}^3$. The notation is as in Fig. \ref{fig:C15rates}.
}
\label{fig:C17rates}
\end{figure}

\begin{figure}[t]
    \includegraphics[scale=0.46]{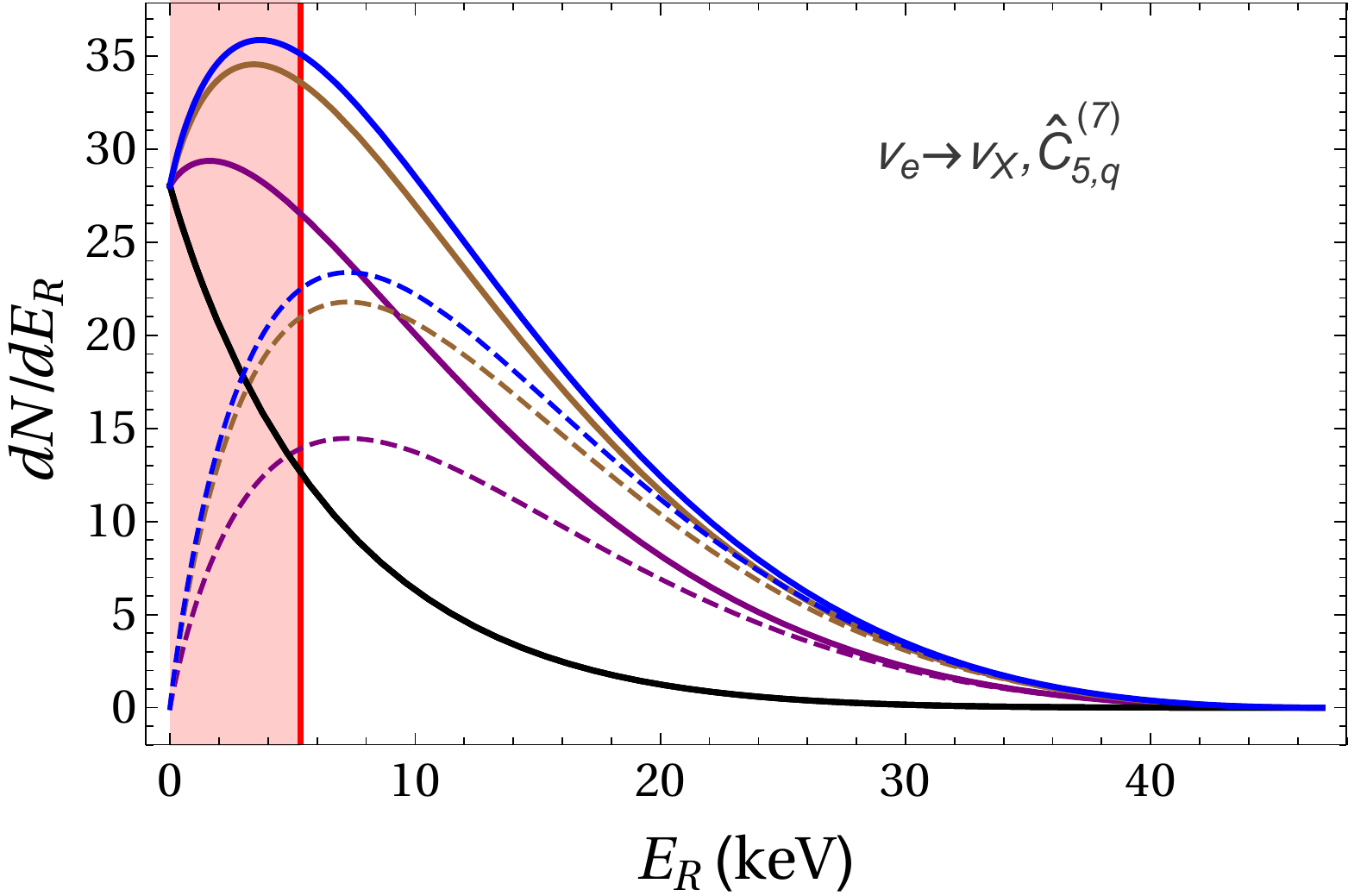}  \hspace{1cm}
    \includegraphics[scale=0.46]{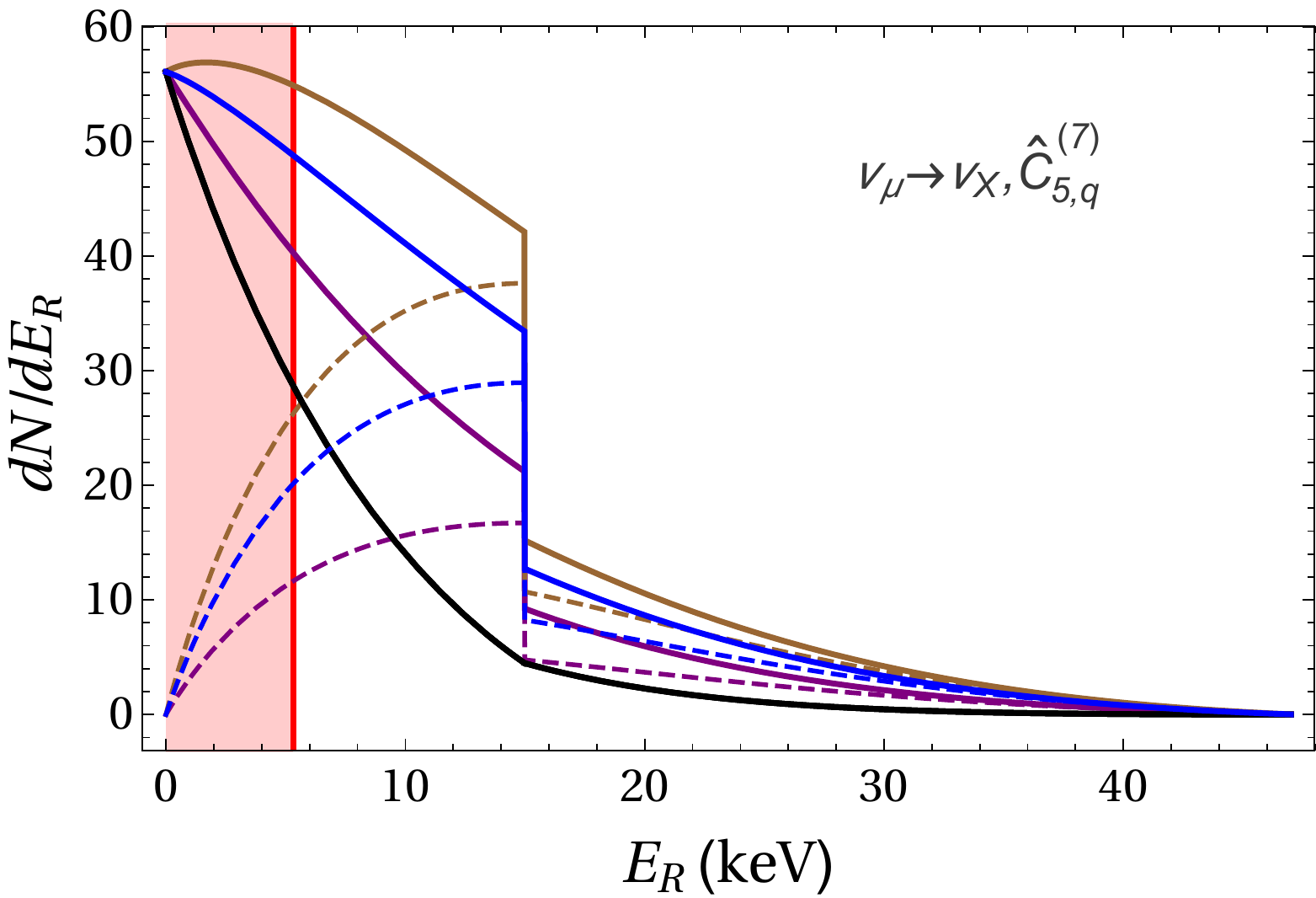}  
    \\[2mm]
       \includegraphics[scale=0.46]{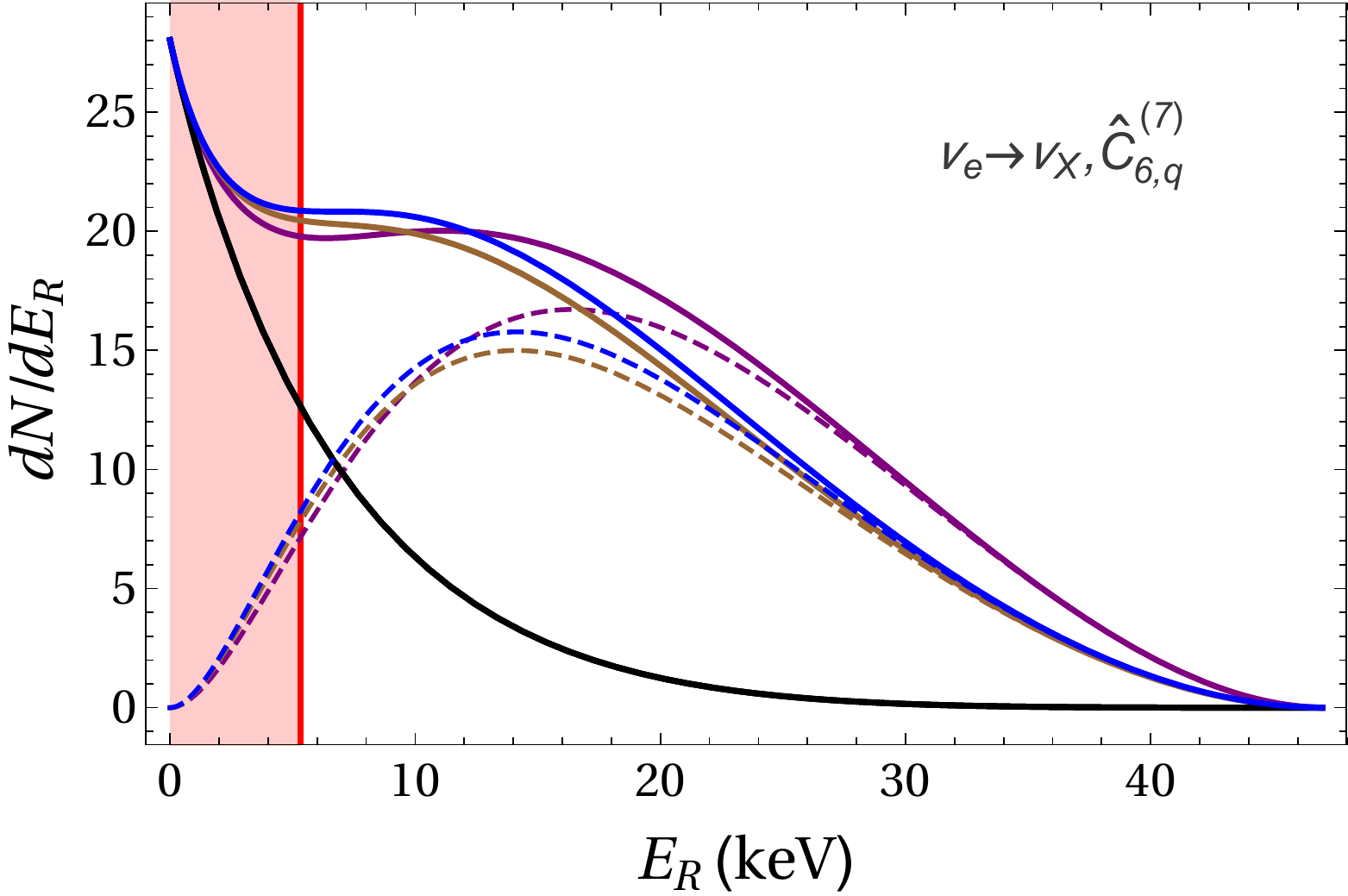}  \hspace{1cm}
    \includegraphics[scale=0.46]{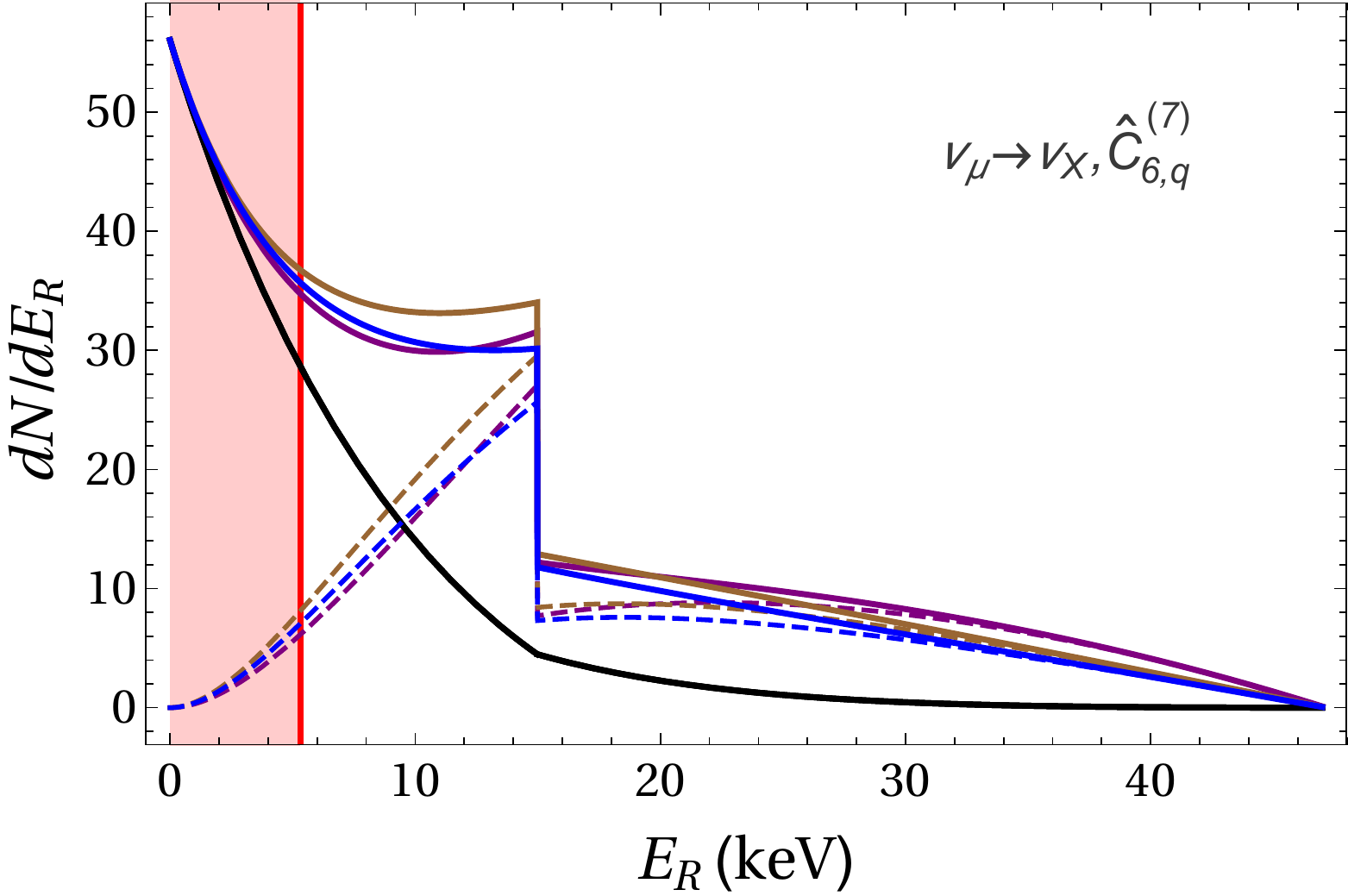} 
    \\[2mm]
        \includegraphics[scale=0.46]{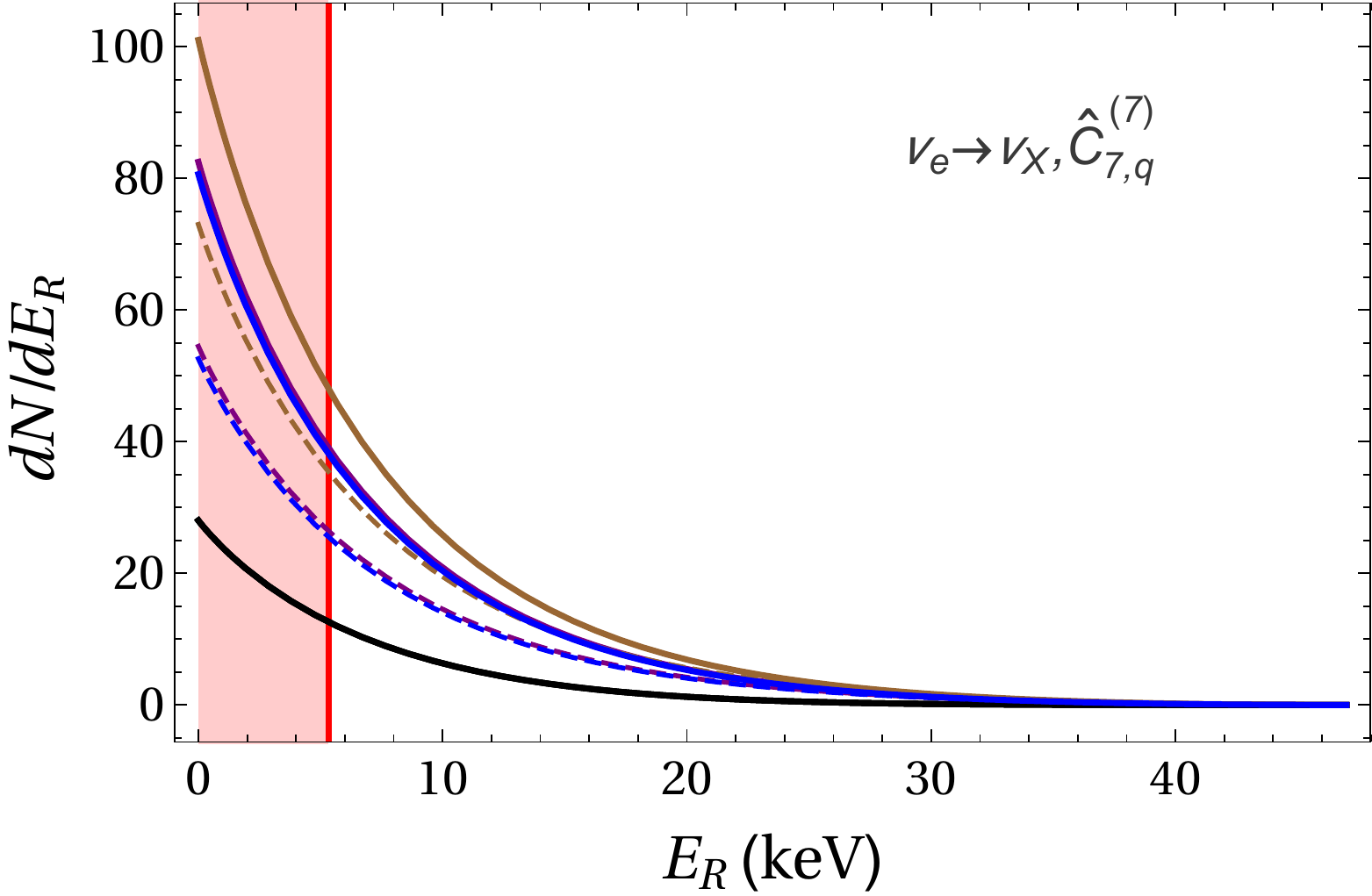}  \hspace{1cm}
    \includegraphics[scale=0.46]{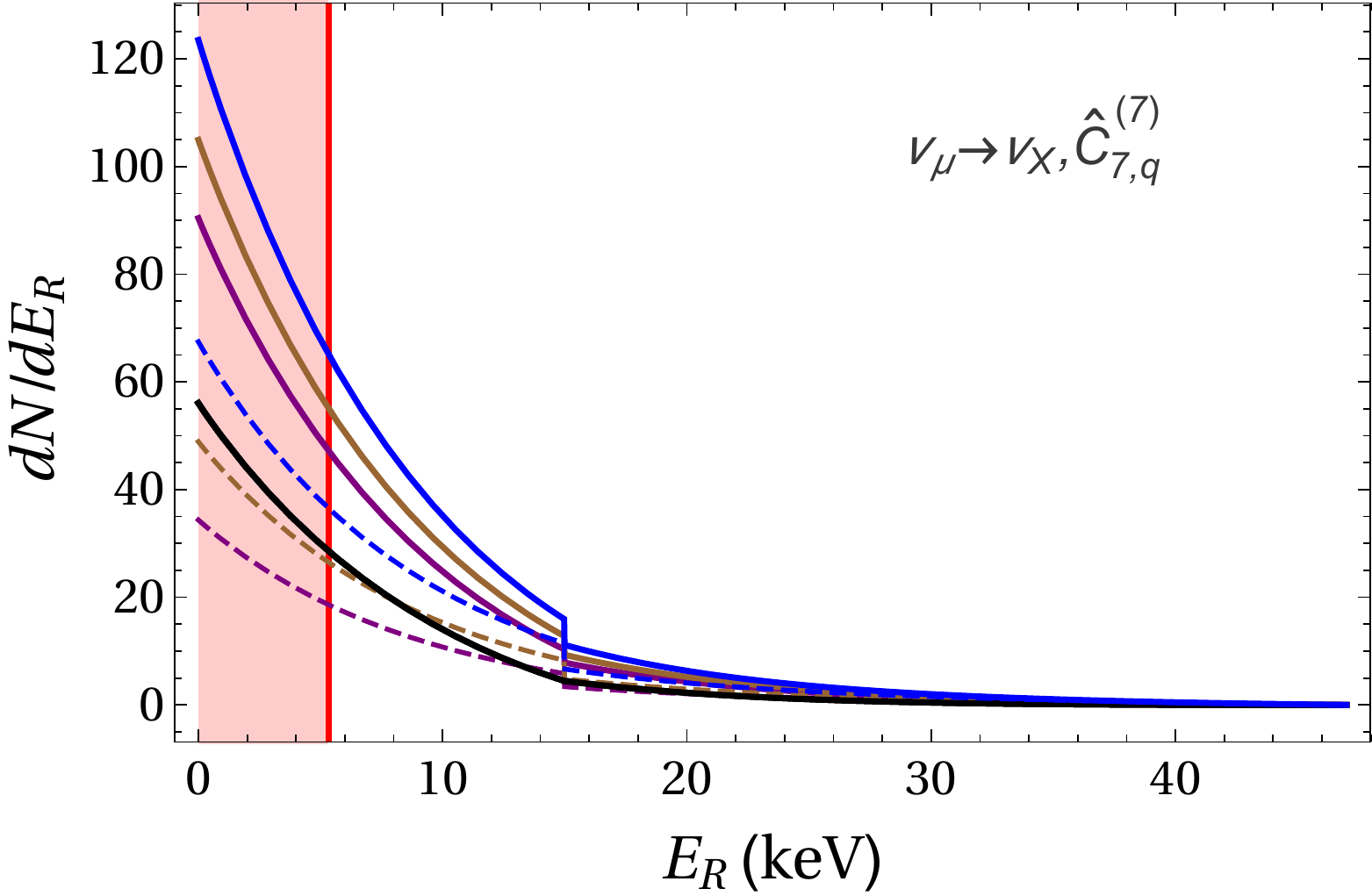} 
\caption{
The $\nu_e\to \nu_X$ (left) and $\nu_\mu\to \nu_X$ scattering rates in COHERENT CsI detector in the presence of NSI operators ${\cal Q}_{5,q}^{(7)}$, ${\cal Q}_{6,q}^{(7)}$, ${\cal Q}_{7,q}^{(7)}$, (top to bottom) setting $\hat C_a^{(7)}=1/\Lambda_{a;{\rm min}}^3$. The notation is as in Fig. \ref{fig:C16rates}.
}
\label{fig:C57:C77:rates}
\end{figure}

 The predicted differential scattering rates, $dN/dE_R$, Eq.~\eqref{eq:totN}, are plotted in Figures \ref{fig:C15rates}-\ref{fig:C57:C77:rates}. For the nuclear response functions, $W_i$, we use the values from \cite{Anand:2013yka}, while the value of nuclear form factors are taken from \cite{Bishara:2017pfq}. In Figures \ref{fig:C15rates}-\ref{fig:C57:C77:rates} we show separately the scattering rates due to the $\nu_e$ (left panels) and $\nu_\mu+\bar \nu_\mu$ (right panels) incoming neutrinos.  The corresponding fluxes are given in Eqs. \eqref{eq:phi:nue}-\eqref{eq:phi:numu}. Note from eq. \eqref{eq:c56(0)} that the operators $\Q_{8,q}^{(7)}$ and  $\Q_{10,q}^{(7)}$ have the same matrix element and will give rise to the same differential scattering rate; same happens for the operators $\Q_{9,q}^{(7)}$ and  $\Q_{11,q}^{(7)}$, see eq. \eqref{eq:c1p1}. 
 
\begin{figure}[t]
    \includegraphics[scale=0.46]{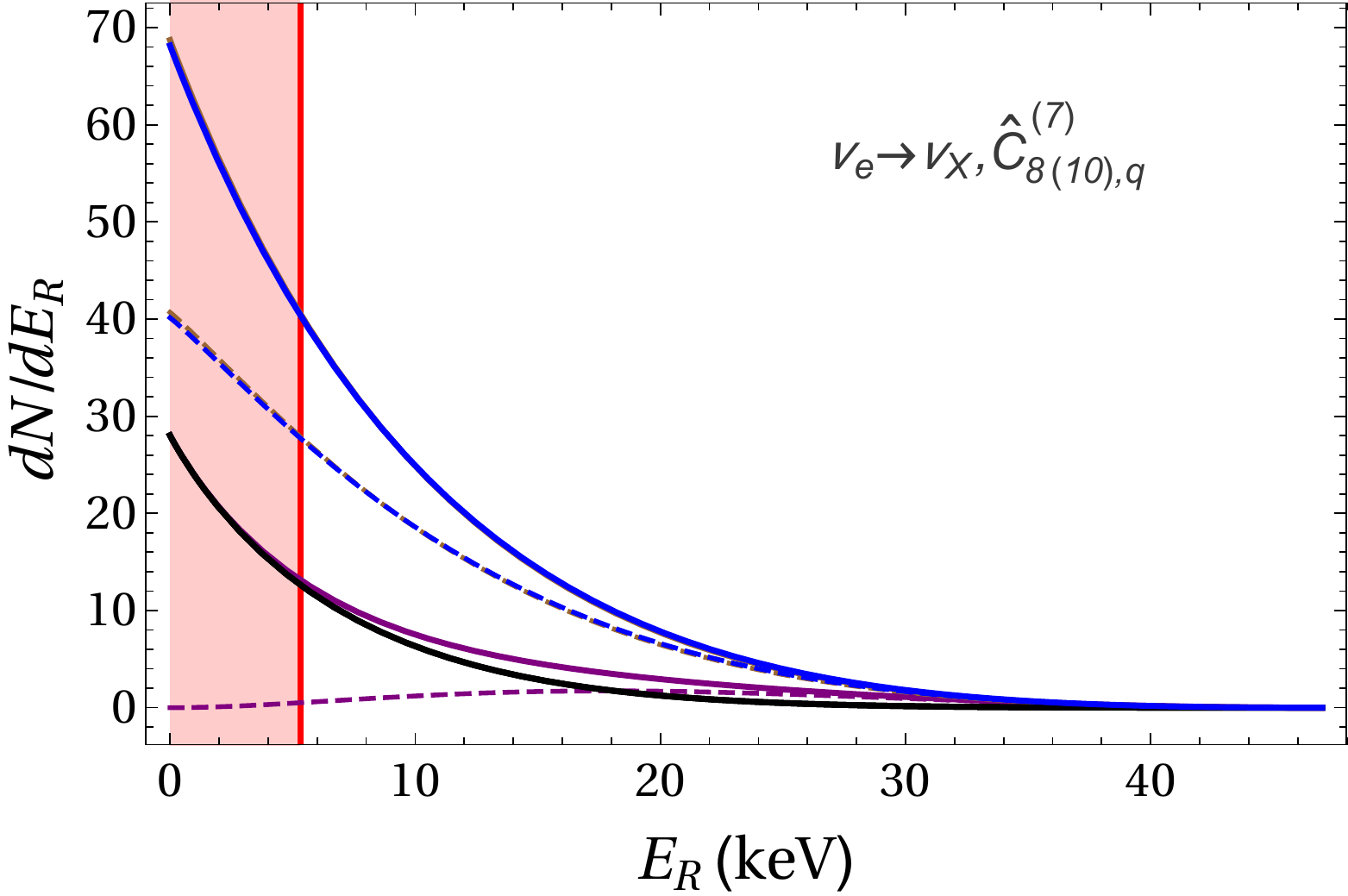}  \hspace{1cm}
    \includegraphics[scale=0.46]{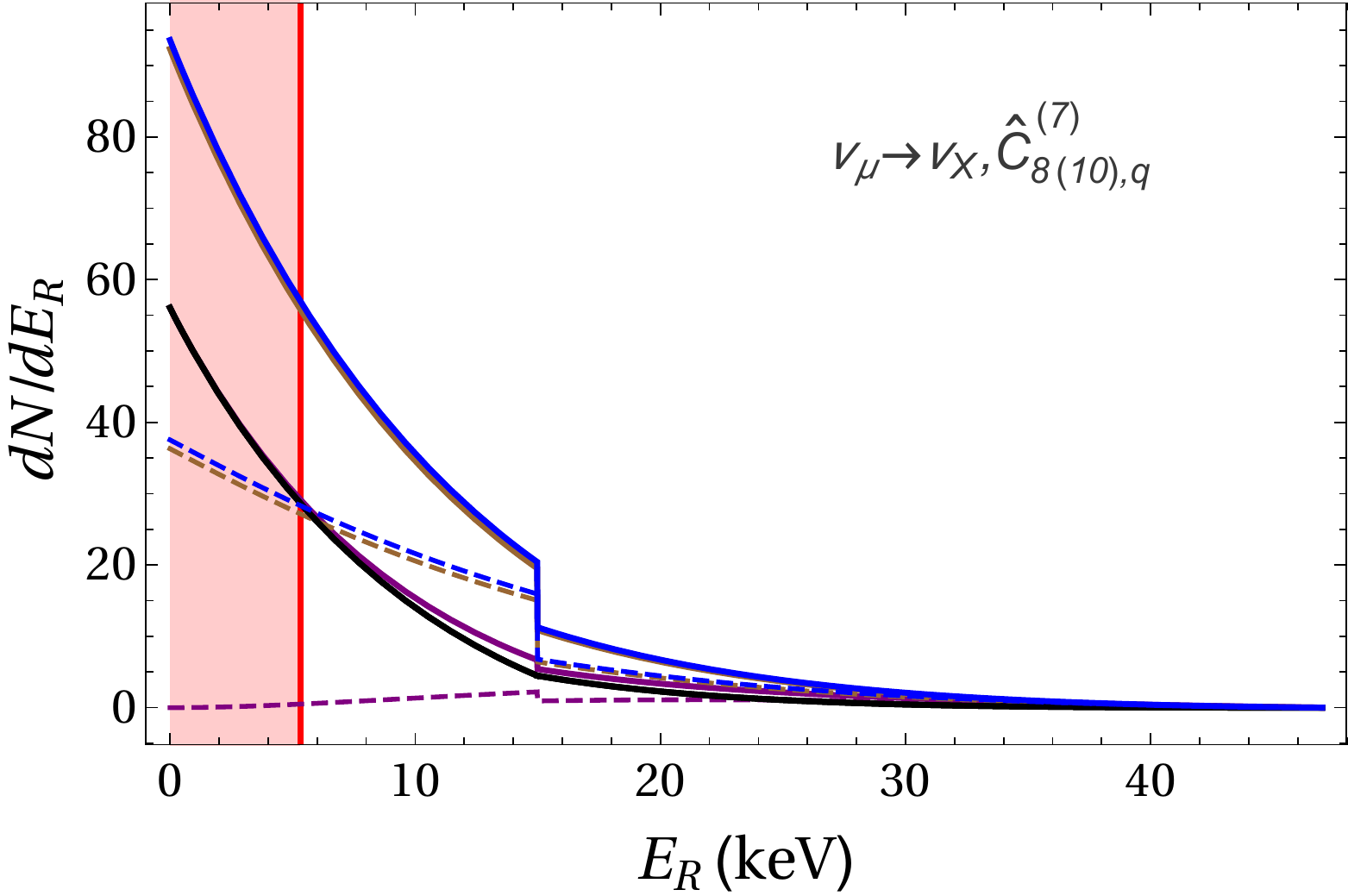}    \\[2mm]
        \includegraphics[scale=0.46]{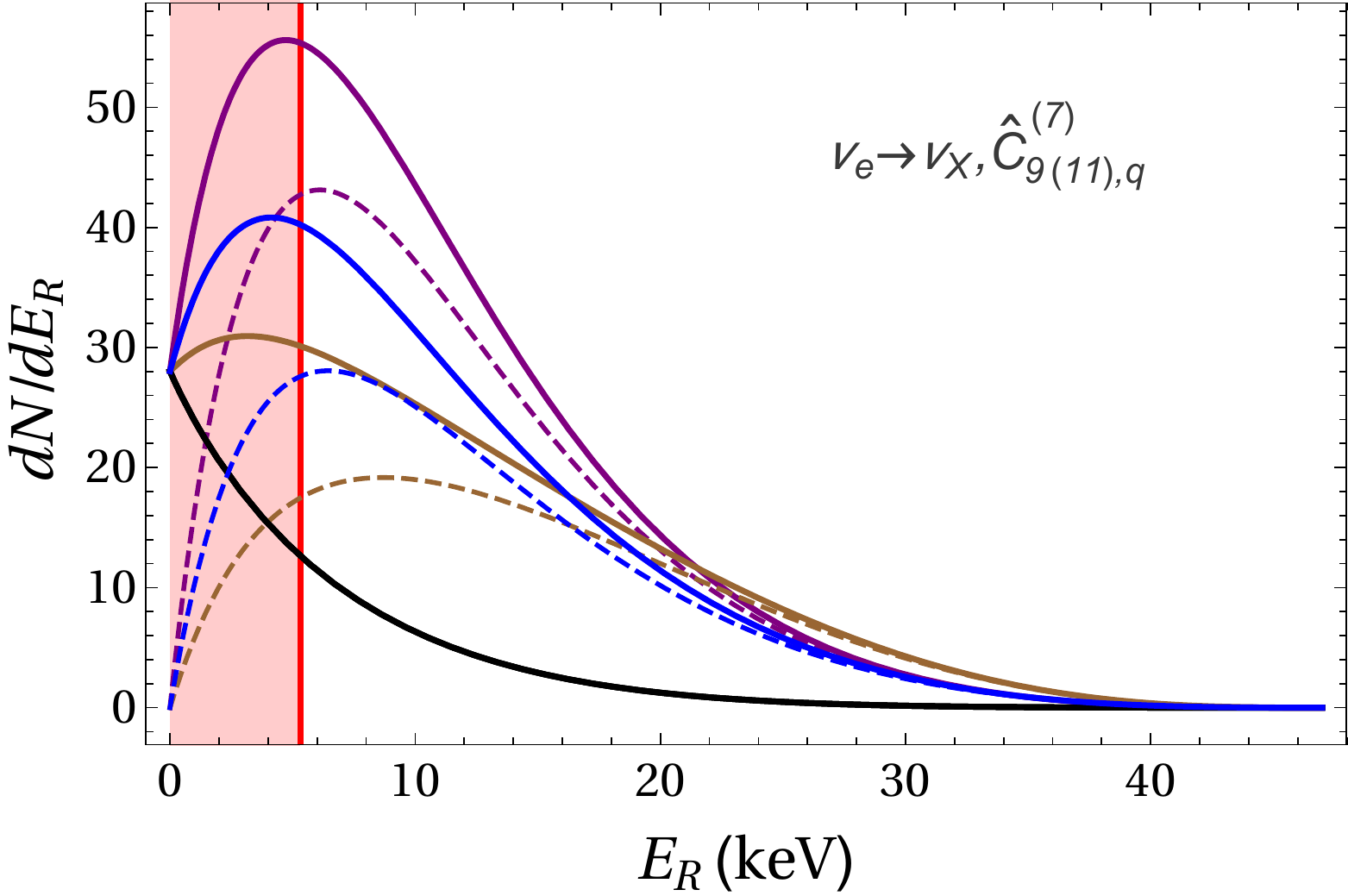}  \hspace{1cm}
    \includegraphics[scale=0.46]{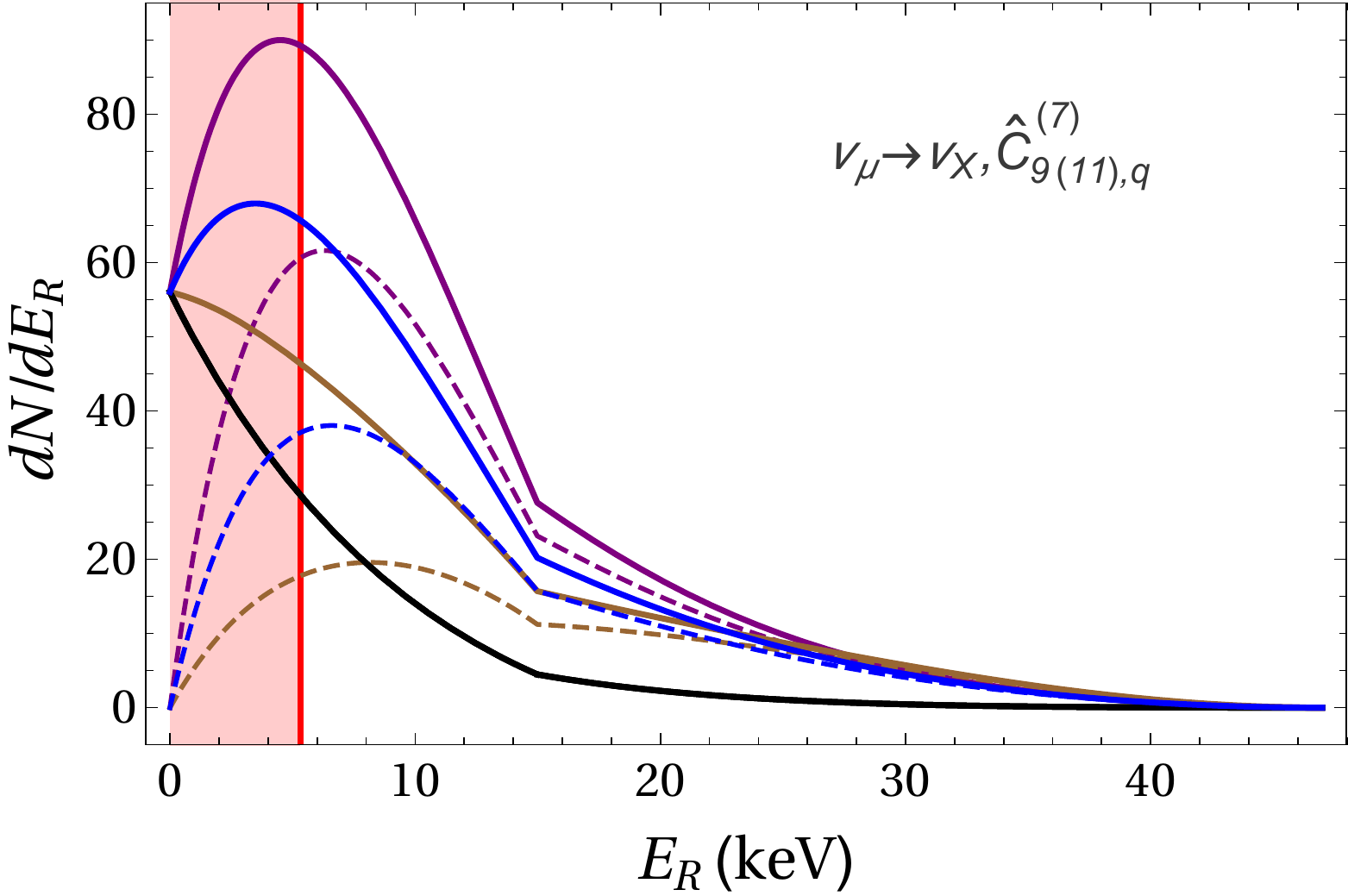} 
\caption{Same as Fig. \ref{fig:C57:C77:rates} but for ${\cal Q}_{8,q}^{(7)}$ (top) and ${\cal Q}_{9,q}^{(7)}$ (bottom).  }
\label{fig:C87rates}
\end{figure}

 The neutrinos are due to stopped muons, which sets the maximal recoil energy, $E_{R, {\rm max}}$, to be around $47$~keV for $\nu_e\to \nu_X$ and $\bar \nu_\nu\to \nu_X$, and about $15$~keV for $\nu_\mu\to \nu_X$ transitions.  The muon neutrino flux is monoenergetic, with $E_\nu\approx 30$ MeV. As a consequence, for $\bar \nu_\mu \to \nu_X$ scattering there is an abrupt drop in the predicted differential rate, $dN/dE_R$, at $E_R\approx 15$~keV, since none of the $\bar \nu_\mu$ can contribute to more energetic recoils. This discontinuity is clearly visible for $\Q_{2,q}^{(6)}$, see Fig.~\ref{fig:C16rates} bottom right, $\Q_{1}^{(7)}, \ldots, Q_{4}^{(7)}$, see Fig. \ref{fig:C17rates} right, and for $\Q_{5,q}^{(7)}, \ldots, \Q_{8(10),q}^{(7)}$, see Figures \ref{fig:C57:C77:rates} right, and Fig. \ref{fig:C87rates} upper right.
 
 For SM prediction there is no such discontinuity in $dN/dE_R$ at $E_R\approx 15$ keV, but rather only a change in the slope of $dN/dE_R$. The spin-independent scattering induced by the SM neutrino interaction with quarks contains the kinematical pre-factor $(2E_\nu^2-\vec q^2)$, see the coefficient of $c_{1,\tau}^{(0)}c_{1,\tau'}^{(0)*}$ in \eqref{eq:RSpp}. This prefactor goes to zero when when the maximal $E_R$ for given value of $E_\nu$ is reached, i.e., when the incoming neutrinos backscatter. This means that the contribution from $\bar \nu_\mu \to \nu_X$ to the SM $dN/de_R$ scattering rate goes to zero at $E_R\approx15$~keV.
 
 In order to obtain the number of events predicted in the CsI[Tl] COHERENT experiment the predictions in the left and the right panels of Figs. \ref{fig:C15rates}-\ref{fig:C57:C77:rates} need to be added up, and then convoluted with the signal acceptance fraction of the detector. At present the acceptance has a lower threshold at around 4.25 keV, denoted as a vertical red line in Figures \ref{fig:C15rates}-\ref{fig:C57:C77:rates}.

From the figures we see that a number of operators lead to a significantly different $E_R$ dependece compared to the SM predictions. The magnetic dipole moment leads to $dN/dE_R$ that has a pole at $\vec q^2=0$, clearly showing a significant increase in the rate at small values of $E_R$, see Fig. \ref{fig:C15rates}. Lowering the energy threshold can thus lead to an increase sensitivity to this NSI operator, as long as the background can be kept low. In other case probing larger recoils may be beneficial. For instance, the scattering matrix element due to $\Q_{1,s}^{(6)}$ is proportional to $F_{1}^{s/N}{}'(0) \vec q^2$ and thus grows with the increased $E_R$, see Fig. \ref{fig:C16rates} (top). A dedicated analysis is required to see to what extend this contribution can be distinguished from the subleading corrections in the SM rate that come from the $q^2$ dependence of the $F_{1}^{u/N}$ and $F_{1}^{d/N}$ form factors, and from the uncertainties in the nuclear response function $W_M$.

\clearpage

A very striking difference in the kinematical dependence of $dN/dE_R$ arises in the case of monoenergetic neutrino beams, as already mentioned above. This can be seen in Figures \ref{fig:C16rates}-\ref{fig:C87rates} (right panels).  Observing experimentally any such discontinuity would be a clear signal for the presence of NSI. The discontinuity if especially pronounced for $\Q_2^{(7)}$, $\Q_4^{(7)}$, Fig. \ref{fig:C17rates}, and $\Q_{6,q}^{(7)}$, Fig. \ref{fig:C57:C77:rates}, since 
these operators contribute to the non-relativistic operator $\op_{1,N}^{(1)}$, see Eqs. \eqref{eq:c1p1}, \eqref{eq:single:FFtilde}. This leads to an additional $\vec q^{\,4}$ dependence in the scattering rate, see Eq. \eqref{eq:RSpp}. For NSI generated by 
$\Q_2^{(7)}$, $\Q_4^{(7)}$, or $\Q_{6,q}^{(7)}$ operators the sensitivity increases for higher $E_R$ recoils. 
$\Q_2^{(7)}$, $\Q_4^{(7)}$, or $\Q_{6q}^{(7)}$ operators the sensitivity increases for higher $E_R$ recoils. This was also illustrated in the main text, in Fig. \ref{fig:HEC47}.

The operators $\Q_1^{(7)}$,  $\Q_3^{(7)}$ and $\Q_{5,q}^{(7)}$ match onto the nonrelativistic operator $\op_{3,N}^{(0)}$ which leads to a $\vec q^2$  prefactor in the scattering rate instead of the $4E_\nu^2-\vec q^2$ one for the SM, see Eq. \eqref{eq:RM}. This different $E_R$ dependence clearly shows in Figures \ref{fig:C17rates} and \ref{fig:C57:C77:rates} (left panels). Similar comment applies to the $\Q_{9,q}^{(7)}$ operator, which matches onto $\op_{2,N}^{(1)}$, which leads to a kinematic prefactor $E_R^2$ in the scattering event rate, cf. Eq. \ref{eq:RSpp}, and thus a very different different recoil dependence compared to the SM rate, see Fig. \ref{fig:C87rates}.

\section{Numerical values of \CE scattering cross sections} \label{app:ratios}
In this appendix we provide numerical expressions for \CE differential cross sections in the presence of a single nonzero NSI operator $\Q_a^{(d)}$. We normalize the NSI cross sections to the SM, so that they take the form
\beq
\label{eq:AppC:ratio}
\frac{\left( d\sigma/dE_R \right)_{\rm NSI}}{\left( d\sigma/dE_R \right)_{\rm SM}}  = 1 + g_{a}^{(d)}(E_\nu,E_R) \hat \C_{a}^{(d)} + f_{a}^{(d)}(E_\nu,E_R) \big|\hat \C_{a}^{(d)}\big|^2\,.
\eeq
These expressions are valid for any neutrino flavor scattering on nuclei, $\nu_\alpha A\to \nu_\beta A$. Here $E_\nu$ is the energy of incoming neutrino, and $E_R$ the recoil energy of the nucleus. 

Below we give the numerical values for the coefficients $g_a^{(d)}$ and $f_a^{(d)}$ for \CE on nuclei $^{23}$Na, Ge, $^{127}$I and Xe. For Germanium and Xenon we calculate the average cross sections for natural abundance of stable isotopes, namely $^{70}$Ge, $^{72}$Ge, $^{73}$Ge, $^{74}$Ge and $^{76}$Ge for Germanium and $^{128}$Xe, $^{129}$Xe, $^{130}$Xe, $^{131}$Xe, $^{132}$Xe and $^{134}$Xe for Xenon. For the numerical evaluation we use the nuclear response functions from \cite{Fitzpatrick:2012ix,Anand:2013yka}. We only quote central values for the coefficients $g_a^{(d)}$ and $f_a^{(d)}$, but comment when these estimates are particularly uncertain.  

Since only dimension 6 operators interfere with the SM amplitude for \CE, these are the only ones that have both $g_a^{(d)}$ and $f_a^{(d)}$ nonzero, while for dimension 5 and dimension 7 operators $g_a^{(d)}=0$.
are the dimension six operators. For dimension 6 operators we provide the $g_a^{(d)}$ and $f_a^{(d)}$ functions both for the NSI notation that uses the $\varepsilon$ parameters and for our notation with the canonically normalized Wilson coefficients. 
In the results we only keep the lowest order in $E_R$ in the expressions of $g_{a,q}^{(d)}(E_\nu,E_R)$ and $f_{a,q}^{(d)}(E_\nu,E_R)$. These quoted results for these functions are thus reliable for recoil energies up to $E_R \sim 10 - 20$ keV, while for higher energies one needs to take into account corrections from higher powers of $E_R$.

\subsection{Numerical results for \CE on $^{23}$Na}\label{sec:Sodium}
To shorten the notation we define the following functions of incoming neutrino energy, $E_\nu$, and nuclear recoil, $E_R$,
\begin{align}
\label{eq:Na:D}
D&= \big[(E_\nu/{\rm MeV})^2 - 10.8 (E_R/{\rm keV})\big]^{-1}, 
\\
\label{eq:Na:RA}
R_A&=
\big[(E_\nu/{\rm MeV})^2 + 10.8 (E_R/{\rm keV})\big] D, \qquad \text{for \CE on ${}^{23}$Na.}
\\
\label{eq:Na:RT}
R_T&=\big[(E_\nu/{\rm MeV})^2 - 7.97 (E_R/{\rm keV})\big] D, 
\end{align}

We first give the results for NSI parametrized by $\varepsilon_i$, Eq. \eqref{eq:LNSI}. Using $\varepsilon_i$, instead of the Wilson coefficients $\hat \C_{1,q (2,q)}^{(6)}$ in Eq. \eqref{eq:AppC:ratio}, the corresponding coefficients are
\begin{align}
g_{\varepsilon}^{uV} &=-12.2, & f_\varepsilon^{uV} &= 37.0,
\\
g_\varepsilon^{dV} &=-12.5, &f_\varepsilon^{dV} &= 39.2,
\\
g_\varepsilon^{sV} &= 3.20\cdot10^{-6}\big(\tfrac{E_R}{\rm keV}\big),
  &f_\varepsilon^{sV} &=2.56\cdot10^{-12}\big(\tfrac{E_R}{\rm keV}\big)^2,
\\
g_\varepsilon^{uA} &= 1.86 \cdot10^{-3} R_A, &f_\varepsilon^{uA} &= 1.34\cdot10^{-3}R_A,
\\  
g_\varepsilon^{dA}&= -6.53\cdot10^{-4} R_A, &f_\varepsilon^{dA}&= 1.65\cdot10^{-4}R_A, 
\\
g_\varepsilon^{sA} &= -7.19\cdot10^{-5}R_A, & f_\varepsilon^{sA} &= 2.00\cdot10^{-6}R_A,
\end{align}
where the function $R_A$ for \CE on ${}^{23}$Na is given in \eqref{eq:Na:RA}.

We give next the results for NSI induced scattering rates for dimension 5, 6 and 7 operators. For dimension 5 operator the nonzero coefficient in the \CE scattering cross section, Eq. \eqref{eq:AppC:ratio}, is for ${}^{23}$Na
\beq
f_1^{(5)}=  6.36 \cdot 10^7\, \big(\tfrac{E_\nu}{\rm MeV}\big)^2 \big(\tfrac{E_R}{\rm keV}\big)^{-1} D\,, 
\eeq
with $D$ given in \eqref{eq:Na:D}. For dimension 6 operators the coefficients in  Eq. \eqref{eq:AppC:ratio} are for \CE on ${}^{23}$Na
\begin{align}
g_ {1,u}^{(6)} &= -1.05\cdot10^6,
&g_ {1,d}^{(6)} &=  -1.08\cdot10^6,
& g_ {1,s}^{(6)} &=  0.276\, \big(\tfrac{E_R}{\rm keV}\big) \,,
\\
f_ {1,u}^{(6)} &=  2.75\cdot10^{11}\,, 
&f_ {1,d}^{(6)} &=  2.92\cdot10^{11}\,, 
&f_ {1,s}^{(6)} &=   1.90\cdot10^{-2}\big(\tfrac{E_R}{\rm keV}\big)^2\,, 
\\
g_ {2,u}^{(6)} &= 1.61\cdot10^{2} R_A \,,
&g_ {2,d}^{(6)} &= -56.3\, R_A\,, 
&g_ {2,s}^{(6)} &= -6.20 \,R_A\,, 
\\
f_ {2,u}^{(6)} &= 9.96\cdot10^{6} \,R_A\,,
&f_ {2,d}^{(6)}&= 1.22\cdot10^{6}\, R_A\,, 
&f_ {2,s}^{(6)} &= 1.49 \, R_A\,, 
\end{align}
with the $R_A$ function given in \eqref{eq:Na:RA}. For dimension 7 operators the coefficients are, for \CE on ${}^{23}$Na, given by
\begin{align}
\label{eq:f17:f27}
f_1^{(7)} &=  -26.3 \, \big(\tfrac{E_R}{\rm keV}\big)^2 D,
&f_2^{(7)} &=  2.44 \cdot 10^{-4}\, \big(\tfrac{E_R}{\rm keV}\big)^2 D, &&
\\
f_3^{(7)} &=  1.34 \cdot 10^9\, \big(\tfrac{E_R}{\rm keV}\big) D\,,
&f_4^{(7)} &=  1.05 \cdot 10^{-3}\, \big(\tfrac{E_R}{\rm keV}\big)^4 D\,, &&
\\
f_ {5,u}^{(7)} &= 8.65\cdot10^7\big(\tfrac{E_R}{\rm keV}\big) D, 
&f_ {5,d}^{(7)} &= 3.95\cdot10^{8}\big( \tfrac{E_R}{\rm keV} \big) D, 
&f_ {5,s}^{(7)} &= 5.79\cdot 10^{8} \big( \tfrac{E_R}{\rm keV} \big) D, 
\\
f_ {6,u}^{(7)} &= 77.3 \big(\tfrac{E_R}{\rm keV}\big)^2 D, 
&f_ {6,d}^{(7)} &= 3.67\cdot10^2\big(\tfrac{E_R}{\rm keV}\big)^2D, 
&f_ {6,s}^{(7)} &= 59.7 \big(\tfrac{E_R}{\rm keV}\big)^2D, 
\\
f_ {7,u}^{(7)} &=2.79\cdot10^2 \,R_T,
&f_ {7,d}^{(7)} &=80.3 \, R_T , 
&f_ {7,s}^{(7)} &= 7.74 \cdot10^{-2} \,R_T, 
\end{align}
while for operators with derivatives on neutrino currents the coefficients are,
\begin{align}
f_ {8(10),u}^{(7)} &= 6.88\cdot10^{10} \big(\tfrac{E_\nu}{\rm MeV}\big)^2 D, 
&f_ {9(11),u}^{(7)} &= 4.30\cdot10^2 \big(\tfrac{E_R}{\rm keV}\big), 
\\
f_ {8(10),d}^{(7)} &=7.29\cdot10^{10} \big(\tfrac{E_\nu}{\rm MeV}\big)^2D, 
&f_ {9(11),d}^{(7)} &= 52.8 \big(\tfrac{E_R}{\rm keV}\big), 
\\
f_ {8(10),s}^{(7)} &= 1.90\cdot10^{-2} \big(\tfrac{E_\nu}{\rm MeV}\big)^2 \big(\tfrac{E_R}{\rm keV}\big)^2 D, 
&f_ {9(11),s}^{(7)} &= 0.642 \big(\tfrac{E_R}{\rm keV}\big). 
\end{align}
The $g$ coefficients are zero for all dimension 7 operators. The $D$ and $R_T$ functions are given in Eqs. \eqref{eq:Na:D} and \eqref{eq:Na:RT}, respectively. Note that for the Rayleigh operators in Eq. \eqref{eq:f17:f27} we used the NDA estimates from Section \ref{sec:Rayleigh:scatt}, which are only very approximate.

\subsection{Numerical results for \CE on Ge}\label{sec:Germanium}
For Germanium we give cross section for natural abundances of Ge in the detector. To shorten the notation we define the following three functions,
\begin{align}
\label{eq:D:Ge}
D&=\big[(E_\nu/{\rm MeV})^2 - 34.2 (E_R/{\rm keV})\big]^{-1}, 
\\
\label{eq:RA:Ge}
R_A&= \big[(E_\nu/{\rm MeV})^2 + 34.2 (E_R/{\rm keV})\big]D, \qquad \text{for \CE on Ge},
\\
\label{eq:RT:Ge}
R_T&=\big[ \big({E_\nu}/{\rm MeV}\big)^2 - 25.3 \big({E_R}/{\rm keV}\big) \big] D.
\end{align}

 We start with the results for NSI parametrized by $\varepsilon_i$, Eq. \eqref{eq:LNSI}. Using $\varepsilon_i$ instead of  $\hat \C_{1,q (2,q)}^{(6)}$ in Eq. \eqref{eq:AppC:ratio}, the corresponding coefficients for \CE on Ge are given by
\begin{align}
g_ {\varepsilon}^{uV} &= -10.9, &f_ {\varepsilon}^{uV} &=  29.7,
\\
g_ {\varepsilon}^{dV} &=-11.8, &f_ {\varepsilon}^{dV} &=  35.0,
\\
g_ {\varepsilon}^{sV} &= 9.33\cdot10^{-6} \big(\tfrac{E_R}{\rm keV}\big), &f_ {\varepsilon}^{sV} &= 2.18\cdot10^{-11} \big(\tfrac{E_R}{\rm keV}\big)^2, 
\\
g_ {\varepsilon}^{uA} &= -8.19\cdot10^{-5} R_A,
&f_ {\varepsilon}^{uA} &=-2.42 \cdot10^{-5} R_A,
\\
g_ {\varepsilon}^{dA} &=2.02 \cdot10^{-4} R_A, 
& f_ {\varepsilon}^{dA}&=-1.46 \cdot10^{-4} R_A, 
\\
g_ {\varepsilon}^{sA} &= -7.15\cdot10^{-6} R_A, \qquad
&f_ {\varepsilon}^{sA}&= -1.85  \cdot10^{-7} R_A.
\end{align}

Next, we give the coefficients in the expression for cross section Eq. \eqref{eq:AppC:ratio} using our notation for the NSI operators. The dimension 5 magnetic dipole operator does not interfere with the SM contribution, and thus only has the quadratic term nonzero. For \CE on Ge we have 
\beq
f_1^{(5)} =  1.44 \cdot 10^7\, \big(\tfrac{E_\nu}{\rm MeV}\big)^2 \big(\tfrac{E_R}{\rm keV}\big)^{-1} D,
\eeq
with $D$ given in \eqref{eq:D:Ge}. For dimension 6 operators the cross section coefficients for \CE on Ge  are given by
\begin{align}
g_ {1,u}^{(6)} &=-9.40 , &f_ {1,u}^{(6)} &=  2.21\cdot10^{11}, 
\\
g_ {1,d}^{(6)} &=  -1.02\cdot10^6, &f_ {1,d}^{(6)} &=  2.60\cdot10^{11}, 
\\
g_ {1,s}^{(6)} &=  0.804 \big(\tfrac{E_R}{\rm keV}\big),
& f_ {1,s}^{(6)} &=  1.62\cdot10^4 \big(\tfrac{E_R}{\rm keV}\big)^2, 
\\
g_ {2,u}^{(6)} &= -7.06 R_A, 
&f_ {2,u}^{(6)} &= -1.80 \cdot 10^5 R_A, 
\\
g_ {2,d}^{(6)} &= 17.4 R_A, 
&f_ {2,d}^{(6)}&= -1.10 \cdot 10^6 R_A, 
\\
g_ {2,s}^{(6)} &= -0.616 R_A, 
& f_ {2,s}^{(6)} & = -1.37 \cdot 10^3 R_A,
\end{align}
where the $R_A$ is given in \eqref{eq:RA:Ge}.

For dimension 7 operators only the $f_a^d$ coefficients in \CE cross section expression are nonzero. For \CE on Ge they are given by
\begin{align}
\label{eq:Rayleigh:Ge}
f_1^{(7)} &=  1.36 \cdot 10^2\, \big(\tfrac{E_R}{\rm keV}\big)^2 D,
&f_2^{(7)} &=  2.27 \cdot 10^{-4}\, \big(\tfrac{E_R}{\rm keV}\big)^2 D, &&
\\
f_3^{(7)} &=  3.60 \cdot 10^9\, \big(\tfrac{E_R}{\rm keV}\big) D,
&f_4^{(7)} &=  1.54 \cdot 10^{-3}\, \big(\tfrac{E_R}{\rm keV}\big)^4 D, &&
\\
f_ {5,u}^{(7)} &= 2.30 \cdot10^8 \big(\tfrac{E_R}{\rm keV}\big) D, 
&f_ {5,d}^{(7)} &= 1.07 \cdot10^9 \big(\tfrac{E_R}{\rm keV}\big) D, 
&f_ {5,s}^{(7)} &= 1.56 \cdot 10^9 \big(\tfrac{E_R}{\rm keV}\big) D, 
\\
f_ {6,u}^{(7)} &= 71.8 \big(\tfrac{E_R}{\rm keV}\big)^2 D,
&f_ {6,d}^{(7)} &=3.41 \cdot10^{2} \big(\tfrac{E_R}{\rm keV}\big)^2 D, 
&f_ {6,s}^{(7)} &= 55.5 \big(\tfrac{E_R}{\rm keV}\big)^2D\,, 
\\
f_ {7,u}^{(7)} &=25.5 R_T,
&f_ {7,d}^{(7)} &=7.41 R_T,
& f_ {7,s}^{(7)} &= 7.14\cdot10^{-3} R_T,
\end{align}
while for dimension 7 operators with derivatives in the neutrino current, the coefficients are
\begin{align}
f_ {8(10),u}^{(7)} &= 5.53 \cdot10^{10} \big(\tfrac{E_\nu}{\rm MeV}\big)^2 D, 
&f_ {9(11),u}^{(7)} &= 24.6 \big(\tfrac{E_R}{\rm keV}\big), 
\\
f_ {8(10),d}^{(7)} &=6.50 \cdot10^{10} \big(\tfrac{E_\nu}{\rm MeV}\big)^2D,
& f_ {9(11),d}^{(7)} &=1.50\cdot10^2\big(\tfrac{E_R}{\rm keV}\big), 
\\
f_ {8(10),s}^{(7)} &= 0.162 \big(\tfrac{E_\nu}{\rm MeV}\big)^2 \big(\tfrac{E_R}{\rm keV}\big)^2 D,
& f_ {9(11),s}^{(7)} &=0.188 \big(\tfrac{E_R}{\rm keV}\big). 
\end{align}
The $D$ and $R_T$ functions for \CE on Ge are given in \eqref{eq:D:Ge} and \eqref{eq:RT:Ge}, respectively. Note that for the Rayleigh operators in Eq. \eqref{eq:Rayleigh:Ge} we used the NDA estimates from Section \ref{sec:Rayleigh:scatt}, which are only very approximate.

\subsection{Numerical results for \CE on $^{127}$I}\label{sec:Iodine}
In order to shorten the notation we define functions,
\begin{align}
\label{eq:D:I}
D&= \big[(E_\nu/{\rm MeV})^2 - 59.6 (E_R/{\rm keV})\big]^{-1},  
\\
\label{eq:RA:I}
R_A&= \big[(E_\nu/{\rm MeV})^2 + 59.6 (E_R/{\rm keV})\big]D, \qquad \text{for \CE on $^{127}$I},
\\
\label{eq:RT:I}
R_T&=\big[(E_\nu/{\rm MeV})^2 -44.0 (E_R/{\rm keV})\big]D. 
\end{align}

The cross section coefficients in Eq. \eqref{eq:AppC:ratio} for \CE on $^{127}$I for dimension 6 NSI operators, using the $\varepsilon_i$ notation, are given by 
\begin{align}
g_ {\varepsilon}^{uV} &=-10.3, &f_ {\varepsilon}^{uV} &= 26.4, \\
g_ {\varepsilon}^{dV} &=-11.5, &f_ {\varepsilon}^{dV} &= 33.0, \\
g_ {\varepsilon}^{sV} &= 1.56\cdot10^{-5}\big(\tfrac{E_R}{\rm keV}\big), 
&f_ {\varepsilon}^{sV} &= 6.05\cdot10^{-11}\big(\tfrac{E_R}{\rm keV}\big)^2\,, 
\\
g_ {\varepsilon}^{uA} &= 3.47\cdot10^{-5} R_A,
&f_ {\varepsilon}^{uA} &= 3.47\cdot10^{-5}R_A, 
\\ 
g_ {\varepsilon}^{dA} &= 6.69\cdot10^{-4} R_A, 
&f_ {\varepsilon}^{dA} &=-2.89 \cdot10^{-4} R_A, 
\\
g_ {\varepsilon}^{sA} &=  1.20 \cdot10^{-5} R_A,
&f_ {\varepsilon}^{sA} &=- 9.22 \cdot10^{-8} R_A,
\end{align}
with $R_A$ for \CE on $^{127}$I given in \eqref{eq:RA:I}.

For dimension 5 operator we have, for \CE on $^{127}$I,
\beq
f_1^{(5)} =  6.80 \cdot 10^6\, \big(\tfrac{E_\nu}{\rm MeV}\big)^2 \big(\tfrac{E_R}{\rm keV}\big)^{-1} D\,, 
\eeq
and $g_1^{(5)}=0$, while for dimension 6 operators the NSI cross section coefficients are given by
\begin{align}
g_ {1,u}^{(6)} &= -8.87\cdot10^5, &f_ {1,u}^{(6)} &=  1.96\cdot10^{11}, 
\\
g_ {1,d}^{(6)} &=  -9.90\cdot10^5, &f_ {1,d}^{(6)} &=  2.45\cdot10^{11}, 
\\
g_ {1,s}^{(6)} &=  1.34 \big(\tfrac{E_R}{\rm keV}\big),
&f_ {1,s}^{(6)} &=   0.449\big(\tfrac{E_R}{\rm keV}\big)^2, 
\\
g_ {2,u}^{(6)} &= 75.0 R_A, &f_ {2,u}^{(6)} &= 3.63\cdot10^{6} R_A, \\
g_ {2,d}^{(6)} &= -57.7 R_A,& f_ {2,d}^{(6)}&= 2.15\cdot10^{6} R_A, \\
g_ {2,s}^{(6)} &= -1.03 R_A, &f_ {2,s}^{(6)} &= 6.854\cdot10^{2} R_A, 
\end{align}
with function $R_A$ for \CE on $^{127}$I given in \eqref{eq:RA:I}. For dimension 7 operators the cross section coefficients for \CE on $^{127}$I are given by
\begin{align}
f_1^{(7)} &=  1.30 \cdot 10^3\, \big(\tfrac{E_R}{\rm keV}\big)^2 D,
&f_2^{(7)} &=  2.41 \cdot 10^{-4}\, \big(\tfrac{E_R}{\rm keV}\big)^2 D, &&
\\
f_3^{(7)} &=  5.75 \cdot 10^9\, \big(\tfrac{E_R}{\rm keV}\big) D,
&f_4^{(7)} &=  2.67\cdot10^{-2}\, \big(\tfrac{E_R}{\rm keV}\big)^4 D, &&
\\
f_ {5,u}^{(7)} &= 3.65\cdot10^{8}\big(\tfrac{E_R}{\rm keV}\big) D,
&f_ {5,d}^{(7)} &= 1.72\cdot10^{9}\big(\tfrac{E_R}{\rm keV}\big) D, 
&f_ {5,s}^{(7)} &= 2.48\cdot 10^{9} \big(\tfrac{E_R}{\rm keV}\big) D, 
\\
f_ {6,u}^{(7)} &= 76.4\big(\tfrac{E_R}{\rm keV}\big)^2 D,
&f_ {6,d}^{(7)} &=5.16\cdot10^{2}\big(\tfrac{E_R}{\rm keV}\big)^2D, 
&f_ {6,s}^{(7)} &= 83.9\big(\tfrac{E_R}{\rm keV}\big)^2 D, 
\\
f_ {7,u}^{(7)} &= 9.06 R_T, 
& f_ {7,d}^{(7)} &= 2.61 R_T, 
& f_ {7,s}^{(7)} &= 2.51\cdot10^{-3} R_T, 
\end{align}
while for dimension 7 operators with derivatives in the neutrino current, the coefficients are
\begin{align}
f_ {8(10),u}^{(7)} &= 4.91\cdot10^{10} \big(\tfrac{E_\nu}{\rm MeV}\big)^2 D, 
&f_ {9(11),u}^{(7)} &= 49.6 \big(\tfrac{E_R}{\rm keV}\big), 
\\
f_ {8(10),d}^{(7)} &=6.12\cdot10^{10} \big(\tfrac{E_\nu}{\rm MeV}\big)^2D, 
&f_ {9(11),d}^{(7)} &=1.81\big(\tfrac{E_R}{\rm keV}\big), 
\\
f_ {8(10),s}^{(7)} &= 0.449 \big(\tfrac{E_\nu}{\rm MeV}\big)^2 \big(\tfrac{E_R}{\rm keV}\big)^2 D, 
&f_ {9(11),s}^{(7)} &= 0.115 \big(\tfrac{E_R}{\rm keV}\big), 
\end{align}
The $D$ and $R_T$ functions for \CE on $^{127}$I are given in \eqref{eq:D:I} and \eqref{eq:RT:I}, respectively.

\subsection{Numerical results for \CE on Xe}\label{sec:Xenon}
Here we show the numerical results for NSI induced corrections to \CE on Xe targets, assuming natural abundances of Xe isotopes (for numerical results of cross section on single isotopes, see \cite{Pirinen:2018gsd}). To shorten the expressions we define three functions for incoming neutrino energy, $E_\nu$, and nuclear recoil energy, $E_R$, 
\begin{align}
\label{eq:D:Xe}
D&=\big[(E_\nu/{\rm MeV})^2 - 61.3 (E_R/{\rm keV})\big]^{-1},
\\
\label{eq:RA:Xe}
R_A&=\big[(E_\nu/{\rm MeV})^2 + 61.3 (E_R/{\rm keV})\big] D, \qquad \text{for \CE on Xe},
\\
\label{eq:RT:Xe}
R_T&=\big[(E_\nu/{\rm MeV})^2 - 44.9 (E_R/{\rm keV})\big] D.
\end{align}

For dimension 6 operators we have, using $\varepsilon_i$, notation for \CE on Xe, 
\begin{align}
g_ {\varepsilon}^{uV} &= -10.2, &f_ {\varepsilon}^{uV} &= 25.9, 
\\
g_ {\varepsilon}^{dV} &= -11.4, &f_ {\varepsilon}^{dV} &= 32.6, 
\\
g_ {\varepsilon}^{sV} &= 1.59\cdot10^{-5} \big(\tfrac{E_R}{\rm keV}),
&f_ {\varepsilon}^{sV} &= 6.31\cdot10^{-11} \big(\tfrac{E_R}{\rm keV})^2,
\\ 
g_ {\varepsilon^{uA}} &= 1.48\cdot10^{-5} R_A, &f_ {\varepsilon^{uA}} &= 4.32\cdot10^{-5} R_A, \\
g_ {\varepsilon^{dA}} &= -3.74\cdot10^{-5} R_A, & f_ {\varepsilon^{dA}} &= 2.75\cdot10^{-5} R_A, \\
g_ {\varepsilon^{sA}} &= 1.35\cdot10^{-6} R_A, & f_ {\varepsilon^{sA}} &= 3.56\cdot10^{-8} R_A, 
\end{align}
with the $R_A$ function given in \eqref{eq:RA:Xe}.

For dimension 5 operator the only nonzero coefficient is for quadratic dependence on the Wilson coefficient. For \CE on Xe we have
\beq
f_1^{(5)} =  6.38 \cdot 10^6\, \big(\tfrac{E_\nu}{\rm MeV}\big)^2 \big(\tfrac{E_R}{\rm keV}\big)^{-1} D\,,
\eeq
with the $D$ function given in \eqref{eq:D:Xe}. For dimension 6 operators the \CE on Xe cross section coefficients are given by 
\begin{align}
g_ {1,u}^{(6)} &= -8.77\cdot10^5, & f_ {1,u}^{(6)} &=  1.92\cdot10^{11}, 
\\
g_ {1,d}^{(6)} &= -9.85\cdot10^5, &f_ {1,d}^{(6)} &= 2.42\cdot10^{11}, 
\\
g_ {1,s}^{(6)} &=  1.37 \big(\tfrac{E_R}{\rm keV}\big),
&f_ {1,s}^{(6)} &=  0.469 \big(\tfrac{E_R}{\rm keV}\big)^2, 
\\
g_ {2,u}^{(6)} &= 1.28 R_A, & f_ {2,u}^{(6)} &= 3.21\cdot 10^4 R_A, \\
g_ {2,d}^{(6)} &= -3.23 R_A, & f_ {2,d}^{(6)}&= 2.05 \cdot 10^5  R_A, \\
g_ {2,s}^{(6)} &= 0.116 R_A, &f_ {2,s}^{(6)} &= 2.64 \cdot 10^5 R_A, 
\end{align}
where the $R_A$ function is given in \eqref{eq:RA:Xe}. For dimension 7 operators only the $f_a^d$ coefficients are nonzero, 
\begin{align}
f_1^{(7)} &=  2.08 \cdot 10^3\, \big(\tfrac{E_R}{\rm keV}\big)^2 D, 
&f_2^{(7)} &=  1.38 \cdot 10^{-4}\, \big(\tfrac{E_R}{\rm keV}\big)^2 D, &&
\\
f_3^{(7)} &=  5.82 \cdot 10^9\, \big(\tfrac{E_R}{\rm keV}\big) D,
&f_4^{(7)} &=  3.08 \cdot 10^{-3}\, \big(\tfrac{E_R}{\rm keV}\big)^4 D, &&
\\
f_ {5,u}^{(7)} &= 3.69 \cdot10^8 \big(\tfrac{E_R}{\rm keV}\big) D, 
&f_ {5,d}^{(7)} &= 1.74 \cdot10^9 \big(\tfrac{E_R}{\rm keV}\big) D, 
& f_ {5,s}^{(7)} &= 2.52 \cdot 10^9 \big(\tfrac{E_R}{\rm keV}\big) D, 
\\
f_ {6,u}^{(7)} &= 43.7 \big(\tfrac{E_R}{\rm keV}\big)^2 D, 
&f_ {6,d}^{(7)} &= 2.08\cdot10^{2} \big(\tfrac{E_R}{\rm keV}\big)^2D, 
&f_ {6,s}^{(7)} &= 33.8 \big(\tfrac{E_R}{\rm keV}\big)^2D, 
\\
f_ {7,u}^{(7)} &=4.97 R_T, 
&f_ {7,d}^{(7)} &=1.43 R_T, 
&f_ {7,s}^{(7)} &= 1.28 \cdot10^{-3}R_T, 
\end{align}
while the operators with derivatives on the neutrino current have the following coefficients for \CE on Xe
\begin{align}
f_ {8(10),u}^{(7)} &= 4.80 \cdot10^{10} \big(\tfrac{E_\nu}{\rm MeV}\big)^2 D\,, &f_ {9(11),u}^{(7)} &= 7.81 \big(\tfrac{E_R}{\rm keV}\big)\,, \\
f_ {8(10),d}^{(7)} &= 6.06 \cdot10^{10} \big(\tfrac{E_\nu}{\rm MeV}\big)^2D\,, &f_ {9(11),d}^{(7)} &=49.8 \big(\tfrac{E_R}{\rm keV}\big)\,, \\
f_ {8(10),s}^{(7)} &= 0.469 \big(\tfrac{E_\nu}{\rm MeV}\big)^2 \big(\tfrac{E_R}{\rm keV}\big)^2 D\,, & f_ {9(11),s}^{(7)} &=6.44\cdot10^{-2} \big(\tfrac{E_R}{\rm keV}\big).
\end{align}

\end{appendix}
\bibliography{neutrino_paper}

\end{document}